\documentclass[journal]{IEEEtran}
\IEEEoverridecommandlockouts
\usepackage{lineno}
\usepackage{cite}
\usepackage{hyperref}
\usepackage{amsmath,amssymb,amsfonts}
\usepackage{amsmath}
\usepackage{amsthm}
\DeclareMathOperator*{\argmax}{argmax}

\usepackage{graphicx}
\usepackage{textcomp}
\usepackage{xcolor}
\usepackage{graphicx}
\usepackage{float}
\usepackage{subfigure}
\usepackage{amsmath}
\usepackage{amsfonts,amssymb}
\usepackage{mathrsfs}
\usepackage{mathtools}
\usepackage{algorithm}
\usepackage{algorithmicx}
\usepackage{algpseudocode}
\usepackage{bm}
\usepackage{multirow}
\usepackage{array}
\usepackage{amssymb}
\usepackage{amsmath}
\usepackage{cite}
\usepackage{url}
\usepackage{xcolor}
\usepackage{cite,graphicx,amsmath,amssymb}
\usepackage{subfigure}
\usepackage{fancyhdr}
\usepackage{mdwmath}
\usepackage{mdwtab}
\usepackage{caption}
\usepackage{amsthm}
\usepackage{setspace}
\usepackage{bm}
\usepackage{algorithm}
\usepackage{algpseudocode}
\usepackage{mathtools}
\usepackage{dsfont}
\usepackage{bbm}
\usepackage{cases}
\usepackage{stfloats}
\usepackage{bbding}
\usepackage{circledsteps}

\newtheorem{remark}{Remark}
\theoremstyle{definition}
\newtheorem{theorem}{Theorem}

\newtheorem{lemma}{Lemma}

\newtheorem{corollary}{Corollary}

\makeatletter
\newcommand{\biggg}{\bBigg@{3}}
\newcommand{\Biggg}{\bBigg@{3.5}}
\makeatother
\usepackage{pifont}
\theoremstyle{definition}
\theoremstyle{definition}
\usepackage{xcolor}
\usepackage{bbm}

\makeatletter
\begin{document}
\title{A Primer on Near-Field Communications for Next-Generation Multiple Access}
\author{Chongjun Ouyang, \IEEEmembership{Member, IEEE}, Zhaolin Wang, \IEEEmembership{Graduate Student Member, IEEE}, Yan Chen, \IEEEmembership{Member, IEEE}, Xidong Mu, \IEEEmembership{Member, IEEE}, and Peiying Zhu, \IEEEmembership{Fellow, IEEE}\\
\vspace{0.2cm}
\emph{(Invited Paper)}
\thanks{Chongjun Ouyang is with the School of Electrical and Electronic Engineering, College of Engineering and Architecture, University College Dublin, D04 V1W8, Ireland, and also with the School of Electronic Engineering and Computer Science, Queen Mary University of London, London, E1 4NS, U.K. (e-mail: chongjun.ouyang@ucd.ie).}
\thanks{Zhaolin Wang is with the School of Electronic Engineering and Computer Science, Queen Mary University of London, London, E1 4NS, U.K. (email: zhaolin.wang@qmul.ac.uk).}
\thanks{Xidong Mu is with the Centre for Wireless Innovation (CWI), Queen's University Belfast, Belfast, BT3 9DT, U.K. (email: x.mu@qub.ac.uk).}
\thanks{Yan Chen and Peiying Zhu are with the Wireless Technology Lab, HUAWEI Technologies Canada Company Ltd., Ottawa, ON K2K 3J1, Canada (e-mail: \{bigbird.chenyan, peiying.zhu\}@huawei.com).}}
\maketitle
\begin{abstract}
Multiple-antenna technologies are advancing toward the development of extremely large aperture arrays and the utilization of extremely high frequencies, driving the progress of next-generation multiple access (NGMA). This evolution is accompanied by the emergence of near-field communications (NFC), characterized by spherical-wave propagation, which introduces additional range dimensions to the channel and enhances system throughput. In this context, a tutorial-based primer on NFC is presented, emphasizing its applications in multiuser communications and multiple access (MA). The following areas are investigated: \romannumeral1) the commonly used near-field channel models are reviewed along with their simplifications under various near-field conditions. \romannumeral2) Building upon these models, the information-theoretic capacity limits of NFC-MA are analyzed, including the derivation of sum-rate capacity and capacity region, and their upper limits for both downlink and uplink scenarios. \romannumeral3) A detailed investigation of near-field multiuser beamforming design is presented, offering low-complexity and effective NFC-MA design methodologies in both the spatial and wavenumber (angular) domains. Throughout these investigations, near-field MA is compared with its far-field counterpart to highlight its superiority and flexibility in terms of interference management, thereby laying the groundwork for achieving NGMA.
\end{abstract}

\section{Introduction}
Multiple access (MA) is a cornerstone of wireless communication systems, enabling the simultaneous servicing of multiple user terminals through available radio resources \cite{goldsmith2005wireless,clerckx2013mimo,vaezi2019multiple}. It addresses two primary challenges: interference management and resource allocation. Since Marconi introduced the principles of frequency-division multiple access (FDMA) in the early 20th century \cite{marcon1900inprovements}, MA techniques have evolved significantly \cite{liu2024road}. Over more than a century of development, MA has transitioned from orthogonal methods to non-orthogonal approaches, expanding from utilizing frequency resources alone to incorporating frequency, time, code, and space domains \cite{liu2017nonorthogonal}. As the foundation for sixth-generation (6G) wireless networks, next-generation multiple access (NGMA) aims to meet new requirements such as terabit rates, extremely low latency, ultra-high reliability, and massive connectivity \cite{saad2019vision}. Additionally, NGMA addresses new scenarios like heterogeneous services and ubiquitous coverage \cite{liu2022evolution}. Achieving these ambitious objectives requires novel tools and techniques \cite{liu2022developing}.

An essential strategy to meet the emerging demands for enhanced rates, coverage, connectivity, and reliability involves signal enhancement and interference mitigation, leading to improved system throughput. Multiple-antenna technologies offer a straightforward solution by providing additional degrees of freedom (DoFs) in the spatial domain for effective interference management and resource allocation \cite{heath2018foundations}. According to Shannon's formula, system throughput scales linearly with the number of antennas, motivating the deployment of larger-aperture arrays. The size and spacing of antennas within an array are inversely related to the wavelength, which enables the feasibility of deploying large antenna arrays at higher frequencies \cite{rappaport2013millimeter,sun2014mimo}. This understanding has guided the evolution of multiple-input multiple-output (MIMO) technologies towards the development of extremely large aperture arrays (ELAAs) and the utilization of extremely high frequencies \cite{bjornson2019massive,de2020non,dreifuerst2023massive}. This transition towards ELAAs and higher frequency bands signifies not just a quantitative increase in antenna size and carrier frequency but also a qualitative paradigm shift from conventional far-field communications (FFC) to near-field communications (NFC) \cite{liu2023near}.

\begin{figure}[!t]
\centering
\includegraphics[width=0.45\textwidth]{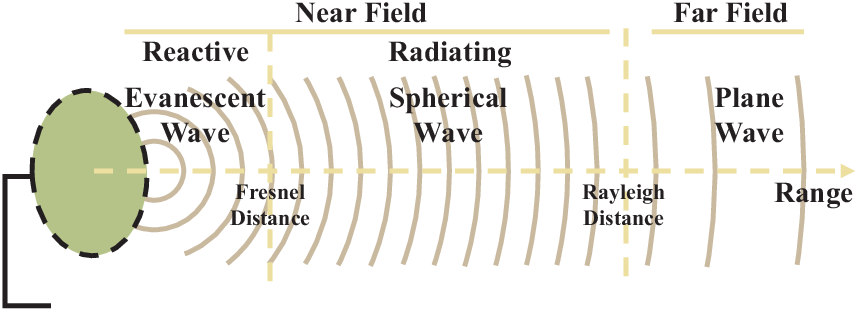}
\caption{Illustration of the flattening of EM waves with range.}
\label{Figure: PW_SW}
\end{figure}

\subsection{Motivation and Contributions}
Far-field electromagnetic (EM) propagation is typically approximated using plane waves, whereas near-field propagation requires precise modeling with spherical waves \cite{balanis2016antenna}, as depicted in {\figurename} {\ref{Figure: PW_SW}}. The boundary between the far-field and near-field regions, known as the \emph{Rayleigh distance}, is calculated as $\frac{2D_{{a}}^2}{\lambda^2}$, where $D_{{a}}$ denotes the array aperture---the maximum distance between any two antennas in the array---and $\lambda$ denotes the signal wavelength. The deployment of ELAAs (augmented $D_{{a}}$) and the utilization of high-frequency bands (diminished $\lambda$) make NFC relevant over distances spanning hundreds of meters, significantly expanding the previously neglected near-field region of traditional wireless systems. For example, consider an antenna array with a dimension of $D_{{a}} = 4$ m---a size feasible for future conformal arrays, such as those deployed on building facades. For signals operating at 3.5 GHz, the Rayleigh distance is 373.3 meters. This distance increases to 2986.7 meters at 28 GHz. This observation underscores how the increased array aperture and the shortened wavelength dramatically extend the near-field region \cite{cui2022near}, motivating our exploration of NGMA in the context of NFC.

Building upon this background, we explore NFC by offering a tutorial-based primer tailored for newcomers to this field. Our focus is on near-field multiuser communications, anticipating the expansion of the near-field region in future networks, which could span several hundred meters and involve numerous user terminals and wireless devices. Investigating near-field multiuser communications is expected to drive the development of \emph{novel spatial-domain techniques for NGMA}. This primer will address three key aspects: the fundamental principles of NFC with an emphasis on \emph{channel modeling}, the information-theoretic limits of NFC with a focus on \emph{channel capacity}, and \emph{beamforming design for NFC}.

Several tutorials, surveys, and overview papers have been published on NFC. For instance, \cite{bjornson2021primer} offers a succinct tutorial on NFC, focusing on channel modeling and signal-to-noise ratio (SNR) analysis in the context of single-user communications. A more exhaustive tutorial review is provided in \cite{liu2023near}, which covers channel modeling, performance analysis, and signal processing techniques. However, most results in these tutorials are specific to single-user communications. Additionally, other articles address various aspects of NFC with different emphases. Works such as \cite{wang2024tutorial} and \cite{lu2023tutorial} prioritize channel modeling, while \cite{gong2022holographic} and \cite{bjornson2024towards} explore electromagnetic (EM) characterizations, and \cite{wang2023rethinking} focuses on near-field integrated sensing and communications. Moreover, \cite{mu2024reconfigurable} investigates the integration of NFC with reconfigurable intelligent surfaces. Although some articles touch upon multiuser communications \cite{lu2023tutorial,bjornson2024towards}, their focus is not specifically on near-field multiuser communications and near-field MA. In summary, while these contributions offer valuable insights into general NFC concepts or specific aspects of NFC technologies, none comprehensively review the characteristics of NFC from the perspective of multiuser communications and MA.

Driven by the significance and broad applicability of NFC in NGMA, and addressing the existing research gaps, this paper serves as a primer on NFC for NGMA. It places particular emphasis on near-field multiuser communications in the spatial domain. The key contributions are outlined as follows.
\begin{enumerate}
  \item[1)] \emph{\textbf{Concise Review of Near-Field Channel Models:}} We provide a succinct review of commonly utilized near-field channel models, encompassing both line-of-sight (LoS) and multipath propagation environments. Each model is explained in its fundamental form, along with several simplification approximations to facilitate performance assessment and beamforming design. In addition to the prevalent spatial channel model, we explore a novel Fourier plane-wave propagation-based method, offering a low-dimensional and sparse representation in the wavenumber or angular domain.   
  \item[2)] \emph{\textbf{Comprehensive Analysis of Information-Theoretic Capacity Limits:}} We conduct an extensive investigation into the information-theoretic capacity limits of near-field LoS multiuser channels. For both \emph{multiple access channels (MACs)} and \emph{broadcast channels (BCs)}, we derive closed-form expressions for the \emph{sum-rate capacity} and characterize the channel \emph{capacity regions}. Additionally, we compare these results with their far-field counterparts. Through numerical and theoretical analyses, we demonstrate the robustness and flexibility of NFC in mitigating IUI. Importantly, we show that NFC approaches the capacity upper bounds for both MAC and BC scenarios.
  \item[3)] \textbf{\emph{Detailed Introduction to Near-Field Multiuser Beamforming Design:}} Our analysis focuses on near-field multiuser beamforming design in both the spatial and wavenumber domains. We introduce the concepts of near-field \emph{range division multiple access (RDMA)} for spatial domain transmission and \emph{wavenumber division multiple access (WDMA)} for the wavenumber domain. For each design paradigm, we present low-complexity algorithmic methods that effectively manage the computational complexity inherent in high-dimensional near-field channels. Theoretical and numerical analyses confirm that NFC substantially enhances system spectral efficiency. Furthermore, we demonstrate that leveraging wavenumber information effectively reduces computational complexity and enhances robustness in near-field beamforming design.
\end{enumerate}
Throughout our discussion, we provide detailed insights into near-field transmission methodologies using both spatially-discrete (SPD) and continuous-aperture (CAP) arrays, each with its unique attributes and challenges. By comprehensively addressing these key aspects, our primer aims to offer a thorough understanding of near-field multiuser communications and its pivotal role in shaping the landscape of NGMA.
\subsection{Organization}
The remainder of this paper is structured as follows. In Section \ref{Section: Channel Modeling of NFC}, we review several commonly utilized near-field channel models. Section \ref{Section: Information-Theoretic Limits of NFC-NGMA} analyzes the information-theoretic capacity limits inherent in near-field multiuser communications. Section \ref{sec4} examines beamforming techniques tailored for near-field multiuser communications, including analyses in both the spatial and wavenumber domains. Finally, Section \ref{Sections: Conclusions} concludes the paper. 

\subsection{Notations}
Throughout this paper, scalars, vectors, and matrices are denoted by non-bold, bold lower-case, and bold upper-case letters, respectively. For the matrix $\mathbf{A}$, $[\mathbf{A}]_{i,j}$, ${\mathbf{A}}^{\mathsf{T}}$, ${\mathbf{A}}^{*}$, and ${\mathbf{A}}^{\mathsf{H}}$ denote the $(i,j)$th entry, transpose, conjugate, and conjugate transpose of $\mathbf{A}$, respectively. For a square matrix $\mathbf{B}$, ${\mathbf{B}}^{\frac{1}{2}}$, ${\mathbf{B}}^{-1}$, ${\mathsf{tr}}(\mathbf{B})$, and $\det(\mathbf{B})$ denote the principal square root, inverse, trace, and determinant of $\mathbf{B}$, respectively. The notation $[\mathbf{a}]_i$ denotes the $i$th entry of vector $\mathbf{a}$, ${\mathsf{diag}}(\mathbf{a})$ returns a diagonal matrix whose diagonal elements are entries of $\mathbf{a}$, and $\lVert \mathbf{a} \rVert_{n}$ represents the $\ell_n$-norm of $\mathbf{a}$. The notations $\lvert a\rvert$, $\lVert \mathbf{a} \rVert$, and $\lVert{\mathbf{A}}\rVert_{F}$ denote the magnitude, norm, and Frobenius norm of scalar $a$, vector $\mathbf{a}$, and matrix $\mathbf{A}$, respectively. The $N\times N$ identity matrix is denoted by ${\mathbf{I}}_N$. The matrix inequalities ${\mathbf{A}}\succeq{\mathbf{0}}$ and ${\mathbf{A}}\succ{\mathbf{0}}$ imply that $\mathbf{A}$ is positive semi-definite and positive definite, respectively. The sets $\mathbbmss{Z}$, $\mathbbmss{R}$, and $\mathbbmss{C}$ stand for the integer, real, and complex spaces, respectively, and notation $\mathbbmss{E}\{\cdot\}$ represents mathematical expectation. ${\mathbbm{1}}_{\mathcal{X}}(x)$ is the indicator function of a subset $x\in{\mathcal{X}}$. The Hadamard product and the ceiling function are represented by $\odot$ and $\lceil \cdot \rceil$, respectively. Finally, ${\mathcal{CN}}({\bm\mu},\mathbf{X})$ is used to denote the circularly-symmetric complex Gaussian distribution with mean $\bm\mu$ and covariance matrix $\mathbf{X}$.

\section{Channel Modeling of NFC}\label{Section: Channel Modeling of NFC}
To begin with, we provide a brief introduction to the basics of NFC, including the different electromagnetic (EM) regions and near-field channel modeling for both SPD and CAP arrays.
\subsection{Electromagnetic Regions}\label{Section: Electromagnetic Regions}
\begin{figure}[!t]
  \centering
    \subfigure[EM field emitted by an antenna element.]{
        \includegraphics[height=0.25\textwidth]{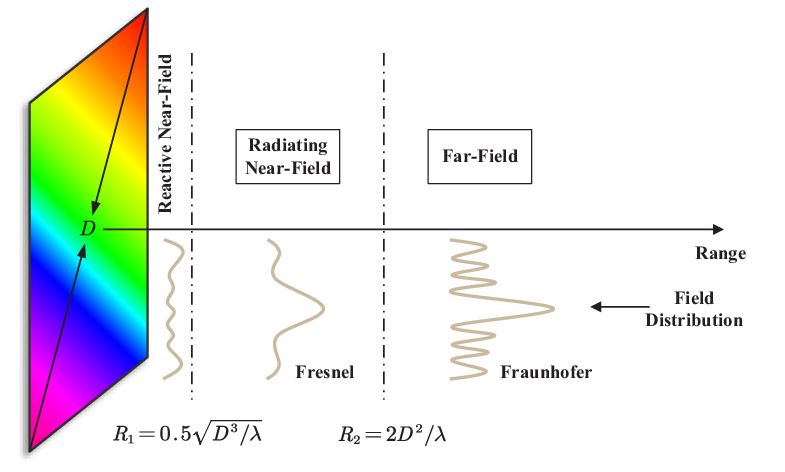}
        \label{Figure: EM_Field_Distribution}	
    }
    \subfigure[EM field emitted by an SPD array.]{
        \includegraphics[height=0.25\textwidth]{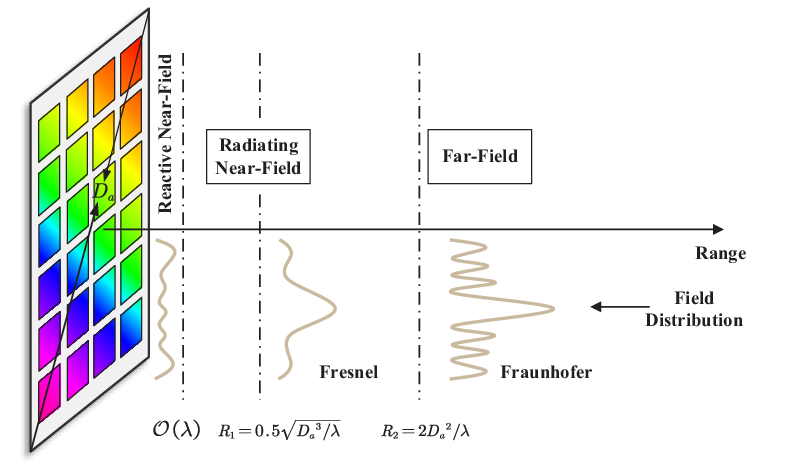}
        \label{Figure: SPD_Array_EM_field}	
    }
  \caption{Illustration of EM field regions and changes of antenna amplitude pattern shape from the reactive near field toward the far field, where $D$/$D_{{a}}$ and $\lambda$ denote the physical dimension of the antenna element/array and signal wavelength, respectively.}
  \label{Figure: EM_field}
\end{figure}

The EM field emitted by \emph{an antenna element} can be divided into three distinct regions: the \emph{far-field} region, the \emph{radiating near-field} region, and the \emph{reactive near-field} region, as illustrated in {\figurename} {\ref{Figure: EM_Field_Distribution}} \cite{balanis2016antenna}. Conventionally, the boundary distinguishing the reactive and radiating near-field regions is estimated at $0.5\sqrt{{D^3}/{\lambda}}$ (known as the \emph{Fresnel distance}), while the boundary between the radiating near-field and far-field regions is estimated at ${2D^2}/{\lambda}$ (referred to as the \emph{Rayleigh or Fraunhofer distance}), where $D$ is the antenna aperture and $\lambda$ is the wavelength \cite{selvan2017fraunhofer}. Within the reactive near-field region, the energy of the EM field oscillates rather than dissipates from the transmitter, predominantly characterized by non-propagating \emph{evanescent waves}. Predicting the behavior of EM waves in this region poses intricate challenges. In contrast, within the radiating near-field and far-field regions, the electric and magnetic fields are perpendicularly in-phase with each other, resulting in the emergence of propagating waves.

In practical antenna design, the apertures of antenna elements typically operate at sub-wavelength scales. Consequently, the reactive and radiating near-field regions of individual antenna elements are often negligible. However, when dealing with ELAAs comprised of numerous SPD antenna elements, the radiating near-field region of the entire array can expand considerably due to the collective influence of all elements. In contrast, the reactive near-field region of an SPD array \emph{remains negligible} in practical communication scenarios, with non-propagating evanescent waves \emph{confined to a small vicinity around each antenna element} \cite{yaghjian1986overview,liu2023near}. Therefore, users would only encounter the reactive region of SPD arrays if their distance from the array is approximately within a few wavelengths---a rare occurrence in practical settings. As a result, the impact of the reactive near-field region is typically negligible when discussing NFC supported by SPD arrays. 

The above arguments imply that for SPD arrays, the boundary distinguishing the reactive and radiating near-field regions should be considered as the \emph{Fresnel distance with respect to (w.r.t.) each antenna element} (i.e., $0.5\sqrt{{D}^3/\lambda}$) rather than the entire array (i.e., $0.5\sqrt{D_{{a}}^3/\lambda}$). On the other hand, the boundary distinguishing the radiating near-field and far-field regions is the \emph{Rayleigh distance w.r.t. to the entire array} (i.e., $2D_{{a}}^2/\lambda$), where $D_{{a}}$ represents the array aperture. For clarity, the EM field boundary of an SPD array is illustrated in {\figurename} {\ref{Figure: SPD_Array_EM_field}}. 
 
Conversely, in NFC systems employing CAP arrays, the reactive near-field region can significantly expand due to the extensive continuous radiating aperture. Therefore, the influence of the reactive near field cannot be overlooked when NFC is supported by CAP arrays \cite{liu2023near}. To ensure effective information transmission within this region, it is imperative to use \emph{dyadic Green's functions} for modeling the propagation environment \cite{yaghjian1980electric}. Note that the EM field boundary of a CAP array resembles that of individual antenna elements, as illustrated in {\figurename} {\ref{Figure: EM_Field_Distribution}}, because a CAP array can be treated as a single antenna element with a larger aperture size \cite{liu2023near}.

At the time of writing, the primary research focus of NFC is on SPD arrays, with particular emphasis on the radiating near field, which predominantly commands current research interests \cite{liu2024near}. This focus aligns with the main thrust of this paper. For brevity and clarity, the term ``radiating near field'' will be referred to as ``near field'' unless explicitly stated otherwise.
\subsection{Near-Field Channel Modeling}
Generally, far-field EM propagation is accurately approximated using plane waves, whereas near-field propagation requires precise modeling with spherical waves \cite{liu2023near,liu2023nearfield}, as illustrated in {\figurename} {\ref{Figure: PW_SW}}. To facilitate the subsequent discussion, we provide a concise overview of spherical-wave channel modeling for NFC.
\subsubsection{LoS Channels}\label{Section: Near-Field Channel Modeling: LoS Channels}
NFC generally operates within the millimeter-wave (mmWave) and sub-terahertz (THz) bands, resulting in channels that are sparsely scattered and dominated by free-space LoS propagation. Additionally, LoS propagation scenarios facilitate theoretical investigations into fundamental performance limits and asymptotic behaviors. Therefore, we commence our discussion with LoS channels.

\begin{figure}[!t]
 \centering
\includegraphics[height=0.3\textwidth]{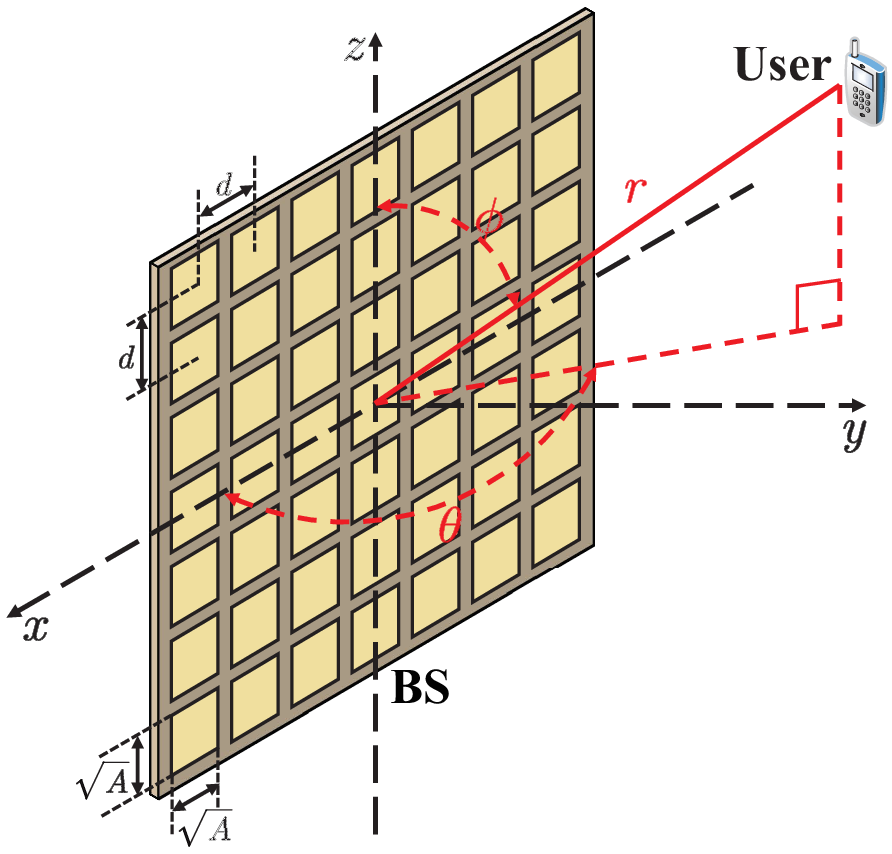}
\caption{Illustration of the array geometry.}
\label{SPD_UPA_Array}
\end{figure}

$\bullet$ \emph{SPD Arrays}:
For clarity, we consider a multiple-input single-output (MISO) channel, where the base station (BS) is equipped with an SPD uniform planar array (UPA) with $M$ antennas. As depicted in {\figurename} {\ref{SPD_UPA_Array}}, the UPA is placed on the $x$-$z$ plane and centered at the origin $[0,0,0]^{\mathsf{T}}$. Here, $M=M_{x}M_{z}$, where $M_{x}$ and $M_{z}$ denote the number of array elements along the $x$- and $z$-axes, respectively. Without loss of generality, we assume that $M_{x}$ and $M_{z}$ are odd numbers with $M_x=2\tilde{M}_x+1$ and $M_z=2\tilde{M}_z+1$. The physical dimensions of each BS array element along the $x$- and $z$-axes are denoted by $\sqrt{A}$, and the inter-element distance is $d$, where $d\geq\sqrt{A}$\footnote{The inter-element distance is constrained by the array aperture size, which is upper bounded by the ratio between the physical dimension and the number of array elements along the $x$- and $z$-axes.}. The central location of the $(m_x,m_z)$th element is denoted by ${\mathbf{s}}_{m_x,m_z}=[m_xd,0,m_zd]^{\mathsf{T}}$, where $m_x\in{\mathcal{M}}_x\triangleq\{0,\pm1,\ldots,\pm\tilde{M}_x\}$ and $m_z\in{\mathcal{M}}_z\triangleq\{0,\pm1,\ldots,\pm\tilde{M}_z\}$. Consequently, the physical dimensions of the UPA along the $x$- and $z$-axes are $L_x\approx M_xd$ and $L_z\approx M_zd$, respectively. Moreover, it is assumed that the user is equipped with a single hypothetical isotropic antenna element with an aperture size of $\frac{\lambda^2}{4\pi}$ to receive incoming signals. Let $r$ denote the propagation distance from the center of the antenna array to the user, and $\theta\in[0,\pi]$ and $\phi\in[0,\pi]$ denote the associated azimuth and elevation angles, respectively. Then, the location of the user can be expressed as ${\mathbf{r}}=[r\Phi,r\Psi,r\Omega]^{\mathsf{T}}$, where $\Phi\triangleq\sin{\phi}\cos{\theta}$, $\Psi\triangleq\sin\phi\sin\theta$, and $\Omega\triangleq\cos{\phi}$.

Due to the near-field behavior, it is imperative to differentiate the power and phase of different antenna elements when modeling the channel response between the BS array and the user. Specifically, the distance between the user and the center of the $(m_x,m_z)$th antenna element is given by
\begin{subequations}\label{General_Distance_Expression}
\begin{align}
r_{m_x,m_z}&=\lVert{\mathbf{r}}-{\mathbf{s}}_{m_x,m_z}\rVert\\
&=r\sqrt{(m_x\varepsilon-\Phi)^2+{\Psi}^2+(m_z\varepsilon-\Omega)^2},
\end{align}
\end{subequations}
where $\varepsilon=\frac{d}{r}$. Note that $r=r_{0,0}$ and, since the antenna element separation $d$ is typically on the order of a wavelength, in practice, we have $\varepsilon\ll1$. The channel response between the user and the $(m_x,m_z)$th antenna element of the BS array can be written as $h_{m_x,m_z}=\sqrt{g_{m_x,m_z}}{\rm{e}}^{-{\rm{j}}\phi_{m_x,m_z}}$, where $g_{m_x,m_z}$ and $\phi_{m_x,m_z}$ are the channel power and phase, respectively. By combining $\{h_{m_x,m_z}\}_{\forall m_x,m_z}$ into a vector, we obtain the channel vector $\mathbf{h}=[h_{m_x,m_z}]_{\forall m_x,m_z}\in{\mathbbmss{C}}^{M\times1}$. Based on \cite{liu2023near}, we model the channel power $g_{m_x,m_z}$ as follows:
\begin{equation}\label{Near-Field_Channel_Power}
g_{m_x,m_z}=\int_{{\mathcal{S}}_{m_x,m_z}}\frac{1}{4\pi\lVert{\mathbf{r}}-{\mathbf{s}}\rVert^2}
\frac{{\mathbf{e}}_y^{\mathsf{T}}({\mathbf{r}}-{\mathbf{s}})}{\lVert{\mathbf{r}}-{\mathbf{s}}\rVert}{\rm{d}}{\mathbf{s}},
\end{equation}
where ${\mathcal{S}}_{m_x,m_z}=[m_xd-{\sqrt{A}}/{2},m_xd+{\sqrt{A}}/{2}]\times[m_zd-{\sqrt{A}}/{2},m_zd+{\sqrt{A}}/{2}]$ denotes the surface region of the $(m_x,m_z)$th antenna element. The term $\frac{1}{4\pi\lVert{\mathbf{r}}-{\mathbf{s}}\rVert^2}$ models the influence of the \emph{free-space path loss}, and $\frac{{\mathbf{e}}_y^{\mathsf{T}}({\mathbf{r}}-{\mathbf{s}})}{\lVert{\mathbf{r}}-{\mathbf{s}}\rVert}$ models the influence of the \emph{projected aperture} of each antenna element, reflected by the projection of the UPA normal vector ${\mathbf{e}}_y=[0,1,0]^{\mathsf{T}}$ onto the wave propagation direction at each local point $\mathbf{s}$ \cite{liu2023near}. 

Since the size of each antenna element, $\sqrt{A}$, is typically on the order of a wavelength, we have $r\gg \sqrt{A}$, which implies that the variation of the complex-valued channel across different points in a given antenna element is negligible \cite{liu2023near}. Specifically, ${\mathbf{s}}\approx{\mathbf{s}}_{m_x,m_z}$ holds true for ${\mathbf{s}}\in{\mathcal{S}}_{m_x,m_z}$, and considering that $\int_{{\mathcal{S}}_{m_x,m_z}}{\rm{d}}{\mathbf{s}}=A$, we get
\begin{subequations}
\begin{align}
g_{m_x,m_z}&\approx\int_{{\mathcal{S}}_{m_x,m_z}}\frac{{\mathbf{e}}_y^{\mathsf{T}}({\mathbf{r}}-{\mathbf{s}}_{m_x,m_z})}{4\pi\lVert{\mathbf{r}}
-{\mathbf{s}}_{m_x,m_z}\rVert^3}{\rm{d}}{\mathbf{s}}\\
&=\frac{{\mathbf{e}}_y^{\mathsf{T}}({\mathbf{r}}-{\mathbf{s}}_{m_x,m_z})A}{4\pi\lVert{\mathbf{r}}-{\mathbf{s}}_{m_x,m_z}\rVert^3}
=\frac{r\Psi A}{4\pi\lVert{\mathbf{r}}-{\mathbf{s}}_{m_x,m_z}\rVert^3}.\label{Near_Field_Model_Channel_Power}
\end{align}
\end{subequations}
Furthermore, applying this approximation to the phase of the channel response for the $(m_x,m_z)$th antenna element, we obtain $\phi_{m_x,m_z}\approx\frac{2\pi}{\lambda}\lVert{\mathbf{r}}-{\mathbf{s}}_{m_x,m_z}\rVert$. Combining these results, we have
\begin{align}\label{SPD_UPA_NFC_Model}
{\mathbf{h}}=\left[\frac{\sqrt{r\Psi A}}{\sqrt{4\pi}\lVert{\mathbf{r}}-{\mathbf{s}}_{m_x,m_z}\rVert^{3/2}}{\rm{e}}^{-{\rm{j}}\frac{2\pi}{\lambda}\lVert{\mathbf{r}}-{\mathbf{s}}_{m_x,m_z}\rVert}\right]_{\forall m_x,m_z}.
\end{align}

The near-field model often employs simplifications for practical ease and clarity. When the near-field user is situated beyond the \emph{uniform-power distance}, the channel power disparity of each link becomes negligible \cite{liu2024near}, leading to
\begin{align}
\frac{\sqrt{r\Psi A}}{\sqrt{4\pi}\lVert{\mathbf{r}}-{\mathbf{s}}_{m_x,m_z}\rVert^{3/2}}\approx
\frac{\sqrt{r\Psi A}}{\sqrt{4\pi}\lVert{\mathbf{r}}-{\mathbf{s}}_{0,0}\rVert^{3/2}}={\frac{\sqrt{\Psi A}}{\sqrt{4\pi} r}},
\end{align} 
and it follows that
\begin{align}\label{SPD_UPA_NFC_Model_USW}
{\mathbf{h}}\approx{\frac{\sqrt{\Psi A}}{\sqrt{4\pi} r}}
\left[{\rm{e}}^{-{\rm{j}}\frac{2\pi}{\lambda}\lVert{\mathbf{r}}-{\mathbf{s}}_{m_x,m_z}\rVert}\right]_{\forall m_x,m_z}.
\end{align}
As discussed in \cite{liu2023nearfield,liu2024near}, the \emph{uniform-power distance} is generally much smaller than the \emph{Rayleigh distance}. Therefore, the approximation in \eqref{SPD_UPA_NFC_Model_USW} is effective for the majority of the area within the near-field region. If the user is situated beyond the \emph{Fresnel distance w.r.t. the entire SPD array}, the channel phase of each link can be simplified using the \emph{Fresnel approximation}\footnote{The \emph{Fresnel approximation} is derived using the fact that $\varepsilon\ll 1$, along with the Maclaurin series expansion $\sqrt{1+x}\approx 1 + \frac{x}{2}-\frac{x^2}{8}$. This distance helps determine the boundary beyond which the approximation remains valid.} \cite{maclaurin1909light}, which yields
\begin{equation}\label{General_Distance_Expression_Maclaurin1}
\begin{split}
\phi_{m_x,m_z}&\approx \frac{2\pi}{\lambda}(r(1-\varepsilon(m_x\Phi+m_z\Omega))-r{\varepsilon^2}/{2}\\
&\times(2m_xm_z\Phi\Omega-m_x^2(1-\Phi^2)-m_z^2(1-\Omega^2))).
\end{split}
\end{equation}
By omitting the bilinear term, we have
\begin{equation}\label{General_Distance_Expression_Maclaurin2}
\begin{split}
\phi_{m_x,m_z}&\approx \frac{2\pi}{\lambda}(r(1-\varepsilon(m_x\Phi+m_z\Omega))\\
&+r{\varepsilon^2}/{2}(m_x^2(1-\Phi^2)+m_z^2(1-\Omega^2))),
\end{split}
\end{equation}
and it follows that
\begin{equation}\label{SPD_UPA_NFC_Model_Fresnel}
\begin{split}
{\mathbf{h}}&\approx{\frac{\sqrt{\Psi A}}{\sqrt{4\pi} r}}\\
&\times\left[{\rm{e}}^{-{\rm{j}}\frac{2\pi r\big(1-\varepsilon(m_x\Phi+m_z\Omega)+\frac{\varepsilon^2(m_x^2(1-\Phi^2)+m_z^2(1-\Omega^2))}{2}\big)}{\lambda}}\right]_{\forall m_x,m_z}.
\end{split}
\end{equation}
By moving the user beyond the \emph{Rayleigh distance}, the quadratic terms in \label{General_Distance_Expression_Maclaurin2} can be further omitted, and the near-field channel model simplifies into the far-field model:
\begin{align}\label{SPD_UPA_FFC_Model}
{\mathbf{h}}\approx{\frac{\sqrt{\Psi A}}{\sqrt{4\pi} r}}\left[{\rm{e}}^{-{\rm{j}}\frac{2\pi}{\lambda}r(1-m_x\varepsilon\Phi-m_z\varepsilon\Omega)}\right]_{\forall m_x,m_z}.
\end{align}
\vspace{-5pt}
\begin{remark}
By comparing \eqref{SPD_UPA_NFC_Model} and \eqref{SPD_UPA_FFC_Model}, we observe that the primary difference between the near-field and far-field models lies in the inclusion of the additional range dimensions $\{r_{m_x,m_z}\}_{\forall m_x,m_z}$ in the near-field channels.
\end{remark}
\vspace{-5pt}
We also note that in many existing research works related to NFC, only the variations in free-space path loss are taken into account, while the impact of the variations in the projected aperture is neglected, e.g., \cite{wu2023multiple}. In this case, the near-field channel is modeled as follows:
\begin{align}\label{SPD_UPA_NFC_Model_No_Pro}
{\mathbf{h}}=\left[\frac{\sqrt{A}}{\sqrt{4\pi}\lVert{\mathbf{r}}-{\mathbf{s}}_{m_x,m_z}\rVert}{\rm{e}}^{-{\rm{j}}\frac{2\pi}{\lambda}\lVert{\mathbf{r}}-{\mathbf{s}}_{m_x,m_z}\rVert}\right]_{\forall m_x,m_z}.
\end{align}
We use ${\mathsf{a}}_{\mathsf{pro}}$ and ${\mathsf{a}}_{\mathsf{npro}}$ to denote the channel gain calculated with and without considering the variations in the projected aperture, respectively. Specifically, we have
\begin{align}
&{\mathsf{a}}_{\mathsf{pro}}=\sum_{m_x,m_z}\frac{{r\Psi A}}{{4\pi}\lVert{\mathbf{r}}-{\mathbf{s}}_{m_x,m_z}\rVert^{3}},\\
&{\mathsf{a}}_{\mathsf{npro}}=\sum_{m_x,m_z}\frac{{ A}}{{4\pi}\lVert{\mathbf{r}}-{\mathbf{s}}_{m_x,m_z}\rVert^{2}}.
\end{align}

\begin{figure}[!t]
 \centering
\setlength{\abovecaptionskip}{0pt}
\includegraphics[width=0.45\textwidth]{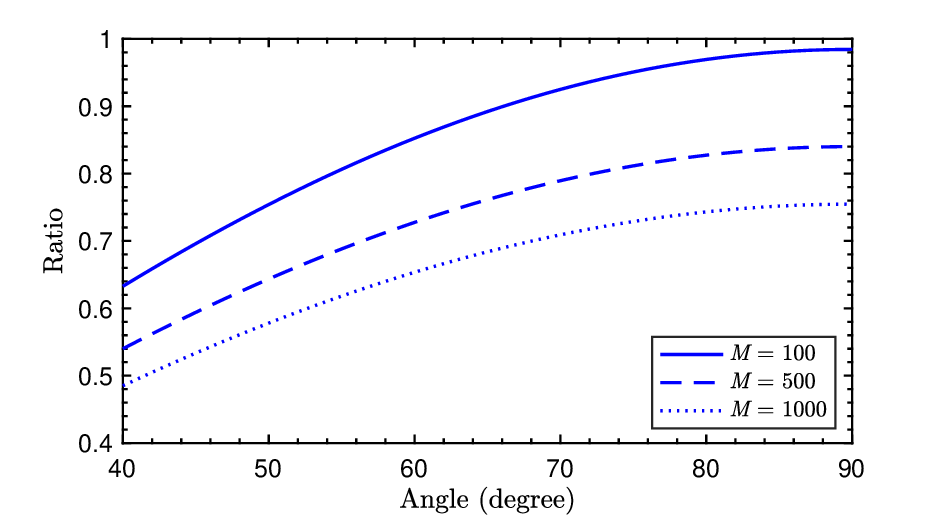}
\caption{Illustration of the impact of the projected aperture. $d=0.0628$ m, $d=\lambda/2$, $A=\frac{\lambda^2}{4\pi}$, $r=10$ m, $M_x=1$, and $M_z=M$.}
\label{projected aperture impacted}
\end{figure}

In the sequel, we provide numerical results to explore the impact of variations in the projected aperture. For simplicity, we consider a linear array placed along the $z$-axis and centered at the origin with $M_x=1$ and $M_z=M$. Additionally, we set the location of the user as ${\mathbf{r}}=[r\cos{\theta},r\sin{\theta},0]^{\mathsf{T}}$. By changing the angle 
$\theta$, the impact of variations in the projected aperture on the channel gain will change accordingly, while the impact of variations in path loss will remain fixed. {\figurename} {\ref{projected aperture impacted}} plots the channel gain ratio $\frac{{\mathsf{a}}_{\mathsf{pro}}}{{\mathsf{a}}_{\mathsf{npro}}}$ in terms of the user direction 
$\theta$. By definition, a smaller value of $\frac{{\mathsf{a}}_{\mathsf{pro}}}{{\mathsf{a}}_{\mathsf{npro}}}$ suggests a more significant impact of the varying projected aperture on the channel gain. As can be seen from this figure, with a decreasing $\theta$ (i.e., as the user deviates away from the normal direction of the array), the impact of the varying projected aperture becomes more pronounced. Furthermore, as the number of antenna elements $M$ increases, the impact of the projected aperture on the channel gain is further emphasized. This is because each newly added antenna is deployed further away from the user, reducing the effective area and thus the channel gain. These results highlight the necessity of considering the varying projected aperture in near-field channel modeling.

$\bullet$ \emph{CAP Arrays}: Near-field channel modeling for CAP arrays is more challenging compared to that for SPD arrays. We next explore the scenario where the transceivers are equipped with a CAP array. In contrast to SPD arrays, CAP arrays support a continuous distribution of source currents, denoted by ${\mathbf{J}}(\mathbf{s})\in{\mathbbmss{C}}^{3\times1}$, where $\mathbf{s}=[s_x,s_y,s_z]^{\mathsf{T}}\in{\mathbbmss{R}}^{3\times1}$ is the source point within the transmit aperture $\mathcal{A}_{\mathsf{S}}$. The electric radiation field $\mathbf{E}(\mathbf{r})\in{\mathbbmss{C}}^{3\times1}$ generated at point ${\mathbf{r}}=[r\Phi,r\Psi,r\Omega]^{\mathsf{T}}\in{\mathbbmss{R}}^{3\times1}$ within the receive aperture $\mathcal{A}_{\mathsf{R}}$ can be expressed as follows \cite{jackson1999classical}:
\begin{align}\label{CAP_Current_Elctric_Field}
\mathbf{E}(\mathbf{r})=\int_{\mathcal{A}_{\mathsf{S}}}{\mathbf{G}}({\mathbf{r}},{\mathbf{s}}){\mathbf{J}}(\mathbf{s}){\rm{d}}{\mathbf{s}},
\end{align}
where ${\mathbf{G}}({\mathbf{r}},{\mathbf{s}})\in{\mathbbmss{C}}^{3\times3}$ represents the \emph{dyadic Green's function} \cite{sarabandi2009dyadic}. Particularly, the dyadic Green's function can be calculated as follows \cite{sarabandi2009dyadic}:
\begin{align}
{\mathbf{G}}({\mathbf{r}},{\mathbf{s}})=-{{\rm{j}}\omega\mu_0}\left({\mathbf{I}}_3+\frac{\lambda}{2\pi}\nabla_{{\mathbf{r}}-{\mathbf{s}}}\nabla_{{\mathbf{r}}-{\mathbf{s}}}\right)
\mathsf{g}({\mathbf{r}},{\mathbf{s}}),
\end{align}
where $\mathsf{g}({\mathbf{r}},{\mathbf{s}})\triangleq\frac{{\rm{e}}^{-{\rm{j}}\frac{2\pi}{\lambda}\lVert {\mathbf{r}}-{\mathbf{s}}\rVert}}{4\pi \lVert {\mathbf{r}}-{\mathbf{s}}\rVert}$ is the \emph{scalar Green's function}, $\mu_0$ is the free-space permeability, and $\omega$ is the angular frequency of the signal. Mathematically, the dyadic Green's function can be further expanded as follows \cite{poon2005degrees}:
\begin{equation}\label{T_Green_Standard}
\begin{split}
{\mathbf{G}}({\mathbf{r}},{\mathbf{s}})&=-{\rm{j}}\eta_0\frac{2\pi}{\lambda}\mathsf{g}({\mathbf{r}},{\mathbf{s}})\left[\left({\mathbf{I}}_3-\hat{\mathbf x}\hat{\mathbf x}^{\mathsf H}\right)+\frac{{\rm{j}}\lambda}{2\pi\lVert{\mathbf{r}}-{\mathbf{s}}\rVert}\right.\\
&\times\left.\left({\mathbf{I}}_3-3\hat{\mathbf x}\hat{\mathbf x}^{\mathsf H}\right)-\frac{\lambda^2}{(2\pi\lVert{\mathbf{r}}-{\mathbf{s}}\rVert)^2}\left({\mathbf{I}}_3-3\hat{\mathbf x}\hat{\mathbf x}^{\mathsf H}\right)\right],
\end{split}
\end{equation}
where $\hat{\mathbf x}=\frac{{\mathbf{r}}-{\mathbf{s}}}{\lVert{\mathbf{r}}-{\mathbf{s}}\rVert}$, $\eta_0=\sqrt{\frac{\mu_0}{\epsilon_0}}$, and $\epsilon_0$ is the free-space permittivity. A scalar simplified version for \eqref{T_Green_Standard} can be also used, which yields \cite{dardari2020communicating,bjornson2021primer}
\begin{equation}\label{T_Green_Standard_Scalar}
\begin{split}
{\mathsf{G}}({\mathbf{r}},{\mathbf{s}})&=-{\rm{j}}\eta_0\frac{2\pi}{\lambda}\mathsf{g}({\mathbf{r}},{\mathbf{s}})\\
&\times\left(1+\frac{{\rm{j}}\lambda}{2\pi\lVert{\mathbf{r}}-{\mathbf{s}}\rVert}
-\frac{\lambda^2}{(2\pi\lVert{\mathbf{r}}-{\mathbf{s}}\rVert)^2}\right).
\end{split}
\end{equation}
The squared magnitude of the parenthesis in \eqref{T_Green_Standard_Scalar} is
\begin{equation}
\begin{split}
&\left\rvert1+\frac{{\rm{j}}\lambda}{2\pi\lVert{\mathbf{r}}-{\mathbf{s}}\rVert}
-\frac{\lambda^2}{(2\pi\lVert{\mathbf{r}}-{\mathbf{s}}\rVert)^2}\right\rvert^2\\
&=1-\frac{\lambda^2}{(2\pi\lVert{\mathbf{r}}-{\mathbf{s}}\rVert)^2}+
\frac{\lambda^4}{(2\pi\lVert{\mathbf{r}}-{\mathbf{s}}\rVert)^4}.
\end{split}
\end{equation}
The dyadic Green's function and its scalar counterpart consist of three terms. The first term corresponds to the radiating near-field and far-field regions, while the remaining two terms correspond to the reactive near-field region. Only the first term contributes to the radiated field as it decreases inversely with the distance $\lVert{\mathbf{r}}-{\mathbf{s}}\rVert$, and hence its power follows the inverse square law. The power of the remaining two terms decreases much faster than $\lVert{\mathbf{r}}-{\mathbf{s}}\rVert^{-2}$ (${\lVert{\mathbf{r}}-{\mathbf{s}}\rVert}^{-4}$ and ${\lVert{\mathbf{r}}-{\mathbf{s}}\rVert}^{-6}$, respectively), so they do not contribute to EM radiation. Using the simplified dyadic Green's function, we can incorporate the influence of the reactive region into the near-field channel models for SPD arrays. Specifically, by considering the evanescent waves, \eqref{Near_Field_Model_Channel_Power} can be adjusted as follows:
\begin{equation}\label{dyadic Green’s function_SPD_Model}
\begin{split}
&g_{m_x,m_z}\approx \frac{r\Psi A}{4\pi\lVert{\mathbf{r}}-{\mathbf{s}}_{m_x,m_z}\rVert^3}\\
&\times \left(1-\frac{\lambda^2}{(2\pi\lVert{\mathbf{r}}-{\mathbf{s}}_{m_x,m_z}\rVert)^2}+
\frac{\lambda^4}{(2\pi\lVert{\mathbf{r}}-{\mathbf{s}}_{m_x,m_z}\rVert)^4}\right).
\end{split}
\end{equation}
Meanwhile, the expression for the channel phase, $\frac{2\pi}{\lambda}\lVert{\mathbf{r}}-{\mathbf{s}}_{m_x,m_z}\rVert$, can be adjusted accordingly.

The near-field LoS channel response for CAP arrays is characterized by Green's function, which can be further simplified under specific conditions. When the near-field user is positioned beyond the \emph{Fresnel distance}, i.e., within the radiating near-field region, the dyadic Green's function can be approximated as follows \cite{poon2005degrees}:
\begin{align}\label{T_Green_Simplify1}
{\mathbf{G}}({\mathbf{r}},{\mathbf{s}})\approx -{\rm{j}}\eta_0\frac{2\pi}{\lambda}\mathsf{g}({\mathbf{r}},{\mathbf{s}})\left({\mathbf{I}}_3-\hat{\mathbf x}\hat{\mathbf x}^{\mathsf H}\right).
\end{align} 
While this approximation is more tractable than the full dyadic Green's function in \eqref{T_Green_Standard}, its matrix form can still pose complexities. As a compromise, many researchers propose employing the scalar Green's function to characterize the EM channel response of CAP arrays, which can be treated as a simplification of ${\mathsf{G}}({\mathbf{r}},{\mathbf{s}})$. In this scenario, the electric radiation field $\mathsf{e}(\mathbf{r})\in{\mathbbmss{C}}$ generated at point $\mathbf{r}\in\mathcal{A}_{\mathsf{R}}$ can be formulated as follows \cite{jackson1999classical}:
\begin{align}\label{CAP_Current_Elctric_Field1}
\mathsf{e}(\mathbf{r})=\int_{\mathcal{A}_{\mathsf{S}}}-{\rm{j}}\eta_0\frac{2\pi}{\lambda}\mathsf{g}({\mathbf{r}},{\mathbf{s}}){\mathsf{j}}(\mathbf{s}){\rm{d}}{\mathbf{s}},
\end{align}
where ${\mathsf{j}}(\mathbf{s})\in{\mathbbmss{C}}$ denotes the continuous distribution of source currents at ${\mathbf{s}}\in{\mathcal{A}_{\mathsf{S}}}$. It is important to note that the scalar Green's function mainly applies to the radiating near-field and far-field regions \cite{rothwell2018electromagnetics}. 

Next, we consider a downlink channel setup where the BS is equipped with a CAP array, whose array geometry is the same as that shown in {\figurename} {\ref{SPD_UPA_Array}}. The BS array is placed on the $x$-$z$ plane and centered at the origin, with the user located at ${\mathbf{r}}=[r\Phi,r\Psi,r\Omega]^{\mathsf{T}}$. When the user is located beyond the \emph{uniform-power distance}, the approximation becomes
\begin{align}
\mathsf{g}({\mathbf{r}},{\mathbf{s}})\approx\frac{{\rm{e}}^{-{\rm{j}}\frac{2\pi}{\lambda}\lVert {\mathbf{r}}-{\mathbf{s}}\rVert}}{4\pi \lVert {\mathbf{r}}-[0,0,0]^{\mathsf{T}}\rVert}=\frac{{\rm{e}}^{-{\rm{j}}\frac{2\pi}{\lambda}\lVert {\mathbf{r}}-{\mathbf{s}}\rVert}}{4\pi r}.
\end{align}
Since $\mathsf{g}({\mathbf{r}},{\mathbf{s}})$ applies to both the radiating near-field and far-field regions beyond the \emph{Fresnel distance}, we can simplify the phase using the \emph{Fresnel approximation}, which yields
\begin{align}
{\rm{e}}^{-{\rm{j}}\frac{2\pi}{\lambda}\lVert {\mathbf{r}}-{\mathbf{s}}\rVert}\approx
{\rm{e}}^{-{\rm{j}}\frac{2\pi}{\lambda}(r-s_x\Phi-s_z\Omega-\frac{2s_xs_z\Phi\Omega-s_x^2(1-\Phi^2)-s_z^2(1-\Omega^2)}{2r})}.
\end{align}
Omitting the bilinear term gives
\begin{align}
{\rm{e}}^{-{\rm{j}}\frac{2\pi}{\lambda}\lVert {\mathbf{r}}-{\mathbf{s}}\rVert}\approx
{\rm{e}}^{-{\rm{j}}\frac{2\pi}{\lambda}(r-s_x\Phi-s_z\Omega+\frac{s_x^2(1-\Phi^2)+s_z^2(1-\Omega^2)}{2r})}.
\end{align}
Taken together, we obtain
\begin{align}\label{Green_Function_Simplify}
\mathsf{g}({\mathbf{r}},{\mathbf{s}})\approx\frac{{\rm{e}}^{-{\rm{j}}\frac{2\pi}{\lambda}(r-s_x\Phi-s_z\Omega+\frac{s_x^2(1-\Phi^2)+s_z^2(1-\Omega^2)}{2r})}}{4\pi r}.
\end{align}
If the user is situated beyond the \emph{Rayleigh distance}, the near-field scalar Green's function simplifies to its far-field counterpart:
\begin{align}\label{CAP_UPA_FFC_Model}
\mathsf{g}({\mathbf{r}},{\mathbf{s}})\approx\frac{{\rm{e}}^{-{\rm{j}}\frac{2\pi}{\lambda}(r-s_x\Phi-s_z\Omega)}}{4\pi r}.
\end{align}

\subsubsection{Multipath Channels}
Next, we explore the near-field multipath channel model, focusing specifically on the SPD array for brevity. More extensive discussions regarding multipath channel models for CAP arrays can be found in \cite{liu2024near}. 

$\bullet$ \emph{Propagation-Based Models}:
The propagation-based multipath model is meticulously constructed by integrating key physical propagation parameters such as directions of arrival and departure, multipath delay, scatterer distribution, and the radar cross-section (RCS) of the scatterers. To elucidate, we focus on a MISO channel, where the BS is equipped with $M$ antennas. Assume that there are $N_{\mathsf{scatterer}}$ scatterers in the nearby environment. The propagation-based near-field channel can be modeled as follows:
\begin{align}
{\mathbf{h}}=\mathbf{h}_{\mathsf{LoS}} + \sum_{n=1}^{N_{\mathsf{scatterer}}}{\beta}_{\mathsf{NLoS}}^{(n)}{\mathbf{h}}_{\mathsf{NLoS}}^{(n)}{{h}}_{\mathsf{NLoS}}^{(n)},
\end{align}
where $\mathbf{h}_{\mathsf{LoS}}\in{\mathbbmss{C}}^{M\times1}$ denotes the LoS component, and $\sum_{n=1}^{N_{\mathsf{scatterer}}}{\beta}_{\mathsf{NLoS}}^{(n)}{\mathbf{h}}_{\mathsf{NLoS}}^{(n)}{{h}}_{\mathsf{NLoS}}^{(n)}$ denotes the non-LoS (NLoS) component. Specifically, ${\mathbf{h}}_{\mathsf{NLoS}}^{(n)}\in{\mathbbmss{C}}^{M\times1}$ represents the near-field channel response between the BS and the $n$th scatterer, ${{h}}_{\mathsf{NLoS}}^{(n)}\in{\mathbbmss{C}}$ denotes the near-field channel response between the $n$th scatterer and the user, and ${\beta}_{\mathsf{NLoS}}^{(n)}\in{\mathbbmss{C}}$ denotes the RCS or the refection coefficient associated with the $n$th scatterer. These near-field channel responses, i.e., $\{{{h}}_{\mathsf{NLoS}}^{(n)}, {\mathbf{h}}_{\mathsf{NLoS}}^{(n)}\}_{n=1}^{N_{\mathsf{scatterer}}}$ and $\mathbf{h}_{\mathsf{LoS}}$, can be modeled using the methods detailed in Section \ref{Section: Near-Field Channel Modeling: LoS Channels}. Following the Swerling-\uppercase\expandafter{\romannumeral1} model \cite{richards2005fundamentals}, each RCS coefficient can be modeled as a complex Gaussian variate. Moreover, due to the sparsely-scattered nature of the near-field channel, we often find $M\gg N_{\mathsf{scatterer}}$. Additionally, the near-field multipath channel is generally LoS-dominated, implying that the channel power of the LoS component $\mathbf{h}_{\mathsf{LoS}}$ is much larger than that of the NLoS component.

The propagation-based channel model, while accurate, entails complexity as it requires the precise distribution of each scatterer. Typically, these models are useful in the deployment or optimization of site-specific radio systems. They serve as indispensable tools in channel evaluation during the critical phases of system design in essential reference cases \cite{imoize2021standard}.

$\bullet$ \emph{Correlation-Based Models}: 
Another significant multipath channel model is the correlation-based model. Unlike the propagation-based model, this approach characterizes the impulse response based on unstructured channel statistics that are independent of the physical parameters of individual multipath rays, such as directions of arrival and departure. Correlation-based channels offer easier simulation compared to propagation-based channels, making them highly preferred for link- and system-level simulations \cite{imoize2021standard}.

\begin{figure}[!t]
 \centering
\includegraphics[width=0.45\textwidth]{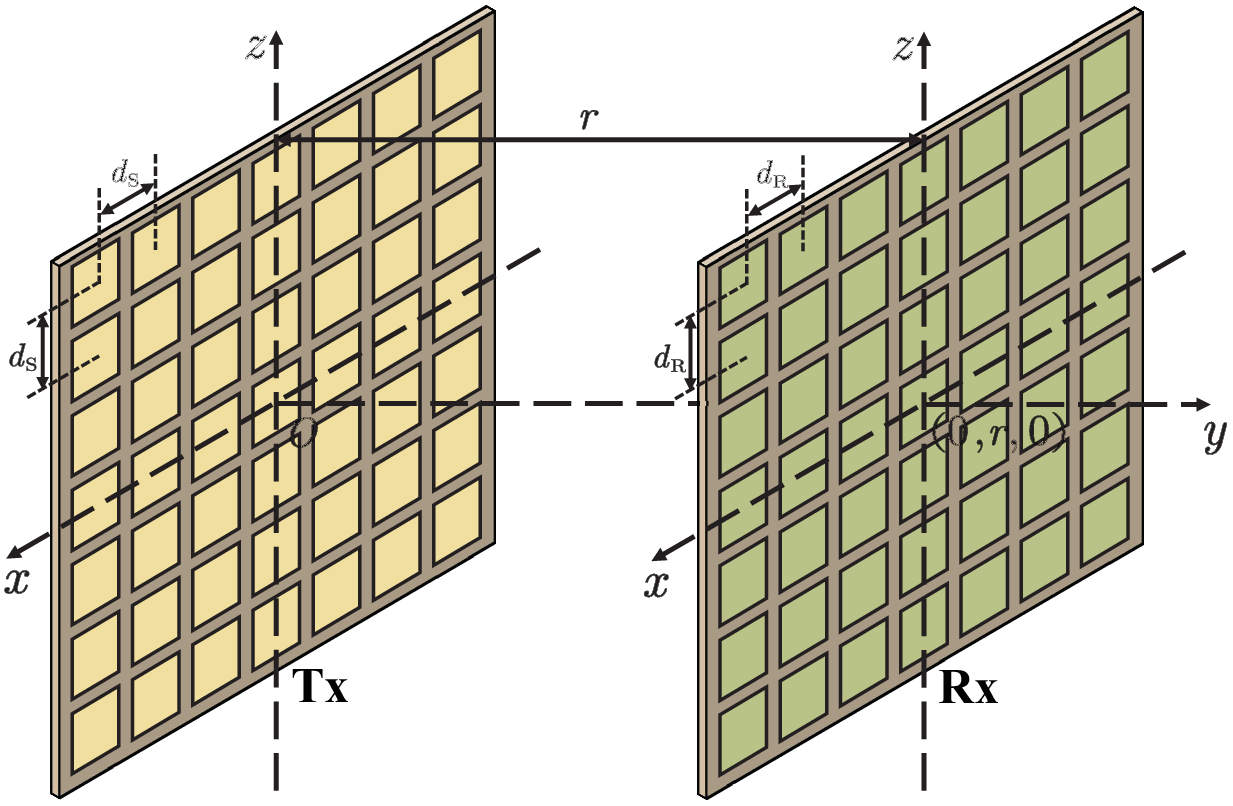}
\caption{Illustration of a near-field MIMO channel.}
\label{near-field MIMO channel}
\end{figure}

In the sequel, we introduce a novel correlation-based model proposed in 2022 for NFC \cite{pizzo2022fourier}. This model is built upon Weyl's decomposition of spherical waves into plane waves \cite{weyl1919ausbreitung} and utilizes scattering matrix theory \cite{kerns1976plane}. For clarity, we assume that both the transmitter and receiver are equipped with UPAs with array geometries similar to that shown in {\figurename} {\ref{SPD_UPA_Array}}. Specifically, the transmit UPA is placed on the $x$-$z$ plane and centered at the origin, while the receive UPA is placed parallel to the $x$-$z$ plane and centered at $[0,r,0]^{\mathsf{T}}$, as illustrated in {\figurename} {\ref{near-field MIMO channel}}. The transmit UPA consists of $M=M_xM_z$ antennas and has physical dimensions of $L_{{\mathsf{S}},x}\times L_{{\mathsf{S}},z}\approx M_x d_{\mathsf{S}} \times M_z d_{\mathsf{S}}$, and the receive UPA comprises $N=N_xN_z$ antennas and has physical dimensions of $L_{{\mathsf{R}},x}\times L_{{\mathsf{R}},z}\approx N_x d_{\mathsf{R}} \times N_z d_{\mathsf{R}}$, where $d_{\mathsf{S}}$ and $d_{\mathsf{R}}$ denote antenna spacing. Let ${\mathbf{s}}_m=[s_x^{(m)},s_y^{(m)},s_z^{(m)}]^{\mathsf{T}}\in{\mathbbmss{R}}^{3\times1}$ denote the coordinates of the $m$th transmit antenna for $m=1,\ldots,M$, and ${\mathbf{r}}_n=[r_x^{(n)},r_y^{(n)},r_z^{(n)}]^{\mathsf{T}}\in{\mathbbmss{R}}^{3\times1}$ denote the coordinates of the $n$th receive antenna for $n=1,\ldots,N$. 

According to the methodology in \cite{pizzo2022fourier}, the spatial impulse response between the $m$th transmit antenna and the $n$th receive antenna can be derived using a \emph{four-dimensional (4D) Fourier plane-wave} representation, given by
\begin{equation}\label{4DFPE_Stadard}
\begin{split}
h_{n,m}&=\frac{1}{(2\pi)^2}\iiiint_{{\mathcal{D}}({\bm\kappa})\times{\mathcal{D}}({\mathbf{k}})}a_{\mathsf{R}}({\mathbf{k}},{\mathbf{r}}_n)\\
&\times H_a(k_x,k_z,{\kappa}_x,{\kappa}_z)a_{\mathsf{S}}({\bm{\kappa}},{\mathbf{s}}_m){\rm{d}}\kappa_x{\rm{d}}\kappa_z{\rm{d}}k_x{\rm{d}}k_z,
\end{split}
\end{equation}
where $a_{\mathsf{S}}({\bm{\kappa}},{\mathbf{s}}_m)={\rm{e}}^{-{\rm{j}}{\bm{\kappa}}^{\mathsf{T}}{\mathbf{s}}_m}$ and $a_{\mathsf{R}}({\mathbf{k}},{\mathbf{r}}_n)={\rm{e}}^{{\rm{j}}{\mathbf{k}}^{\mathsf{T}}{\mathbf{r}}_n}$ represent the source response and the receive response, respectively. Here, ${\bm{\kappa}}=[\kappa_x,\kappa_y,\kappa_z]^{\mathsf{T}}\in{\mathbbmss{R}}^{3\times1}$ with $\kappa_y=\gamma(\kappa_x,\kappa_z)=\sqrt{(2\pi/\lambda)^2-\kappa_x^2-\kappa_z^2}$, and ${\mathbf{k}}=[k_x,k_y,k_z]^{\mathsf{T}}\in{\mathbbmss{R}}^{3\times1}$ with $k_y=\gamma(k_x,k_z)=\sqrt{(2\pi/\lambda)^2-k_x^2-k_z^2}$. The integration region in \eqref{4DFPE_Stadard} satisfies
\begin{align}
{\mathcal{D}}({\bm\kappa})&=\{(\kappa_x,\kappa_z)\in{\mathbbmss{R}}^{2}:\kappa_x^2+\kappa_z^2\leq (2\pi/\lambda)^2\},\\
{\mathcal{D}}({\mathbf{k}})&=\{(k_x,k_z)\in{\mathbbmss{R}}^{2}:k_x^2+k_z^2\leq (2\pi/\lambda)^2\}.
\end{align}
It is worth noting that ${\rm{e}}^{-{\rm{j}}{\bm{\kappa}}^{\mathsf{T}}{\mathbf{s}}_m}$ represents the incident plane wave from ${\mathbf{s}}_m$ along the direction $\frac{\bm\kappa}{\lVert\bm\kappa\rVert}$, while ${\rm{e}}^{{\rm{j}}{\mathbf{k}}^{\mathsf{T}}{\mathbf{r}}_n}$ represents the received plane wave at ${\mathbf{r}}_n$ from the direction $\frac{\mathbf{k}}{\lVert\mathbf{k}\rVert}$, hence the name \emph{4D Fourier plane-wave representation}. The function $H_a(k_x,k_z,{\kappa}_x,{\kappa}_z)$ represents the angular or wavenumber response from the transmit direction $\frac{\bm\kappa}{\lVert\bm\kappa\rVert}$ to the receive direction $\frac{\mathbf{k}}{\lVert\mathbf{k}\rVert}$, which is determined by the \emph{scattering environment} and the \emph{array geometry}. When considering only NLoS components, $H_a(k_x,k_z,{\kappa}_x,{\kappa}_z)$ can be characterized as a zero-mean complex Gaussian random variable. However, when both LoS and NLoS components are taken into account, the Fourier coefficient $H_a(k_x,k_z,{\kappa}_x,{\kappa}_z)$ can be decomposed as follows \cite{pizzo2020holographic,zhang2023fundamental}:
\begin{equation}
\begin{split}
H_a(k_x,k_z,{\kappa}_x,{\kappa}_z)&=\bar{H}_a(k_x,k_z,{\kappa}_x,{\kappa}_z)\\
&+\tilde{H}_a(k_x,k_z,{\kappa}_x,{\kappa}_z).
\end{split}
\end{equation}
Here, $\bar{H}_a(k_x,k_z,{\kappa}_x,{\kappa}_z)$ represents the coefficient of the LoS component, which is a deterministic complex constant, and $\tilde{H}_a(k_x,k_z,{\kappa}_x,{\kappa}_z)$ represents the coefficient of the NLoS component, which is a zero-mean complex Gaussian random variable. 

Upon dividing the integration region ${\mathcal{D}}({\bm\kappa})\times{\mathcal{D}}({\mathbf{k}})$ of $h_{n,m}$ uniformly with angular sets ${\mathcal{W}}_{\mathsf{S}}(\ell_x,\ell_z)=\left\{\left[\frac{2\pi \ell_x}{L_{{\mathsf{S}},x}},\frac{2\pi (\ell_x+1)}{L_{{\mathsf{S}},x}}\right]\times \left[\frac{2\pi \ell_z}{L_{{\mathsf{S}},z}},\frac{2\pi (\ell_z+1)}{L_{{\mathsf{S}},z}}\right]\right\}$ and ${\mathcal{W}}_{\mathsf{R}}(q_x,q_z)=\left\{\left[\frac{2\pi q_x}{L_{{\mathsf{R}},x}},\frac{2\pi (q_x+1)}{L_{{\mathsf{R}},x}}\right]\times \left[\frac{2\pi q_z}{L_{{\mathsf{R}},z}},\frac{2\pi (q_z+1)}{L_{{\mathsf{R}},z}}\right]\right\}$, we obtain
\begin{equation}\label{4DFPE_Appr1}
\begin{split}
&h_{n,m}=\frac{1}{(2\pi)^2}\sum_{(\ell_x,\ell_z)\in{\mathcal{E}}_{\mathsf{S}}}\sum_{(q_x,q_z)\in{\mathcal{E}}_{\mathsf{R}}}
\iiiint_{{\mathcal{W}}_{\mathsf{S}}(\ell_x,\ell_z)\times {\mathcal{W}}_{\mathsf{R}}(q_x,q_z)}\\
&\times {\mathbbm{1}}_{\mathcal{D}(\mathbf{k})}({\mathcal{W}}_{\mathsf{R}}(q_x,q_z))a_{\mathsf{R}}(\mathbf{k},{\mathbf{r}}_n)H_a(k_x,k_z,{\kappa}_x,{\kappa}_z)\\ &\times a_{\mathsf{S}}({\bm{\kappa}},{\mathbf{s}}_m){\mathbbm{1}}_{\mathcal{D}(\bm\kappa)}({\mathcal{W}}_{\mathsf{S}}(\ell_x,\ell_z)){\rm{d}}\kappa_x{\rm{d}}\kappa_z{\rm{d}}k_x{\rm{d}}k_z,
\end{split}
\end{equation}
where ${\mathcal{E}}_{\mathsf{S}}\triangleq\{(\ell_x,\ell_z)\in{\mathbbmss{Z}}^2:(\frac{\ell_x \lambda}{L_{{\mathsf{S}},x}})^2+(\frac{\ell_z \lambda}{L_{{\mathsf{S}},z}})^2\leq 1\}$ and ${\mathcal{E}}_{\mathsf{R}}\triangleq\{(q_x,q_z)\in{\mathbbmss{Z}}^2:(\frac{q_x \lambda}{L_{{\mathsf{R}},x}})^2+(\frac{q_z \lambda}{L_{{\mathsf{R}},z}})^2\leq 1\}$. The cardinality of ${\mathcal{E}}_{\mathsf{S}}$ and ${\mathcal{E}}_{\mathsf{R}}$ can be evaluated by
\begin{subequations}
\begin{align}
\label{25(a)}
n_{\mathsf{S}}&\triangleq \lvert {\mathcal{E}}_{\mathsf{S}} \rvert= \left\lceil\frac{\pi}{\lambda^2}L_{{\mathsf{S}},x}L_{{\mathsf{S}},z}\right\rceil
+o\left(\frac{L_{{\mathsf{S}},x}L_{{\mathsf{S}},z}}{\lambda^2}\right),\\
n_{\mathsf{R}}&\triangleq \lvert {\mathcal{E}}_{\mathsf{R}} \rvert= \left\lceil\frac{\pi}{\lambda^2}L_{{\mathsf{R}},x}L_{{\mathsf{R}},z}\right\rceil
+o\left(\frac{L_{{\mathsf{R}},x}L_{{\mathsf{R}},z}}{\lambda^2}\right).
\end{align}
\end{subequations}
Applying the first mean-value theorem over each partition gives rise to the Fourier series expansion:
\begin{equation}\label{4DFPE_Appr2}
\begin{split}
h_{n,m}&\approx\sum_{(\ell_x,\ell_z)\in{\mathcal{E}}_{\mathsf{S}}}\sum_{(q_x,q_z)\in{\mathcal{E}}_{\mathsf{R}}}
\phi_{\mathsf{R}}(q_x,q_z,{\mathbf{r}}_n)\\
&\times{\rm{e}}^{{\rm{j}}\gamma(\frac{2\pi (q_x+1/2)}{L_{{\mathsf{R}},x}},\frac{2\pi (q_z+1/2)}{L_{{\mathsf{R}},z}})r_y^{(n)}}\tilde{H}(q_x,q_z,\ell_x,\ell_z)\\
&\times  {\rm{e}}^{-{\rm{j}}\gamma(\frac{2\pi (\ell_x+1/2)}{L_{{\mathsf{S}},x}},\frac{2\pi (\ell_z+1/2)}{L_{{\mathsf{S}},z}})s_y^{(m)}}
\phi_{\mathsf{S}}^{*}(\ell_x,\ell_z,{\mathbf{s}}_m),
\end{split}
\end{equation}
where $\phi_{\mathsf{R}}(q_x,q_z,{\mathbf{r}}_n)={\rm{e}}^{{\rm{j}}(\frac{2\pi q_x}{L_{{\mathsf{R}},x}}r_x^{(m)}+\frac{2\pi q_z}{L_{{\mathsf{R}},z}}r_z^{(m)})}$, $\phi_{\mathsf{S}}(\ell_x,\ell_z,{\mathbf{s}}_m)={\rm{e}}^{{\rm{j}}(\frac{2\pi \ell_x}{L_{{\mathsf{S}},x}}s_x^{(m)}+\frac{2\pi \ell_z}{L_{{\mathsf{S}},z}}s_z^{(m)})}$, and 
\begin{equation}
\begin{split}
&\tilde{H}(q_x,q_z,\ell_x,\ell_z)=\frac{1}{(2\pi)^2}
\iiiint_{{\mathcal{W}}_{\mathsf{S}}(\ell_x,\ell_z)\times {\mathcal{W}}_{\mathsf{R}}(q_x,q_z)}\\
&\times {\mathbbm{1}}_{\mathcal{D}(\mathbf{k})}({\mathcal{W}}_{\mathsf{R}}(q_x,q_z))H_a(k_x,k_z,{\kappa}_x,{\kappa}_z){\rm{e}}^{{\rm{j}}(\frac{\pi r_x^{(m)}}{L_{{\mathsf{R}},x}}+\frac{\pi r_z^{(m)}}{L_{{\mathsf{R}},z}})}\\
&\times {\mathbbm{1}}_{\mathcal{D}(\bm\kappa)}({\mathcal{W}}_{\mathsf{S}}(\ell_x,\ell_z)) {\rm{e}}^{-{\rm{j}}(\frac{\pi s_x^{(m)}}{L_{{\mathsf{S}},x}}+\frac{\pi s_z^{(m)}}{L_{{\mathsf{S}},z}})}{\rm{d}}\kappa_x{\rm{d}}\kappa_z{\rm{d}}k_x{\rm{d}}k_z.
\end{split}
\end{equation}
When only considering the NLoS components, we have $\tilde{H}(q_x,q_z,\ell_x,\ell_z)\sim{\mathcal{CN}}(0,\sigma^2(q_x,q_z,\ell_x,\ell_z))$, where the calculation method for $\sigma^2(q_x,q_z,\ell_x,\ell_z)$ is detailed in \cite[Appendix]{pizzo2022fourier}. Furthermore, based on the concept of definite integral, the approximations in \eqref{4DFPE_Appr1} and \eqref{4DFPE_Appr2} are asymptotically accurate in the regime of  
\begin{align}
\min\left\{\frac{L_{{\mathsf{S}},x}}{ \lambda},\frac{L_{{\mathsf{S}},z}}{ \lambda},\frac{L_{{\mathsf{R}},x}}{ \lambda},\frac{L_{{\mathsf{R}},z}}{ \lambda}\right\}\gg 1.
\end{align}
Moreover, the approximation retains all channel information when the antenna separation is at most half-wavelength, i.e., $\max\{d_{\mathsf{S}},d_{\mathsf{R}}\}\leq \frac{\lambda}{2}$ \cite{pizzo2020spatially,pizzo2022nyquist}. In other words, no information loss occurs when
\begin{align}
N\geq \frac{4 L_{{\mathsf{R}},x} L_{{\mathsf{R}},z}}{\lambda^2}\geq n_{\mathsf{R}},
M\geq \frac{4 L_{{\mathsf{S}},x} L_{{\mathsf{S}},z}}{\lambda^2}\geq n_{\mathsf{S}}.
\end{align}

\begin{figure}[!t]
 \centering
\includegraphics[width=0.45\textwidth]{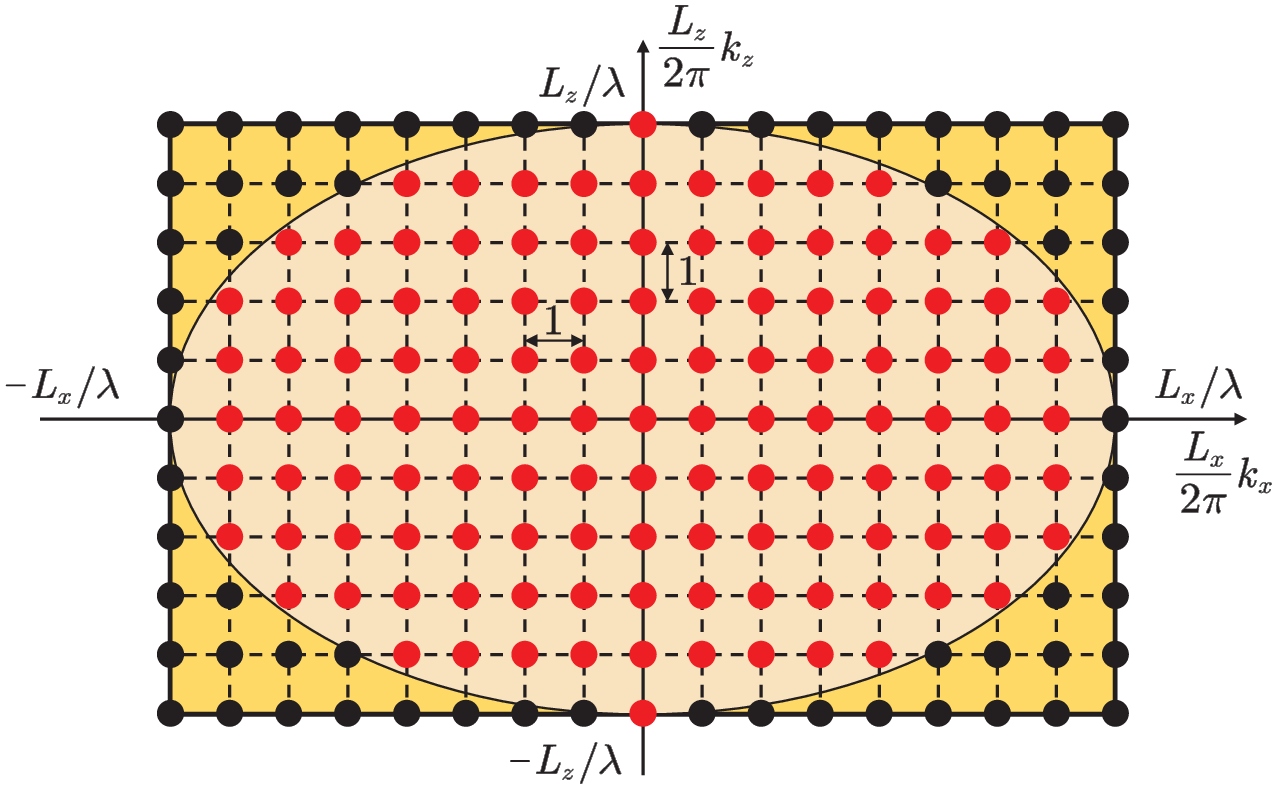}
\caption{The 2D lattice with elliptical wavenumber support. $L_x\in\{L_{{\mathsf{S}},x},L_{{\mathsf{R}},x}\}$ and $L_z\in\{L_{{\mathsf{S}},z},L_{{\mathsf{R}},z}\}$ are the aperture sizes along the $x$- and $z$-axes. $k_x$ and $k_z$ are integers.}
\label{4DFPE}
\end{figure}

\begin{figure}[!t]
  \centering
    \subfigure[Channel coupling between the transmit angular domain to the receive angular domain.]{
        \includegraphics[width=0.48\textwidth]{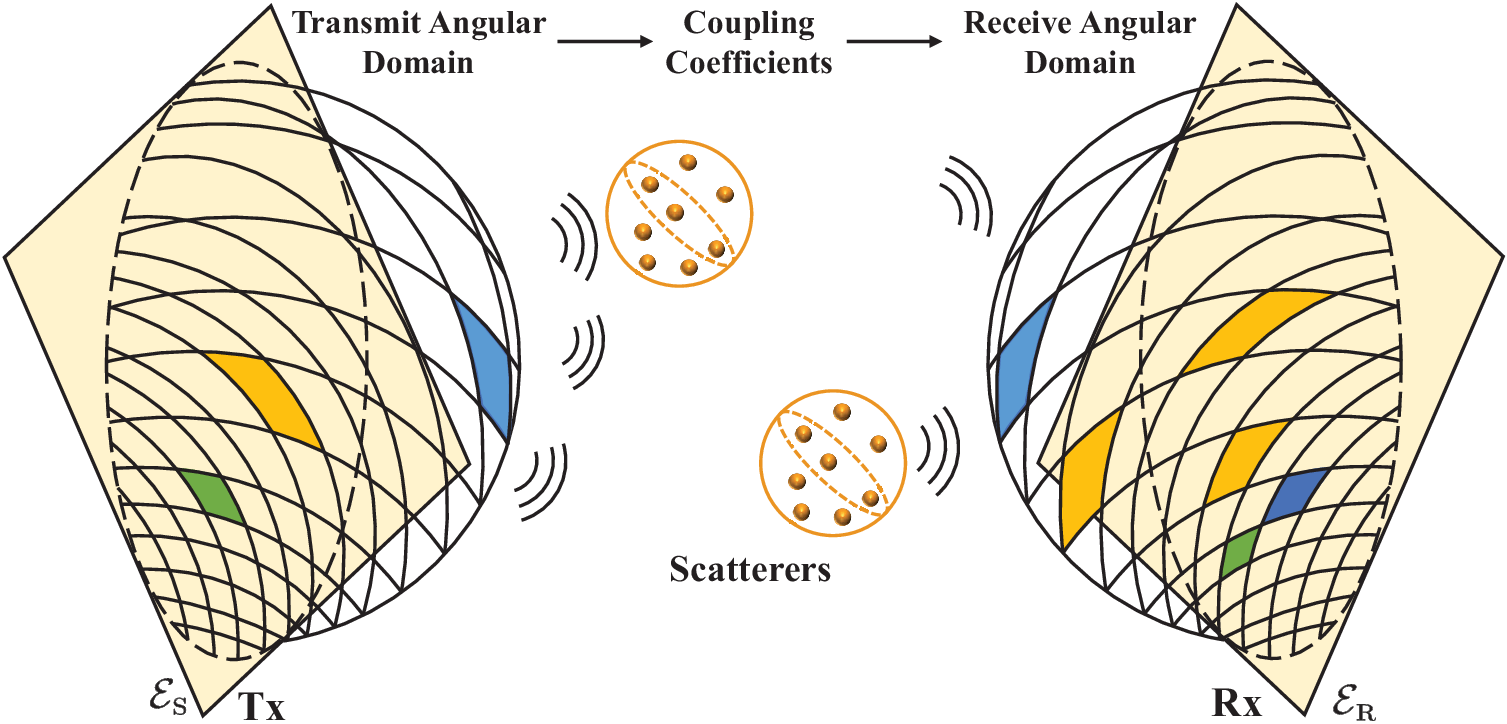}
        \label{Figure: Wavenumber_Domain_Channel}	
    }
    \subfigure[Power transfer from the transmit angular domain to the receive angular domain.]{
        \includegraphics[height=0.3\textwidth]{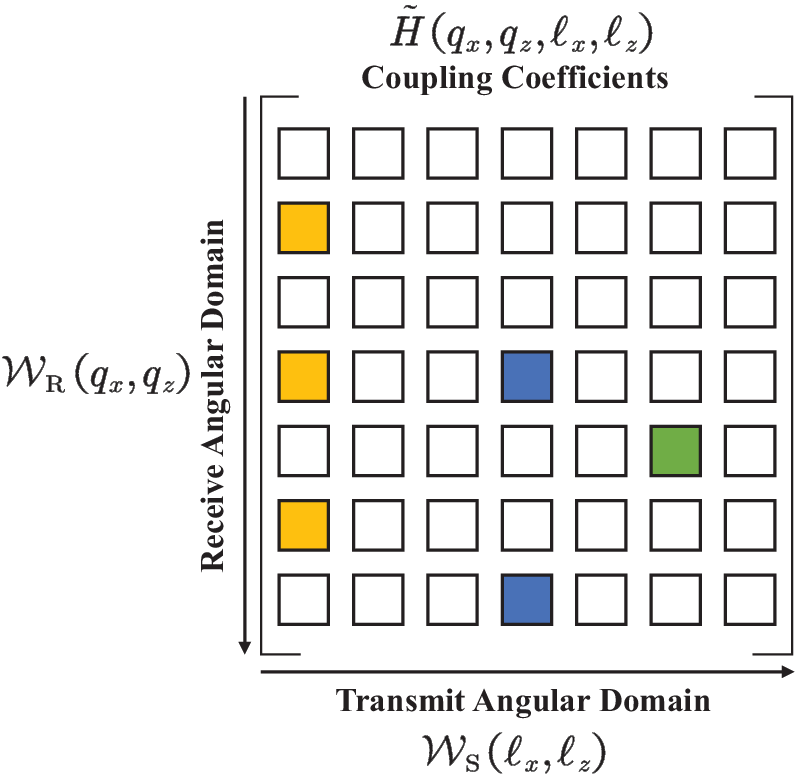}
        \label{Figure: Wavenumber_Coupling}	
    }
  \caption{Illustration of the Fourier plane-wave series expansion-based model in the angular or wavenumber domain.}
  \label{Figure: Fourier plane-wave series}
\end{figure}

The discretization in \eqref{4DFPE_Appr2} is akin to the transition from Fourier integrals to Fourier series for time-domain signals \cite{pizzo2020spatially}. Since the angular response is non-zero solely within specific wavenumber regions, viz. ${\mathcal{D}}({\bm\kappa})\times{\mathcal{D}}({\mathbf{k}})$, the discretized responses are confined within lattice ellipses, viz. ${\mathcal{E}}_{\mathsf{S}}\times{\mathcal{E}}_{\mathsf{R}}$, as illustrated in {\figurename} {\ref{4DFPE}}. Subsequent to discretization, the continuous incident plane wave ${\rm{e}}^{-{\rm{j}}{\bm{\kappa}}^{\mathsf{T}}{\mathbf{s}}_m}$ and received plane wave ${\rm{e}}^{{\rm{j}}{\mathbf{k}}^{\mathsf{T}}{\mathbf{r}}_n}$ are replaced by their discretized counterparts. Furthermore, the continuous angular response $H_a(k_x,k_z,{\kappa}_x,{\kappa}_z)$ is replaced by a sequence $\{\tilde{H}(q_x,q_z,\ell_x,\ell_z)\}$, describing \emph{the channel coupling} between each pair of transmit and receive angular sets designated by the aforementioned discretized plane waves, viz. $({\mathcal{W}}_{\mathsf{S}}(\ell_x,\ell_z),{\mathcal{W}}_{\mathsf{R}}(q_x,q_z))$, as depicted in {\figurename} {\ref{Figure: Wavenumber_Domain_Channel}}. 

For clarity, in {\figurename} {\ref{Figure: Wavenumber_Coupling}}, the coupling coefficients are arranged in matrix form. Three distinct angular sets, activated by the source (orange, blue, and green), transfer their radiated power to six angular sets at the receiver. The resulting random channel coupling coefficients are modeled as zero-mean circularly-symmetric complex-Gaussian random variables. Their variance represents the discrete angular power distribution of the channel, indicating the proportion of power transferred from the transmit angular domain ${\mathcal{W}}_{\mathsf{S}}(\ell_x,\ell_z)$ to the receive angular domain ${\mathcal{W}}_{\mathsf{R}}(q_x,q_z)$. The strength of the coupling coefficients varies and depends on both array apertures and the scattering mechanism. Due to their inherent Gaussian-distributed complex gains, the Fourier plane-wave series expansion-based model exhibits the characteristics of correlated Rayleigh fading, as detailed below.

Using the fact that $r_y^{(n)}=r$ ($\forall n$) and $s_y^{(m)}=0$ ($\forall m$), the channel matrix can be represented as follows:
\begin{equation}
\begin{split}
\mathbf{H}&=[h_{n,m}]\approx \sqrt{MN}\sum_{(\ell_x,\ell_z)\in{\mathcal{E}}_{\mathsf{S}}}\sum_{(q_x,q_z)\in{\mathcal{E}}_{\mathsf{R}}}
{\bm\phi}_{\mathsf{R}}(q_x,q_z)\\
&\times {\rm{e}}^{{\rm{j}}\gamma(\frac{2\pi (q_x+1/2)}{L_{{\mathsf{R}},x}},\frac{2\pi (q_z+1/2)}{L_{{\mathsf{R}},z}})r}\tilde{H}(q_x,q_z,\ell_x,\ell_z)\\ 
&\times {\rm{e}}^{-{\rm{j}}\gamma(\frac{2\pi (\ell_x+1/2)}{L_{{\mathsf{S}},x}},\frac{2\pi (\ell_z+1/2)}{L_{{\mathsf{S}},z}})\times0}
{\bm\phi}_{\mathsf{S}}^{\mathsf{H}}(\ell_x,\ell_z),
\end{split}
\end{equation}
where ${\bm\phi}_{\mathsf{R}}(q_x,q_z)=\frac{1}{\sqrt{N}}[\phi_{\mathsf{R}}(q_x,q_z,{\mathbf{r}}_n)]_{\forall n}^{\mathsf{T}}\in{\mathbbmss{C}}^{N\times1}$ and ${\bm\phi}_{\mathsf{S}}(\ell_x,\ell_z)=\frac{1}{\sqrt{M}}[\phi_{\mathsf{S}}(\ell_x,\ell_z,{\mathbf{s}}_m)]_{\forall m}^{\mathsf{T}}\in{\mathbbmss{C}}^{M\times1}$. Denoting ${\bm\Phi}_{\mathsf{S}}\in{\mathbbmss{C}}^{M\times n_{\mathsf{S}}}$ and ${\bm\Phi}_{\mathsf{R}}\in{\mathbbmss{C}}^{N\times n_{\mathsf{R}}}$ as the deterministic matrices collecting the $n_{\mathsf{S}}$ and $n_{\mathsf{R}}$ vectors $\{{\bm\phi}_{\mathsf{S}}(\ell_x,\ell_z)\}$ and $\{{\bm\phi}_{\mathsf{R}}(q_x,q_z)\}$, respectively, we have ${\bm\Phi}_{\mathsf{S}}^{\mathsf{H}}{\bm\Phi}_{\mathsf{S}}={\mathbf{I}}_{n_{\mathsf{S}}}$ and ${\bm\Phi}_{\mathsf{R}}^{\mathsf{H}}{\bm\Phi}_{\mathsf{R}}={\mathbf{I}}_{n_{\mathsf{R}}}$. Consequently, the channel matrix can be further expressed as follows:
\begin{equation}
\begin{split}
\mathbf{H}={\bm\Phi}_{\mathsf{R}}{\bm\Gamma}_{\mathsf{R}} ({\tilde{\mathbf{H}}} \odot \tilde{\bm\Omega}) {\bm\Gamma}_{\mathsf{S}}^{\mathsf{H}} {\bm\Phi}_{\mathsf{S}}^{\mathsf{H}},
\end{split}
\end{equation}
where ${\bm\Gamma}_{\mathsf{R}}={\mathsf{diag}}([{\rm{e}}^{{\rm{j}}\gamma(\frac{2\pi (q_x+1/2)}{L_{{\mathsf{R}},x}},\frac{2\pi (q_z+1/2)}{L_{{\mathsf{R}},z}})r}]_{(q_x,q_z)\in{\mathcal{E}}_{\mathsf{R}}}^{\mathsf{T}})\in{\mathbbmss{C}}^{n_{\mathsf{R}} \times n_{\mathsf{R}}}$, ${\bm\Gamma}_{\mathsf{S}}={\mathsf{diag}}([{\rm{e}}^{{\rm{j}}\gamma(\frac{2\pi (\ell_x+1/2)}{L_{{\mathsf{S}},x}},\frac{2\pi (\ell_z+1/2)}{L_{{\mathsf{S}},z}})\times0}]_{(\ell_x,\ell_z)\in{\mathcal{E}}_{\mathsf{S}}}^{\mathsf{T}})\in{\mathbbmss{C}}^{n_{\mathsf{S}} \times n_{\mathsf{S}}}$, $\tilde{\bm\Omega}=[\sqrt{MN}\sigma(q_x,q_z,\ell_x,\ell_z)]_{\forall (q_x,q_z)\in{\mathcal{E}}_{\mathsf{R}}, (\ell_x,\ell_z)\in{\mathcal{E}}_{\mathsf{S}}}\in{\mathbbmss{C}}^{n_{\mathsf{R}}\times n_{\mathsf{S}}}$, and ${\tilde{\mathbf{H}}}\in{\mathbbmss{C}}^{n_{\mathsf{R}}\times n_{\mathsf{S}}}$ consists of $n_{\mathsf{R}} n_{\mathsf{S}}$ independent and identically distributed (i.i.d.) standard complex Gaussian random variables. Generally, ${\tilde{\mathbf{H}}} \odot \tilde{\bm\Omega}\in{\mathbbmss{C}}^{n_{\mathsf{R}}\times n_{\mathsf{S}}}$ is also referred to as the angle- or wavenumber-domain channel, which is semi-unitarily equivalent to $\mathbf{H}$ but with significantly lower dimensions since $n_{\mathsf{R}}\ll N$ and $n_{\mathsf{S}}\ll M$. Moreover, the matrix $\tilde{\bm\Omega}\odot\tilde{\bm\Omega}$ reflects the strength of the channel coupling from the transmit angular domain to the receive angular domain, exhibiting strong sparsity, especially under non-isotropic propagation conditions. If the transceivers are equipped with linear arrays instead of planar arrays, the resultant channel matrix is derived by selecting a submatrix from $\mathbf{H}$ through adjustments to ${\bm\Phi}_{\mathsf{R}}{\bm\Gamma}_{\mathsf{R}}$ and ${\bm\Gamma}_{\mathsf{S}}^{\mathsf{H}} {\bm\Phi}_{\mathsf{S}}^{\mathsf{H}}$. In the scenario where the receiver is equipped with a single antenna, the corresponding channel vector can be obtained by selecting a specific row of $\mathbf{H}$.

This transformation reveals that only a finite number of plane waves are necessary to convey the fundamental channel information between the two arrays, emphasizing the lower-dimensional angular representation in EM channels. This results in a significant reduction in channel dimensions, which in turn significantly lowers the complexity involved in tasks such as channel estimation, optimal signaling, and coding.
\section{Information-Theoretic Limits of NFC-NGMA}\label{Section: Information-Theoretic Limits of NFC-NGMA}
By definition, MA involves the simultaneous communications of multiple users \cite{el1980multiple}. When $K$ users communicate concurrently with a receiver, this configuration is termed a multiple access channel (MAC). Conversely, if the configuration is inverted such that one transmitter broadcasts independent messages simultaneously to $K$ users, it transforms into a broadcast channel (BC). MAC and BC correspond to uplink and downlink MA transmission, respectively. In the following, we explore the information-theoretic limits of these two channels in the context of NFC, with an emphasis on the limits of channel capacity, i.e., capacity region and sum-rate capacity. For brevity, the subsequent system descriptions will focus on the two-user scenario, with the understanding that these descriptions can be trivially extended to cases involving more than two users. Additionally, as a theoretical exploration of fundamental performance limits and asymptotic behaviors, we consider basic free-space LoS propagation scenarios.
\subsection{Capacity Limits for SPD Arrays}
\begin{figure}[!t]
 \centering
\includegraphics[height=0.3\textwidth]{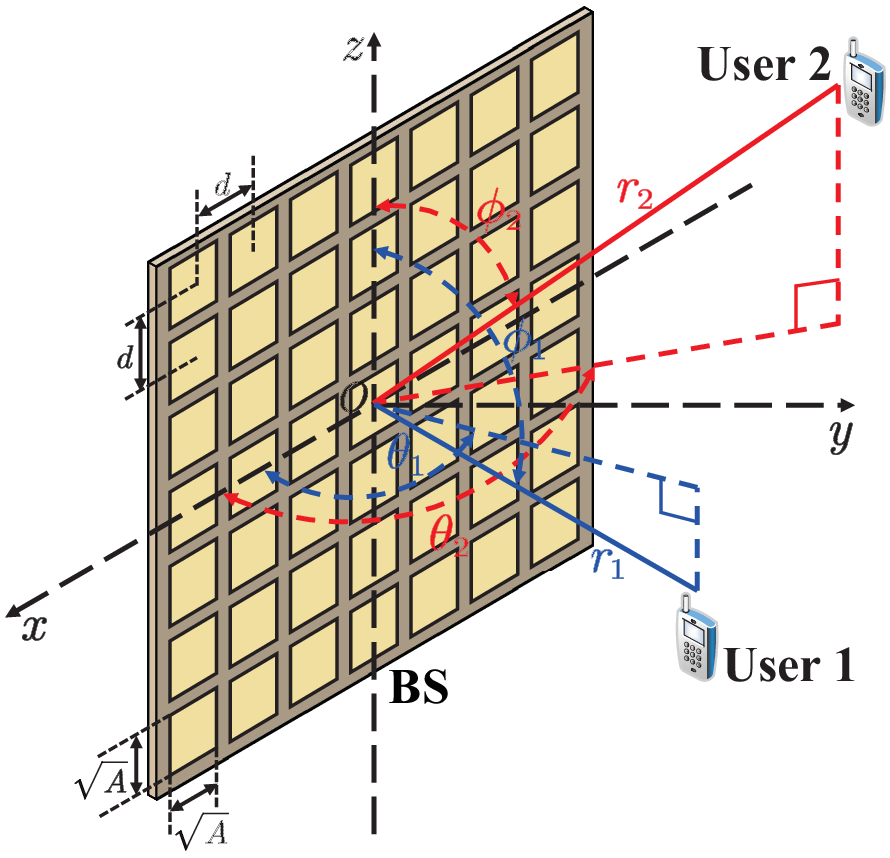}
\caption{Illustration of the array geometry of a two-user channel.}
\label{LoS_3D_Model_MU}
\end{figure}

We begin our exploration of capacity limits with the LoS channel relying on SPD arrays, where the system layout is sketched in {\figurename} {\ref{LoS_3D_Model_MU}}. Specifically, each user $k=1,2$ is equipped with a single hypothetical isotropic antenna element, and the BS is equipped with an $M$-antenna UPA that follows the structure depicted in {\figurename} {\ref{SPD_UPA_Array}}. Here, $M=M_{x}M_{z}$, where $M_{x}$ and $M_{z}$ represent the number of antenna elements along the $x$- and $z$-axes, respectively. For brevity, we assume that $M_{x}$ and $M_{z}$ are odd numbers with $M_x=2\tilde{M}_x+1$ and $M_z=2\tilde{M}_z+1$. The physical dimensions of each BS array element along the $x$- and $z$-axes are denoted by $\sqrt{A}$, and the inter-element distance is $d$, where $d\geq\sqrt{A}$. The central location of the $(m_x,m_z)$th element is given by ${\mathbf{s}}_{m_x,m_z}=[m_xd,0,m_zd]^{\mathsf{T}}$, where $m_x\in{\mathcal{M}}_x\triangleq\{0,\pm1,\ldots,\pm\tilde{M}_x\}$ and $m_z\in{\mathcal{M}}_z\triangleq\{0,\pm1,\ldots,\pm\tilde{M}_z\}$. Let $r_k$ denote the propagation distance from the center of the antenna array to user $k=1,2$, and $\theta_k\in[0,\pi]$ and $\phi_k\in[0,\pi]$ denote the associated azimuth and elevation angles, respectively. The location of user $k$ can then be expressed as ${\mathbf{r}}_k=r_k[\Phi_k,\Psi_k,\Omega_k]^{\mathsf{T}}$, where $\Phi_k\triangleq\sin{\phi_k}\cos{\theta_k}$, $\Psi_k\triangleq\sin\phi_k\sin\theta_k$, and $\Omega_k\triangleq\cos{\phi_k}$. Utilizing the channel model described in \eqref{SPD_UPA_NFC_Model}, the near-field channel vector of user $k=1,2$ is given by
\begin{align}\label{SPD_UPA_NFC_LoS_Model_User}
{\mathbf{h}}_k=\left[\frac{\sqrt{r_k\Psi_k A}}{\sqrt{4\pi}\lVert{\mathbf{r}}_k-{\mathbf{s}}_{m_x,m_z}\rVert^{3/2}}{\rm{e}}^{-{\rm{j}}\frac{2\pi}{\lambda}\lVert{\mathbf{r}}_k-{\mathbf{s}}_{m_x,m_z}\rVert}\right]_{\forall m_x,m_z}.
\end{align}
In the subsequent analyses, we assume perfect knowledge of the channel at both the transmitter and the receivers.
\subsubsection{Preliminaries}\label{Section: Capacity Limits for LoS Channels: Preliminaries}
Before characterizing the capacity limits, we introduce some preliminary concepts that will be instrumental in the subsequent analyses. 

$\bullet$ \emph{Discussion on Channel Gain}: As per \eqref{SPD_UPA_NFC_LoS_Model_User}, the channel gain of user $k=1,2$ can be calculated as follows:
\begin{equation}\label{SU_NFC_Channel_Gain}
\begin{split}
{\mathsf{a}}_k&=\lVert{\mathbf{h}}_k\rVert^2=\frac{r_k\Psi_k A}{4\pi}\sum_{m_x=-\tilde{M}_x}^{\tilde{M}_x}\sum_{m_z=-\tilde{M}_z}^{\tilde{M}_z}
\frac{1}{\lVert{\mathbf{r}}_k-{\mathbf{s}}_{m_x,m_z}\rVert^{3}}\\
&=\frac{\Psi_k A}{4\pi r_k^2}\sum_{m_x,m_z}
\frac{1}{((m_x\varepsilon_k-\Phi_k)^2+{\Psi}_k^2+(m_z\varepsilon_k-\Omega_k)^2)^{\frac{3}{2}}}
\end{split}
\end{equation}
with $\varepsilon_k\triangleq\frac{d}{r_k}\ll1$. For clarity, we define the following function:
\begin{align}
f_{\mathsf{gain}}^{(k,\ell)}(x,z)\triangleq((x-\Phi_k)^2+{\Psi}_k^2+(z-\Omega_k)^2)^{-\frac{\ell}{2}},
\end{align}
within the rectangular area ${\mathcal{H}}^{(k)}=\{\left.(x,z)\right|-\frac{M_x\varepsilon_k}{2}\leq x\leq\frac{M_x\varepsilon_k}{2},-\frac{M_z\varepsilon_k}{2}\leq z\leq\frac{M_z\varepsilon_k}{2}\}$. This area is further partitioned into $M_xM_z$ sub-rectangles, each with an area of $\varepsilon_k^2$. Given $\varepsilon_k\ll 1$, we have $f_{\mathsf{gain}}^{(k,\ell)}(x,z)\approx f_{\mathsf{gain}}^{(k,\ell)}(m_x\varepsilon_k,m_z\varepsilon_k)$ for $\forall (x,z)\in\{\left.(x,z)\right|(m_x-\frac{1}{2})\varepsilon_k\leq x\leq(m_x+\frac{1}{2})\varepsilon_k,(m_z-\frac{1}{2})\varepsilon_k\leq z\leq(m_z+\frac{1}{2})\varepsilon_k,m_x\in{\mathcal{M}}_x,m_z\in{\mathcal{M}}_z\}$. Using the concept of double integrals, we obtain
\begin{align}\label{Received_SNR_Analog_SU_Near_Field_Trans_Pre}
\sum_{m_x,m_z}f_{\mathsf{gain}}^{(k,\ell)}(m_x\varepsilon_k,m_z\varepsilon_k)\varepsilon_k^2\approx
\iint_{{\mathcal{H}}^{(k)}}f_{\mathsf{gain}}^{(k,\ell)}(x,z){\rm{d}}x{\rm{d}}z.
\end{align}
Inserting \eqref{Received_SNR_Analog_SU_Near_Field_Trans_Pre} into \eqref{SU_NFC_Channel_Gain} gives 
\begin{equation}\label{SU_NFC_Channel_Gain_Result}
\begin{split}
{\mathsf{a}}_k\approx\frac{\Psi_k A}{4\pi d^2}\iint_{{\mathcal{H}}^{(k)}}f_{\mathsf{gain}}^{(k,3)}(x,z){\rm{d}}x{\rm{d}}z,
\end{split}
\end{equation}
which, together with the integral identities \cite[Eq. (2.264.5)]{gradshteyn2014table} and \cite[Eq. (2.284)]{gradshteyn2014table}, yields
\begin{equation}\label{SU_NFC_Channel_Gain_Result_Final}
\begin{split}
{\mathsf{a}}_k\approx\frac{A}{4\pi d^2}\sum_{x\in{\mathcal{X}}_k}\sum_{z\in{\mathcal{Z}}_k}\arctan\left(\frac{xz}{\Psi_k\sqrt{\Psi_k^2+x^2+z^2}}\right),
\end{split}
\end{equation}
where ${\mathcal{X}}_k=\{\frac{M_x}{2}\varepsilon_k\pm\Phi_k\}$ and ${\mathcal{Z}}_k=\{\frac{M_z}{2}\varepsilon_k\pm\Psi_k\}$. By letting $M_x,M_z\rightarrow\infty$, the asymptotic channel gain of user $k$ is obtained as follows:
\begin{equation}\label{SU_NFC_Channel_Gain_Result_Asymptotic}
\begin{split}
{\mathsf{a}}_k\rightarrow\frac{\lim\limits_{x,z\rightarrow\infty}\arctan\left(\frac{xz/\Psi_k}{\sqrt{\Psi_k^2+x^2+z^2}}\right)}{4\pi d^2/(4A)}=\frac{4A\frac{\pi}{2}}{4\pi d^2}=\frac{\xi_{\mathsf{r}}}{2},
\end{split}
\end{equation}
where $\xi_{\mathsf{r}}\triangleq\frac{A}{d^2}\leq1$ represents the array occupation ratio. 

\begin{figure}[!t]
 \centering
\includegraphics[height=0.3\textwidth]{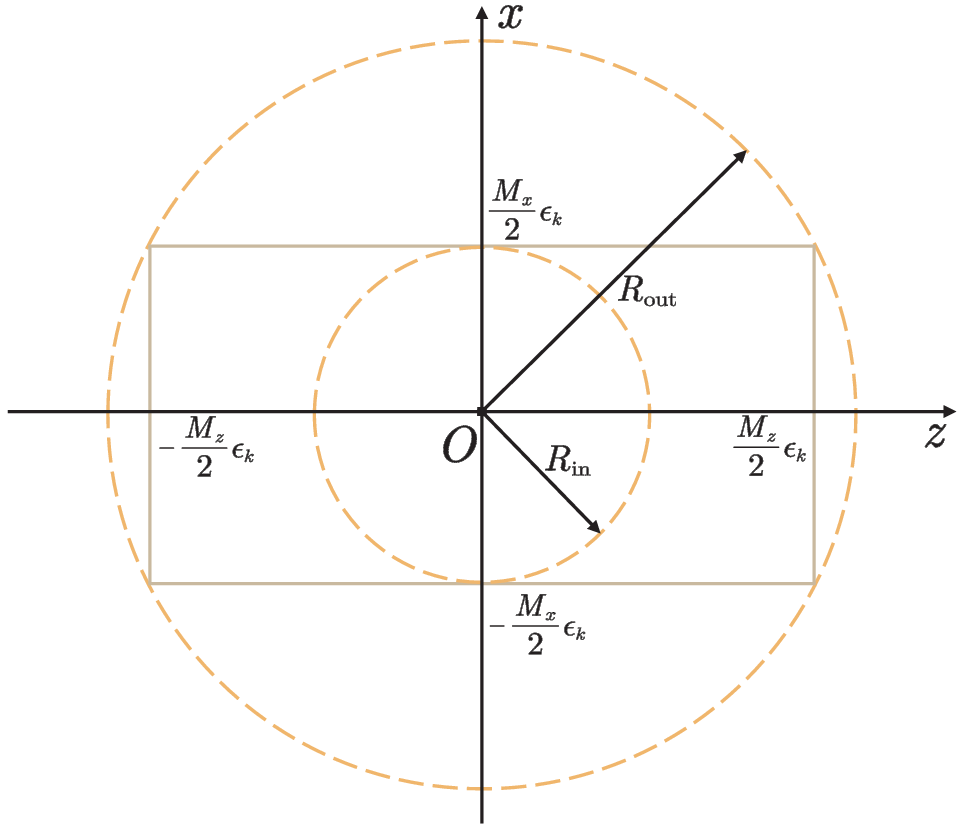}
\caption{ The inscribed and circumscribed disks of the rectangular region ${\mathcal{H}}^{(k)}=\{\left.(x,z)\right|-\frac{M_x\varepsilon_k}{2}\leq x\leq\frac{M_x\varepsilon_k}{2},-\frac{M_z\varepsilon_k}{2}\leq z\leq\frac{M_z\varepsilon_k}{2}\}$.}
\label{Integral_Region_Reactive}
\end{figure}

As discussed in Section \ref{Section: Electromagnetic Regions}, the reactive region of an SPD array corresponds to the area covered by each of its antenna elements, typically expanding only a few wavelengths. However, in the asymptotic analysis, the number of antenna elements is assumed to be infinite, prompting the question of whether the influence of the reactive region or evanescent waves can still be ignored. To address this, we utilize the Green’s function-based SPD channel model presented in \eqref{dyadic Green’s function_SPD_Model}. In this context, the channel gain needs to be adjusted for the reactive terms, which leads to the following expression:
\begin{equation}\label{SU_NFC_Channel_Gain_Result_Evanescent}
\begin{split}
{\mathsf{a}}_k&\approx\frac{\Psi_k A}{4\pi d^2}\iint_{{\mathcal{H}}^{(k)}}f_{\mathsf{gain}}^{(k,3)}(x,z){\rm{d}}x{\rm{d}}z\\
&-\left(\frac{\lambda}{2\pi r_k}\right)^2\frac{\Psi_k A}{4\pi d^2}\iint_{{\mathcal{H}}^{(k)}}f_{\mathsf{gain}}^{(k,5)}(x,z){\rm{d}}x{\rm{d}}z\\
&+\left(\frac{\lambda}{2\pi r_k}\right)^4\frac{\Psi_k A}{4\pi d^2}\iint_{{\mathcal{H}}^{(k)}}f_{\mathsf{gain}}^{(k,7)}(x,z){\rm{d}}x{\rm{d}}z.
\end{split}
\end{equation}
While \eqref{SU_NFC_Channel_Gain_Result_Evanescent} can be solved analytically, the resulting expression is intricate and may not provide significant insights. Therefore, for brevity, we present only the asymptotic results as $M_x,M_z\rightarrow\infty$. Since $\Phi_k\in[0,1]$ and $\Omega_k\in[0,1]$, \eqref{Received_SNR_Analog_SU_Near_Field_Trans_Pre} simplifies to the following form as $M_x,M_z\rightarrow\infty$:
\begin{align}
\iint_{{\mathcal{H}}^{(k)}}f_{\mathsf{gain}}^{(k,\ell)}(x,z){\rm{d}}x{\rm{d}}z\approx 
\iint_{{\mathcal{H}}^{(k)}}\frac{{\rm{d}}x{\rm{d}}z}{(x^2+z^2+\Psi_k^2)^{\frac{\ell}{2}}},
\end{align}
where the integration region is bounded by two disks with radius ${\mathfrak{R}}_{2,k}= \frac{\epsilon_k}{2}\min\{M_x,M_z\}$ and ${\mathfrak{R}}_{1,k}= \frac{\epsilon_k}{2}\sqrt{M_x^2+M_z^2}$, respectively, as illustrated in {\figurename} {\ref{Integral_Region_Reactive}}. By defining the function: $\hat{f}_{\mathsf{gain}}^{(k,\ell)}({\mathfrak{R}})\triangleq\int_{0}^{2\pi}\int_{0}^{{\mathfrak{R}}}\frac{\rho{\rm{d}}\rho{\rm{d}}\vartheta}
{(\Psi_k^2+\rho^2)^{\ell/2}}$, we have
\begin{align}
\hat{f}_{\mathsf{gain}}^{(k,\ell)}({\mathfrak{R}})=
\frac{\pi}{\frac{\ell}{2}-1}(\Psi_k^{2-\ell}-(\Psi_k^2+{\mathfrak{R}}^2)^{1-\frac{\ell}{2}}),
\end{align}
and it follows that
\begin{align}
\hat{f}_{\mathsf{gain}}^{(k,\ell)}({\mathfrak{R}}_{2,k})\leq
\iint_{{\mathcal{H}}^{(k)}}\frac{{\rm{d}}x{\rm{d}}z}{(x^2+z^2+\Psi_k^2)^{\frac{\ell}{2}}}\leq
\hat{f}_{\mathsf{gain}}^{(k,\ell)}({\mathfrak{R}}_{1,k}).
\end{align}
As $M_x,M_z\rightarrow\infty$, we have ${\mathfrak{R}}_{1,k},{\mathfrak{R}}_{2,k}\rightarrow\infty$ and 
\begin{equation}
\begin{split}
\lim_{M_x,M_z\rightarrow\infty}\hat{f}_{\mathsf{gain}}^{(k,\ell)}({\mathfrak{R}}_{1,k})
&=\lim_{M_x,M_z\rightarrow\infty}\hat{f}_{\mathsf{gain}}^{(k,\ell)}({\mathfrak{R}}_{2,k})\\
&=\frac{\pi}{(\frac{\ell}{2}-1)\Psi_k^{\ell-2}},~\ell>2
\end{split}
\end{equation}
which, together with the Squeeze Theorem, yields
\begin{equation}\label{Channel_Gain_Ganeral_Basic}
\lim_{M_x,M_z\rightarrow\infty}\iint_{{\mathcal{H}}^{(k)}}\frac{{\rm{d}}x{\rm{d}}z}{(x^2+z^2+\Psi_k^2)^{\frac{\ell}{2}}}=\frac{\pi}{(\frac{\ell}{2}-1)\Psi_k^{\ell-2}}.
\end{equation}
Upon substituting \eqref{Channel_Gain_Ganeral_Basic} into \eqref{SU_NFC_Channel_Gain_Result_Evanescent}, we obtain the asymptotic channel gain as follows:
\begin{align}
{\mathsf{a}}_k^{\mathsf{SPD},{\mathsf{eva}}}=\frac{\xi_{\mathsf{r}}}{2}\left(1-\frac{1}{3}\left(\frac{\lambda/2\pi}{r_k\Psi_k}\right)^2+
\frac{1}{5}\left(\frac{\lambda/2\pi}{r_k\Psi_k}\right)^4\right).\label{SU_NFC_Channel_Gain_Result_Evanescent2}
\end{align}

By comparing \eqref{SU_NFC_Channel_Gain_Result_Evanescent2} with \eqref{SU_NFC_Channel_Gain_Result_Asymptotic}, we observe that the consideration of evanescent waves introduces additional range dimensions to the asymptotic channel gain. To further understand the impact of evanescent waves, let us examine their effect. It is clear that the additional two reactive terms in \eqref{SU_NFC_Channel_Gain_Result_Evanescent2} will diminish as the propagation distance increases. When $r_k\Psi_k=\lambda$, the value of $\frac{{\mathsf{a}}_k^{\mathsf{SPD},{\mathsf{eva}}}}{{\xi_{\mathsf{r}}}/{2}}$ is approximately 0.9917. Hence, the last two terms in \eqref{SU_NFC_Channel_Gain_Result_Evanescent2} can be disregarded when studying the asymptotic performance in practical systems where $r_k\Psi_k\gg \lambda$. In other words, the influence of the reactive region can be overlooked even when considering an infinitely large number of antennas. Consequently, in the subsequent analyses, we will omit the reactive terms presented in \eqref{SU_NFC_Channel_Gain_Result_Evanescent2}.

Consider an uplink channel where $p_k$ denotes the power transmitted by user $k$. In the large limit of $M$, the received power at the BS approaches $\frac{\xi_{\mathsf{r}}}{2}p_k$. This result intuitively aligns with the fact that, with an infinitely large array, only $\frac{\xi_{\mathsf{r}}}{2}$ of the total transmitted power will be captured by the BS. This ratio increases with the array occupation $\xi_{\mathsf{r}}$, which indicates that a larger array aperture can capture more signal power. When $\xi_{\mathsf{r}}=1$, half of the power transmitted by an isotropic source is captured, while the other half does not reach the array \cite{hu2018beyond}. Next, we analyze the channel gain under the far-field model as described in \eqref{SPD_UPA_FFC_Model}, where the channel gain of user $k$ is given by ${\frac{{\Psi_k A}}{{4\pi} r_k^2}}M$, which scales linearly with $M$. This means that for a given transmit power, increasing the number of antenna elements can increase the received power to any desired level, potentially exceeding the transmit power. This behavior breaks the law of conservation of energy. The reason for this discrepancy is that, when $M$ tends to infinity, the planar-wave propagation-based far-field model fails to capture the exact physical properties of near-field propagation.

$\bullet$ \emph{Discussion on Channel Correlation}: The correlation between channels is defined by the inner product of the normalized channels, i.e.,
\begin{align}\label{Channel_Correlation_SPD}
\rho_{\mathsf{u}}\triangleq \frac{\lvert {\mathbf{h}}_1^{\mathsf{H}}{\mathbf{h}}_2\rvert}{\lVert{\mathbf{h}}_1\rVert\lVert{\mathbf{h}}_1\rVert}\in[0,1].
\end{align}
For far-field channels, we have \cite{bjornson2016massive,bjornson2017massive}
\begin{align}\label{Channel_Correlation_Far_Field}
\lim_{M\rightarrow\infty}\rho_{\mathsf{u}}=\left\{
\begin{array}{ll}
1             & {\mathbf{a}_1}= {\mathbf{a}_2}\\
0           & {\mathbf{a}_1}\ne {\mathbf{a}_2}
\end{array} \right.,
\end{align}
where $\mathbf{a}_k=[\Phi_k,\Psi_k,\Omega_k]^{\mathsf{T}}$ represents the directional vector of user $k=1,2$. The result in \eqref{Channel_Correlation_Far_Field} suggests that in the far-field region, the channels of users situated in \emph{different directions} tend toward orthogonality as $M\rightarrow\infty$. This property can be harnessed to mitigate IUI, which inspires the concept of \emph{space-division multiple access (SDMA)} \cite{suard1998uplink}, more specifically, \emph{angle-division multiple access (ADMA)} \cite{liu2024road}. However, when two far-field users are in the same direction, their channels become parallel, which makes it challenging to distinguish between users and results in significant IUI. In contrast, for near-field channels, it follows that \cite{liu2023nearfield,zhao2023modeling,zhao2024performance,zhao2024near}
\begin{align}\label{Channel_Correlation_Near_Field}
\lim_{M\rightarrow\infty}\rho_{\mathsf{u}}\left\{
\begin{array}{ll}
=1             & {\mathbf{r}_1}= {\mathbf{r}_2}\\
\ll 1           & {\mathbf{r}_1}\ne {\mathbf{r}_2}
\end{array} \right..
\end{align}
This suggests that the near-field channels become approximately orthogonal for arbitrary users in \emph{different locations}. Unlike the asymptotic orthogonality observed only for far-field users in \emph{different directions}, this advantageous property of near-field channels arises from the introduction of the range dimension in \eqref{SPD_UPA_NFC_Model} due to spherical-wave propagation. By leveraging the insights from \eqref{Channel_Correlation_Near_Field}, SDMA in NFC can comprehensively mitigate IUI, even when users share the \emph{same direction}. As a result, SDMA in NFC manifests as ``\emph{range division multiple access (RDMA)}'' \cite{liu2024road,liu2024near,wu2023multiple}, which encompasses ADMA as a special case. Details of beamforming design for RDMA will be discussed in Section \ref{sec4}.
\subsubsection{Multiple Access Channel}\label{Section: Capacity Limits for LoS Channels: Multiple Access Channel}
The MAC refers to the scenario where all users simultaneously send their respective messages to the BS \cite{suard1998uplink,cheng1993gaussian}. The received signal vector at the BS is given by
\begin{align}\label{MAC_Channel}
{\mathbf{y}}={\mathbf{h}}_1\sqrt{p_1}s_{1}+{\mathbf{h}}_2\sqrt{p_2}s_{2}+{\mathbf{n}},
\end{align}
where $s_{k}\in{\mathbbmss{C}}$ is the signal sent by user $k=1,2$ with zero mean and unit variance, $p_k$ is the associated transmit power, and ${\mathbf{n}}\sim{\mathcal{CN}}({\mathbf{0}},\sigma^2{\mathbf{I}}_M)$ is the additive white Gaussian noise (AWGN) with $\sigma^2$ denoting the noise power. 

$\bullet$ \emph{Sum-Rate Capacity}:
The sum-rate capacity of the MAC is given by \cite{yu2004iterative}:
\begin{align}\label{Sum_Capacity_MAC}
{\mathsf{C}}_{\mathsf{MAC}}=\max_{\{p_k\}_{k=1}^{2}:0\leq p_k\leq P_k}\log_2\det\left({\mathbf{I}}_M+\sum\nolimits_{k=1}^{2}\frac{p_k}{\sigma^2}{\mathbf{h}}_k{\mathbf{h}}_k^{\mathsf{H}}\right),
\end{align}
where $P_k$ is the power budget of user $k$. The MAC sum-rate capacity can be achieved by using point-to-point Gaussian codes and successive interference cancellation (SIC) decoding in some message decoding order along with time sharing \cite{el2011network}. Besides, \emph{the sum-rate capacity is always the same regardless of the decoding order}. Based on \eqref{Sum_Capacity_MAC}, the sum-rate capacity is achieved when each user transmits at maximum power, i.e., $p_k=P_k$ \cite{suard1998uplink,yu2004iterative}. As a result, the sum-rate capacity can be simplified as follows:
\begin{align}\label{Sum_Capacity_MAC1}
{\mathsf{C}}_{\mathsf{MAC}}&=\log_2\det\left({\mathbf{I}}_M+\frac{P_{\mathsf{m}}}{\sigma^2}[{\mathbf{h}}_1,{\mathbf{h}}_2][{\mathbf{h}}_1, {\mathbf{h}}_2]^{\mathsf{H}}\right),
\end{align}
where we assume $P_1=P_2=P_{\mathsf{m}}$ for simplicity. It follows from the Sylvester's theorem \cite{zhang2017matrix} that 
\begin{subequations}\label{Sum_Capacity_MAC2}
\begin{align}
{\mathsf{C}}_{\mathsf{MAC}}&=\log_2\det\left({\mathbf{I}}_2+\frac{P_{\mathsf{m}}}{\sigma^2}[{\mathbf{h}}_1,{\mathbf{h}}_2]^{\mathsf{H}}[{\mathbf{h}}_1 , {\mathbf{h}}_2]\right)\\
&=\log_2\det\left({\mathbf{I}}_2+\frac{P_{\mathsf{m}}}{\sigma^2}\left[ \begin{matrix}
	{\mathsf{a}}_1&		{\mathbf{h}}_1^{\mathsf{H}}{\mathbf{h}}_2\\
	{\mathbf{h}}_2^{\mathsf{H}}{\mathbf{h}}_1&		{\mathsf{a}}_2\\
\end{matrix} \right]\right),
\end{align}
\end{subequations}
which simplifies further to the following form:
\begin{align}\label{Sum_Capacity_MAC2}
{\mathsf{C}}_{\mathsf{MAC}}=
\log_2\left(1+\frac{P_{\mathsf{m}}}{\sigma^2}({\mathsf{a}}_1+{\mathsf{a}}_2)+\frac{P_{\mathsf{m}}^2}{\sigma^4}{\mathsf{a}}_1{\mathsf{a}}_2(1-\rho_{\mathsf{u}}^2)\right).
\end{align}
An upper bound for ${\mathsf{C}}_{\mathsf{MAC}}$ can be obtained by assuming no IUI, i.e., $\rho_{\mathsf{u}}=0$, which yields
\begin{align}
{\mathsf{C}}_{\mathsf{MAC}}\leq {\mathsf{C}}_{\mathsf{MAC}}^{\mathsf{ub}}=\sum\nolimits_{k=1}^{2}\log_2\left(1+\frac{P_{\mathsf{m}}}{\sigma^2}{\mathsf{a}}_k\right).
\end{align}

Using the results in \eqref{SU_NFC_Channel_Gain_Result_Asymptotic}, the near-field MAC sum-rate capacity and its upper bound satisfy
\begin{align}
&\lim_{M\rightarrow\infty}{\mathsf{C}}_{\mathsf{MAC}}=\log_2\left(1+\frac{P_{\mathsf{m}}}{\sigma^2}\xi_{\mathsf{r}}
+\frac{P_{\mathsf{m}}^2}{\sigma^4}\frac{\xi_{\mathsf{r}}^2}{4}(1-\rho_{\mathsf{u}}^2)\right),\label{MAC_Sum_Capacity_Near_Field1}\\
&\lim_{M\rightarrow\infty}{\mathsf{C}}_{\mathsf{MAC}}^{\mathsf{ub}}=2\log_2\left(1+\frac{P_{\mathsf{m}}}{\sigma^2}\frac{\xi_{\mathsf{r}}}{2}\right),
\end{align}
both of which are constants and thus adhere to the law of conservation of energy. By taking into account the condition that $\lim_{M\rightarrow\infty}\rho_{\mathsf{u}}\ll1$, we have
\begin{subequations}\label{MAC_Sum_Capacity_Near_Field2}
\begin{align}
\lim_{M\rightarrow\infty}{\mathsf{C}}_{\mathsf{MAC}}&\approx \log_2\left(1+\frac{P_{\mathsf{m}}}{\sigma^2}\xi_{\mathsf{r}}+\frac{P_{\mathsf{m}}^2}{\sigma^4}\frac{\xi_{\mathsf{r}}^2}{4}\right)\\
&=2\log_2\left(1+\frac{P_{\mathsf{m}}}{\sigma^2}\frac{\xi_{\mathsf{r}}}{2}\right)=\lim_{M\rightarrow\infty}{\mathsf{C}}_{\mathsf{MAC}}^{\mathsf{ub}}.
\end{align}
\end{subequations}
For comparison, we also discuss the asymptotic sum-rate capacity and its upper bound under the far-field model, which yields
\begin{subequations}\label{MAC_Sum_Capacity_Far_Field}
\begin{align}
&\lim_{M\rightarrow\infty}{\mathsf{C}}_{\mathsf{MAC}}\simeq\left\{
\begin{array}{ll}
\log_2(1+\frac{MP_{\mathsf{m}}}{\sigma^2}\sum_{k=1}^{2}{\frac{{\Psi_k A}}{{4\pi} r_k^2}})            & {\mathbf{a}_1}= {\mathbf{a}_2}\\
\sum\nolimits_{k=1}^{2}\log_2(1+\frac{MP_{\mathsf{m}}}{\sigma^2}{\frac{{\Psi_k A}}{{4\pi} r_k^2}})           & {\mathbf{a}_1}\ne {\mathbf{a}_2}
\end{array} \right.,\\
&\lim_{M\rightarrow\infty}{\mathsf{C}}_{\mathsf{MAC}}^{\mathsf{ub}}\simeq\sum\nolimits_{k=1}^{2}\log_2\left(1+\frac{MP_{\mathsf{m}}}{\sigma^2}{\frac{{\Psi_k A}}{{4\pi} r_k^2}}\right).
\end{align}
\end{subequations}
Under the far-field model, the sum-rate capacity and its upper bound scale with ${\mathcal{O}}(\log_2(1+M))$, which potentially approach infinity with increasing $M$. This leads to the violation of the law of conservation of energy. Furthermore, upon comparing \eqref{MAC_Sum_Capacity_Near_Field2} with \eqref{MAC_Sum_Capacity_Far_Field}, we draw the following conclusions.
\vspace{-5pt}
\begin{remark}\label{MAC_RDMA_vs_ADMA}
The result in \eqref{MAC_Sum_Capacity_Near_Field2} indicates that in the limit of large $M$, the near-field sum-rate capacity can approach its upper bound as long as the two users are located in \textbf{different locations}. Conversely, the result in \eqref{MAC_Sum_Capacity_Far_Field} suggests that the far-field sum-rate capacity can only approach its upper bound when the two users are located in \textbf{different directions}. This advantageous property arises due to the additional range dimensions introduced by spherical-wave propagation, which underscores \textbf{the superiority of near-field RDMA over far-field ADMA}. 
\end{remark}
\vspace{-5pt}
\begin{figure}[!t]
 \centering
\includegraphics[width=0.48\textwidth]{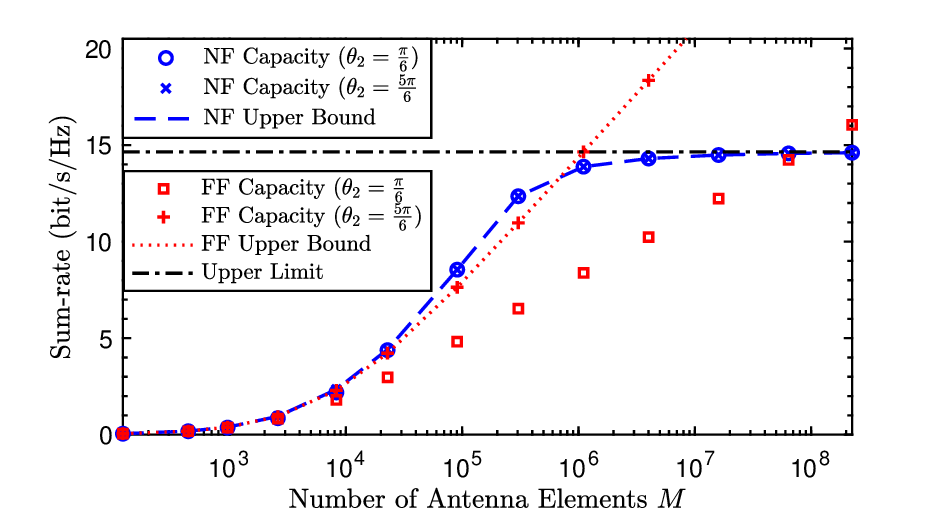}
\caption{Comparison of the MAC sum-rate capacity for different channel models versus the number of antennas $M$, where ``NF'' and ``FF'' refer to ``near-field'' and ``far-field'', respectively. $d=0.0628$ m, $d=\lambda/2$, $A=\frac{\lambda^2}{4\pi}$, $r_1=15$ m, $\theta_1=\frac{\pi}{6}$, $\phi_1=\frac{\pi}{3}$, $r_2=20$ m, $\phi_2=\frac{\pi}{3}$, $M_x=M_z=\sqrt{M}$, and $\frac{P_{\mathsf{m}}}{\sigma^2}=30$ dB.}
\label{MAC_Sum_Rate}
\end{figure}
To further validate our results, we present the sum-rate capacity and its upper bound for the considered channel models versus the number of antenna elements in {\figurename} {\ref{MAC_Sum_Rate}}. In our simulation setups, the capacity upper bounds for different values of $\theta_2$ are identical. As observed, the sum-rate capacity under the far-field model increases unboundedly with $M$, which can exceed the upper limit $2\log_2(1+\frac{P_{\mathsf{m}}}{\sigma^2}\frac{\xi_{\mathsf{r}}}{2})$ for sufficiently large $M$. In contrast, {\figurename} {\ref{MAC_Sum_Rate}} confirms that when $M\rightarrow\infty$, the sum-rate capacity and its upper bound obtained for the near-field model approach the finite upper limit, which adheres to the law of conservation of energy. Furthermore, when users are located in different directions, it can be seen from the figure that the sum-rate capacity under both far-field and near-field models closely aligns with their respective upper bounds. This is because the channel correlation is nearly zero when the BS is equipped with an ELAA. However, when users are positioned in the same direction, this alignment is only observed for the near-field channel model, as discussed in Remark \ref{MAC_RDMA_vs_ADMA}.

$\bullet$ \emph{Capacity Region}:
\begin{figure}[!t]
 \centering
\includegraphics[width=0.48\textwidth]{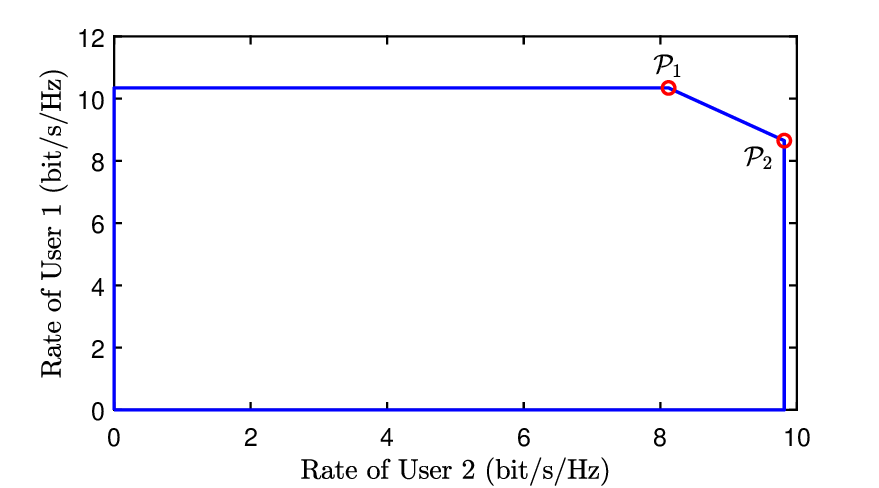}
\caption{An illustrated example of the capacity region for a two-user MISO-MAC. ${\mathbf{h}}_1=[1+{\rm{j}};2+{\rm{j}}]^{\mathsf{T}}$, ${\mathbf{h}}_2=[1-{\rm{j}};2+{\rm{j}}]^{\mathsf{T}}$, and $\frac{P_{\mathsf{m}}}{\sigma^2}=20$ dB.}
\label{MAC_Region}
\end{figure}
Having obtained the sum-rate capacity, we now study the capacity region of near-field MAC. The capacity region of a two-user MISO-MAC comprises all achievable rate pairs $({\mathsf{R}}_{1},{\mathsf{R}}_{2})$ such that \cite{el2011network}
\begin{subequations}\label{Two_User_MAC_Capacity_Region}
\begin{align}
{\mathsf{R}}_{1}&\leq \log_2\left(1+\frac{P_{\mathsf{m}}}{\sigma^{2}}{\mathsf{a}}_1\right),\\
{\mathsf{R}}_{2}&\leq \log_2\left(1+\frac{P_{\mathsf{m}}}{\sigma^{2}}{\mathsf{a}}_2\right),\\
{\mathsf{R}}_{1}+{\mathsf{R}}_{2}&\leq \log_2\det\left({\mathbf{I}}_M+\frac{P_{\mathsf{m}}}{\sigma^{2}}[{\mathbf{h}}_1,{\mathbf{h}}_2][{\mathbf{h}}_1, {\mathbf{h}}_2]^{\mathsf{H}}\right),
\end{align}
\end{subequations}
where ${\mathsf{R}}_{k}$ denotes the achievable rate of user $k=1,2$. An example illustrating the MAC capacity region is shown in {\figurename} {\ref{MAC_Region}}. This region can be achieved through SIC decoding along with time sharing. Notably, the corner point ${\mathcal{P}}_1$ and ${\mathcal{P}}_2$ are attained by the SIC decoding orders $2\rightarrow1$ and $1\rightarrow2$, respectively, whereas the line segment connecting ${\mathcal{P}}_1$ and ${\mathcal{P}}_2$ is achieved via a time sharing between ${\mathcal{P}}_1$ and ${\mathcal{P}}_2$. For ${\mathcal{P}}_1$, the achievable rates are given by \cite{el2011network}
\begin{subequations}
\begin{align}
{\mathsf{R}}_{1}&= \log_2(1+P_{\mathsf{m}}/\sigma^{2}{\mathsf{a}}_1),\\
{\mathsf{R}}_{2}&= \log_2\det\Big({\mathbf{I}}_M+\frac{P_{\mathsf{m}}}{\sigma^2}{\mathbf{h}}_2{\mathbf{h}}_2^{\mathsf{H}}\Big({\mathbf{I}}_M+\frac{P_{\mathsf{m}}}{\sigma^2}
{\mathbf{h}}_1{\mathbf{h}}_1^{\mathsf{H}}\Big)^{-1}\Big).
\end{align}
\end{subequations}
Using Sylvester's theorem and the Woodbury matrix identity, ${\mathsf{R}}_{2}$ can be expressed as follows \cite{zhang2017matrix}:
\begin{subequations}
\begin{align}
{\mathsf{R}}_{2}&=\log_2\left(1+\frac{P_{\mathsf{m}}}{\sigma^2}\lVert{\mathbf{h}}_2\rVert^2-\frac{\frac{P_{\mathsf{m}}^2}{\sigma^4}\lvert{\mathbf{h}}_1^{\mathsf{H}}{\mathbf{h}}_2\rvert^2}
{1+\frac{P_{\mathsf{m}}}{\sigma^2}\lVert{\mathbf{h}}_1\rVert^2}\right)\\
&=\log_2\left(\frac{1+\frac{P_{\mathsf{m}}}{\sigma^2}({\mathsf{a}}_1+{\mathsf{a}}_2)+\frac{P_{\mathsf{m}}^2}{\sigma^4}{\mathsf{a}}_1{\mathsf{a}}_2(1-\rho_{\mathsf{u}}^2)}
{1+\frac{P_{\mathsf{m}}}{\sigma^2}{\mathsf{a}}_1}\right).
\end{align}
\end{subequations}
As for ${\mathcal{P}}_2$, we have \cite{el2011network}
\begin{subequations}
\begin{align}
{\mathsf{R}}_{1}&= \log_2\left(\frac{1+\frac{P_{\mathsf{m}}}{\sigma^2}({\mathsf{a}}_1+{\mathsf{a}}_2)+\frac{P_{\mathsf{m}}^2}{\sigma^4}{\mathsf{a}}_1{\mathsf{a}}_2(1-\rho_{\mathsf{u}}^2)}
{1+\frac{P_{\mathsf{m}}}{\sigma^2}{\mathsf{a}}_2}\right),\\
{\mathsf{R}}_{2}&= \log_2(1+P_{\mathsf{m}}/\sigma^{2}{\mathsf{a}}_2).
\end{align}
\end{subequations}
Note that each corner point can be expressed as a function of the channel gain and the channel correlation factor. Following similar steps used in deriving \eqref{MAC_Sum_Capacity_Near_Field1}, we can derive the limitations of these two corner points and obtain the asymptotic capacity region. Using the approximation $\lim_{M\rightarrow\infty}\rho_{\mathsf{u}}\ll1$, we have
\begin{subequations}
\begin{align}\label{MAC_LoS_Capacity_Region_Limiting}
&\lim_{M\rightarrow\infty}{\mathcal{P}}_1\approx \Big(\log_2\Big(1+\frac{P_{\mathsf{m}}}{\sigma^2}\frac{\xi_{\mathsf{r}}}{2}\Big),\log_2\Big(1+\frac{P_{\mathsf{m}}}{\sigma^2}\frac{\xi_{\mathsf{r}}}{2}\Big)\Big),\\
&\lim_{M\rightarrow\infty}{\mathcal{P}}_2\approx \Big(\log_2\Big(1+\frac{P_{\mathsf{m}}}{\sigma^2}\frac{\xi_{\mathsf{r}}}{2}\Big),\log_2\Big(1+\frac{P_{\mathsf{m}}}{\sigma^2}\frac{\xi_{\mathsf{r}}}{2}\Big)\Big).
\end{align}
\end{subequations}
As for the capacity region under the far-field model, we have
\begin{equation}\label{FF_MAC_LoS_Capacity_Region_Limiting1}
\begin{split}
&\lim_{M\rightarrow\infty}{\mathcal{P}}_1\simeq \lim_{M\rightarrow\infty}{\mathcal{P}}_2\\
&\simeq \Big(\log_2\Big(1+\frac{MP_{\mathsf{m}}}{\sigma^2}{\frac{{\Psi_1 A}}{{4\pi} r_1^2}}\Big),
\log_2\Big(1+\frac{MP_{\mathsf{m}}}{\sigma^2}{\frac{{\Psi_2 A}}{{4\pi} r_2^2}}\Big)\Big)
\end{split}
\end{equation}
for ${\mathbf{a}_1}\ne {\mathbf{a}_2}$, and
\begin{subequations}
\begin{align}\label{FF_MAC_LoS_Capacity_Region_Limiting2}
&\lim_{M\rightarrow\infty}{\mathcal{P}}_1\simeq \Big(\log_2\Big(1+\frac{MP_{\mathsf{m}}}{\sigma^2}{\frac{{\Psi_1 A}}{{4\pi} r_1^2}}\Big),\log_2\Big(1+\frac{\Psi_2r_1^2}{\Psi_1r_2^2}\Big)\Big),\\
&\lim_{M\rightarrow\infty}{\mathcal{P}}_2\simeq \Big(\log_2(1+\frac{\Psi_1r_2^2}{\Psi_2r_1^2}\Big),\log_2\Big(1+\frac{MP_{\mathsf{m}}}{\sigma^2}{\frac{{\Psi_2 A}}{{4\pi} r_2^2}}\Big)\Big)
\end{align}
\end{subequations}
for ${\mathbf{a}_1}= {\mathbf{a}_2}$. We note that
\begin{align}\label{FF_MAC_LoS_Capacity_Region_Limiting2}
\lim_{M\rightarrow\infty}\frac{\log_2(1+\frac{\Psi_2r_1^2}{\Psi_1r_2^2})}{\log_2(1+\frac{MP_{\mathsf{m}}}{\sigma^2}{\frac{{\Psi_1 A}}{{4\pi} r_1^2}})}=
\lim_{M\rightarrow\infty}\frac{\log_2(1+\frac{\Psi_1r_2^2}{\Psi_2r_1^2})}{\log_2(1+\frac{MP_{\mathsf{m}}}{\sigma^2}{\frac{{\Psi_2 A}}{{4\pi} r_2^2}})}=0.
\end{align}
This suggests that the capacity region under the far-field model approximately converges to a triangular shape for ${\mathbf{a}_1}= {\mathbf{a}_2}$.
\vspace{-5pt}
\begin{remark}\label{MAC_Region_RDMA_vs_ADMA}
The result in \eqref{MAC_LoS_Capacity_Region_Limiting} reveals that in the limit of large $M$, the near-field MAC capacity region approximately transforms from a pentagon to a rectangle, which is attributed to the near-zero IUI as $M\rightarrow\infty$. It is worth noting that this convergence behavior also holds for the far-field capacity region if the two users are situated in different directions (${\mathbf{a}_1}\ne {\mathbf{a}_2}$), as per \eqref{FF_MAC_LoS_Capacity_Region_Limiting1}. However, when two far-field users are located in the same direction, the capacity region asymptotically transitions into a triangular shape due to the strong channel correlation. In this case, the capacity region nearly degenerates to that achieved by orthogonal MA techniques, such as TDMA. This comparison underscores the flexibility and robust interference management capabilities of near-field RDMA compared to its far-field counterparts. The ability to maintain a rectangular capacity region even in scenarios where users are situated in the same direction makes near-field RDMA a promising approach for efficient and interference-resilient communication systems.
\end{remark}
\vspace{-5pt}
\begin{figure}[!t]
 \centering
    \subfigure[$\theta_2=\frac{\pi}{6}$.]{
        \includegraphics[width=0.48\textwidth]{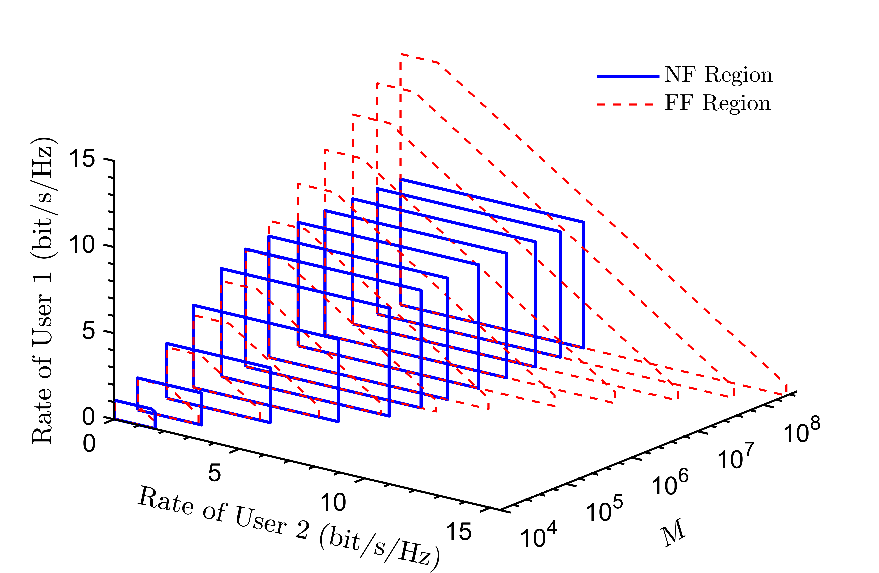}
        \label{Figure: MAC_Capacity_Region_LoS1}
    }
    \subfigure[$\theta_2=\frac{5\pi}{6}$.]{
        \includegraphics[width=0.48\textwidth]{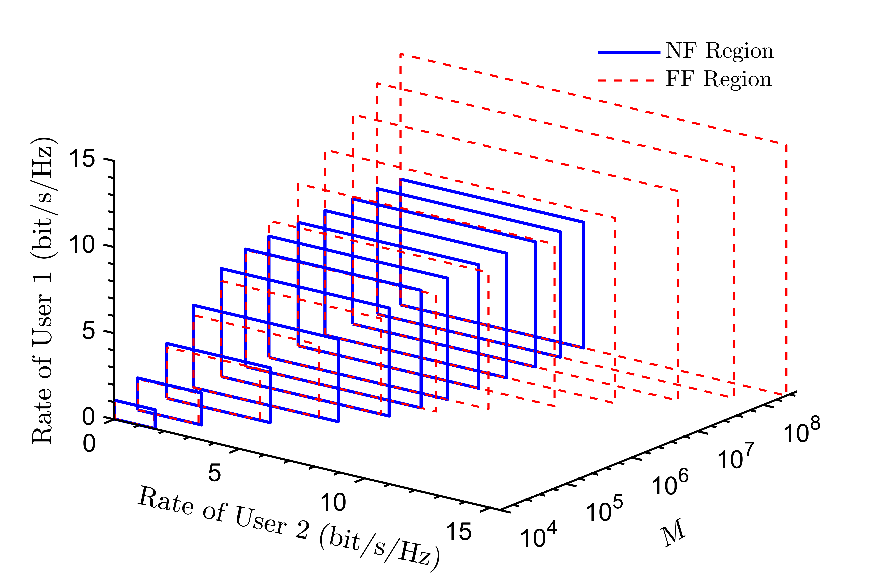}
        \label{Figure: MAC_Capacity_Region_LoS2}	
    }
\caption{Comparison of the MAC capacity regions for different channel models versus the number of antennas $M$, where ``NF'' and ``FF'' refer to ``near-field'' and ``far-field'', respectively. $d=0.0628$ m, $d=\lambda/2$, $A=\frac{\lambda^2}{4\pi}$, $r_1=15$ m, $\theta_1=\frac{\pi}{6}$, $\phi_1=\frac{\pi}{3}$, $r_2=20$ m, $\phi_2=\frac{\pi}{3}$, $M_x=M_z=\sqrt{M}$, and $\frac{P_{\mathsf{m}}}{\sigma^2}=30$ dB.}
\label{Figure: MAC_Capacity_Region_LoS}
\end{figure}
To further verify our results, we present the capacity regions under both the far-field and near-field models versus the number of antenna elements in {\figurename} {\ref{Figure: MAC_Capacity_Region_LoS}}. As observed, the capacity region under the far-field model exhibits an unbounded widening as $M$ increases, whereas the capacity region under the near-field model converges to its finite limit. This behavior validates the asymptotic results derived for the near-field channel model, which adheres to the principle of conservation of energy. Furthermore, when users are situated in different directions, both the far-field and near-field capacity regions approximate rectangles. However, when users are positioned in the same direction, this alignment is not observed for the far-field channel model, as discussed in Remark \ref{MAC_Region_RDMA_vs_ADMA}.

$\bullet$ \emph{Extension to Cases of $K>2$}: We next extend the above results to a more general case where the number of users is larger than $2$, i.e., $K>2$. In this case, the channel capacity is also achieved using point-to-point Gaussian codes and SIC decoding along with time sharing. Particularly, the sum-rate capacity can be expressed as follows \cite{yu2004iterative}:
\begin{equation}\label{Sum_Capacity_MAC_General}
{\mathsf{C}}_{\mathsf{MAC}}=\max_{\{p_k\}_{k=1}^{K}:0\leq p_k\leq P_k}\log_2\det\left({\mathbf{I}}_M+\sum\nolimits_{k=1}^{K}\frac{p_k}{\sigma^2}{\mathbf{h}}_k{\mathbf{h}}_k^{\mathsf{H}}\right),
\end{equation}
where $P_k$ is the power budget of user $k$. The sum-rate capacity is achieved when each user transmits at maximum power, i.e., $p_k=P_k$, which yields
\begin{equation}\label{Sum_Capacity_MAC_General1}
{\mathsf{C}}_{\mathsf{MAC}}=\log_2\det\left({\mathbf{I}}_M+\sum\nolimits_{k=1}^{K}\frac{P_k}{\sigma^2}{\mathbf{h}}_k{\mathbf{h}}_k^{\mathsf{H}}\right).
\end{equation}
Using the Sylvester's theorem, we rewrite \eqref{Sum_Capacity_MAC_General1} as follows:
\begin{equation}\label{Sum_Capacity_MAC_General2}
{\mathsf{C}}_{\mathsf{MAC}}=\log_2\det\left({\mathbf{I}}_K+\frac{1}{\sigma^2}{\mathbf{H}}_{\mathsf{MAC}}^{\mathsf{H}}{\mathbf{H}}_{\mathsf{MAC}}\right),
\end{equation}
where ${\mathbf{H}}_{\mathsf{MAC}}=[\sqrt{P_1}{\mathbf{h}}_1,\ldots,\sqrt{P_K}{\mathbf{h}}_K]\in{\mathbbmss{C}}^{M\times K}$. Each element of the following matrix 
\begin{equation}
{\mathbf{H}}_{\mathsf{MAC}}^{\mathsf{H}}{\mathbf{H}}_{\mathsf{MAC}}=\left[ \begin{matrix}
	P_1\lVert{\mathbf{h}}_1\rVert^2&		\cdots&		\sqrt{P_1P_K}{\mathbf{h}}_1^{\mathsf{H}}{\mathbf{h}}_K\\
	\vdots&		\ddots&		\vdots\\
	\sqrt{P_1P_K}{\mathbf{h}}_K^{\mathsf{H}}{\mathbf{h}}_1&		\cdots&		P_K\lVert{\mathbf{h}}_K\rVert^2\\
\end{matrix} \right] 
\end{equation}
can be written as a function of the channel gain $\mathsf{a}_k=\lVert{\mathbf{h}}_k\rVert^2$ and the channel correlation factor $\frac{\lvert {\mathbf{h}}_k^{\mathsf{H}}{\mathbf{h}}_{k'}\rvert}{\lVert{\mathbf{h}}_k\rVert\lVert{\mathbf{h}}_{k'}\rVert}$, and thus the sum-rate capacity for $K>2$ can be analyzed using the same method as in analyzing \eqref{Sum_Capacity_MAC2}. Further details are omitted here for brevity.

Regarding the capacity region, any achievable rate $K$-tuple $({\mathsf{R}}_1,{\mathsf{R}}_2,\ldots,{\mathsf{R}}_K)$ corresponding to the users must satisfy \cite{suard1998uplink}
\begin{equation}\label{Gaussian_Vector_MAC_Rate_Region}
\begin{split}
&\left\{({\mathsf{R}}_1,\ldots,{\mathsf{R}}_K):\sum\nolimits_{k\in{\mathscr{K}}}{\mathsf{R}}_k\right.\\
&\left.\leq\log_2\det\Big({\mathbf{I}}+\frac{1}{\sigma^2}\sum\nolimits_{k\in{\mathscr{K}}}P_k{\mathbf{h}}_k{\mathbf{h}}_k^{\mathsf{H}}\Big),\forall {\mathscr{K}}\subseteq{\mathcal{K}}\right\},
\end{split}
\end{equation}
where ${\mathcal{K}}\triangleq\{1,2,\ldots,K\}$. The convex hull of the region delimited by \eqref{Gaussian_Vector_MAC_Rate_Region} is the capacity region, which corresponds to a $K$-dimensional polyhedron. SIC decoding achieves each of the $K!$ corner points, where users' signals are successively decoded and subtracted from the received signal \cite{el2011network}. For each corner point, the corresponding rates can be expressed as a function of the channel gain and channel correlation factor, which can be analyzed in the same way as described in discussing \eqref{Two_User_MAC_Capacity_Region}.

\subsubsection{Broadcast Channel}
If we reverse the MAC and have one BS broadcasting simultaneously to all users, it becomes the BC. The received signal at user $k=1,2$ is given by
\begin{align}\label{Downlink_Signal}
y_k={\mathbf{h}}_k^{\mathsf{H}}{\mathbf{x}}+{{n}}_k,
\end{align}
where ${\mathbf{x}}\in{\mathbbmss{C}}^{M\times1}$ is the transmitted signal vector, and $n_k\sim{\mathcal{CN}}(0,\sigma_k^2)$ is AWGN with noise power $\sigma_k^2$. For simplicity, we assume $\sigma_k^2=\sigma^2$ for $k=1,2$. The transmitted signal is constructed as follows \cite{viswanath2003sum}:
\begin{align}\label{Downlink_Transmit_Signal}
{\mathbf{x}}=\sum\nolimits_{k=1}^{2}{\mathbf{w}}_ks_k,
\end{align}
where $s_k\in{\mathbbmss{C}}$ represents the normalized coded data symbol for user $k$, and ${\mathbf{w}}_k\in{\mathbbmss{C}}^{M\times1}$ denotes the associated transmit beamforming vector. After receiving $y_k$, user $k$ will exploit a decoder to recover the private message encoded in $s_k$ from $y_k$. The sum-rate capacity is obtained through the following maximization \cite{vishwanath2003duality,jindal2005sum,viswanath2003sum}:
\begin{equation}\label{Sum_Rate_Capacity_BC}
\begin{split}
{\mathsf{C}}_{\mathsf{BC}}&=\max_{\sum_{k=1}^{2}\lVert{\mathbf{w}}_k\rVert^2\leq P_{\mathsf{b}}}\log_2(1+\lvert{\mathbf{h}}_1^{\mathsf{H}}{\mathbf{w}}_1\rvert^2/\sigma^2)\\
&+\log_2\left(1+\frac{\frac{1}{\sigma^2}\lvert{\mathbf{h}}_2^{\mathsf{H}}{\mathbf{w}}_2\rvert^2}
{1+\frac{1}{\sigma^2}\lvert{\mathbf{h}}_2^{\mathsf{H}}{\mathbf{w}}_1\rvert^2}\right),
\end{split}
\end{equation}
where $P_{\mathsf{b}}$ is the power budget.
\subsubsection*{Dirty-Paper Coding} The capacity region of the multiple-antenna Gaussian BC is achieved through successive dirty-paper coding (DPC) with Gaussian codebooks \cite{costa1983writing,weingarten2006capacity}. Without loss of generality, we consider the encoding order $\varepsilon(2)\rightarrow\varepsilon(1)$ with $\{\varepsilon(k)\}_{k=1}^{2}=\{k\}_{k=1}^{2}$. For user $\varepsilon(k)$, the dirty-paper encoder treats the interference caused by user $\varepsilon(k')$ for $k'>k$ as known non-causally and his decoder treats the interference from user $\varepsilon(k')$ for $k'<k$ as additional noise. By applying DPC and by using minimum Euclidean distance decoding at each user, the rate achieved by user $\varepsilon(k)$ is given by \cite{vishwanath2003duality,jindal2005sum,viswanath2003sum}
\begin{align}
{\mathsf{R}}_{\varepsilon(k)}=\log_2\left(1+\frac{\frac{1}{\sigma^2}\lvert{\mathbf{h}}_{\varepsilon(k)}^{\mathsf{H}}{\mathbf{w}}_{\varepsilon(k)}\rvert^2}
{1+\sum_{k'<k}\frac{1}{\sigma^2}\lvert{\mathbf{h}}_{\varepsilon(k)}^{\mathsf{H}}{\mathbf{w}}_{\varepsilon(k')}\rvert^2}\right).
\end{align}
The expression in \eqref{Sum_Rate_Capacity_BC} holds for the specific encoding order $2\rightarrow1$. According to \cite{vishwanath2003duality,jindal2005sum,viswanath2003sum}, \emph{the sum-rate capacity is always the same regardless of the encoding order}.

$\bullet$ \emph{Sum-Rate Capacity}: 
Problem \eqref{Sum_Rate_Capacity_BC} is a non-convex problem. Thus, numerically finding the maximum is a challenging task. However, in \cite{viswanath2003sum,vishwanath2003duality}, a duality is shown to exist between the uplink and downlink, which reveals that the dirty-paper capacity region for the BC equals the capacity region of the dual MAC (described in \eqref{MAC_Channel}). This implies that
\begin{align}
{\mathsf{C}}_{\mathsf{BC}}&=\max_{\{\varepsilon(k)\}_{k=1}^{2},\sum_{k=1}^{2}\lVert{\mathbf{w}}_{\varepsilon(k)}\rVert^2\leq P_{\mathsf{b}}}\sum\nolimits_{k=1}^{2}{\mathsf{R}}_{\varepsilon(k)}\\
&=\max_{p_k\geq0,\sum_{k=1}^{2}p_k= P_{\mathsf{b}}}\log_2\det\left({\mathbf{I}}_M+\sum_{k=1}^{2}\frac{p_k}{\sigma^2}{\mathbf{h}}_k{\mathbf{h}}_k^{\mathsf{H}}\right).\label{BC_Sum_Capacity}
\end{align}
Note that problem \eqref{BC_Sum_Capacity} is convex and can be optimally solved using the sum power iterative water-filling method \cite{jindal2005sum}. After obtaining the optimized $\{p_k\}_{k=1}^{2}$, we can employ the dirty-paper encoding order and the transformation outlined in \cite[Equs. (8)–(10)]{vishwanath2003duality} and \cite[Appendix A]{jindal2005sum} to recover corresponding downlink beamforming vectors that achieve the same rates on a per-user basis, thus also in terms of sum-rate. This methodology readily extends to scenarios with more than two users \cite{jindal2005sum}. More specifically, given $p_1$ and $p_2$, the transmit beamforming vectors corresponding to the DPC encoding order $2\rightarrow1$ are given as follows:
\begin{align}
{\mathbf{w}}_1&=\frac{\sqrt{p_1}({\mathbf{I}}_M+\frac{p_2}{\sigma^2}{\mathbf{h}}_2{\mathbf{h}}_2^{\mathsf{H}})^{-1}{\mathbf{h}}_1}
{\sqrt{{\mathbf{h}}_1^{\mathsf{H}}({\mathbf{I}}_M+\frac{p_2}{\sigma^2}{\mathbf{h}}_2{\mathbf{h}}_2^{\mathsf{H}})^{-1}{\mathbf{h}}_1}},\\
{\mathbf{w}}_2&=\frac{\sqrt{p_2(1+\frac{1}{\sigma^2}\lvert{\mathbf{h}}_2^{\mathsf{H}}{\mathbf{w}}_1\rvert^2)}}{\lVert{\mathbf{h}}_2\rVert}{\mathbf{h}}_2.
\end{align}
It can be further verified that
\begin{subequations}
\begin{align}
\sum_{k=1}^{2}\lVert{\mathbf{w}}_k\rVert^2&=p_2(1+\frac{1}{\sigma^2}\lvert{\mathbf{h}}_2^{\mathsf{H}}{\mathbf{w}}_1\rvert^2)+{\mathbf{w}}_1^{\mathsf{H}}{\mathbf{w}}_1\\
&=p_2+{\mathsf{tr}}(({p_2}/{\sigma^2}{\mathbf{h}}_2{\mathbf{h}}_2^{\mathsf{H}}+{\mathbf{I}}_M){\mathbf{w}}_1{\mathbf{w}}_1^{\mathsf{H}})\\
&=p_2+{\mathbf{w}}_1^{\mathsf{H}}({p_2}/{\sigma^2}{\mathbf{h}}_2{\mathbf{h}}_2^{\mathsf{H}}+{\mathbf{I}}_M){\mathbf{w}}_1\\
&=p_1+p_2\leq P_{\mathsf{b}},
\end{align}
\end{subequations}
which satisfies the power budget. Additionally, for the DPC encoding order $1\rightarrow2$, we have
\begin{align}
{\mathbf{w}}_1&=\frac{\sqrt{p_1(1+\frac{1}{\sigma^2}\lvert{\mathbf{h}}_1^{\mathsf{H}}{\mathbf{w}}_2\rvert^2)}}{\lVert{\mathbf{h}}_1\rVert}{\mathbf{h}}_1,\\
{\mathbf{w}}_2&=\frac{\sqrt{p_2}({\mathbf{I}}_M+\frac{p_1}{\sigma^2}{\mathbf{h}}_1{\mathbf{h}}_1^{\mathsf{H}})^{-1}{\mathbf{h}}_2}
{\sqrt{{\mathbf{h}}_2^{\mathsf{H}}({\mathbf{I}}_M+\frac{p_1}{\sigma^2}{\mathbf{h}}_1{\mathbf{h}}_1^{\mathsf{H}})^{-1}{\mathbf{h}}_2}}.
\end{align}
Following the steps in deriving \eqref{Sum_Capacity_MAC2}, we rewrite ${\mathsf{C}}_{\mathsf{MAC}}^{\mathsf{d}}\triangleq\log_2\det\Big({\mathbf{I}}_M+\sum_{k=1}^{2}\frac{p_k}{\sigma^2}{\mathbf{h}}_k{\mathbf{h}}_k^{\mathsf{H}}\Big)$ as follows:
\begin{equation}\label{BC_Sum_Capacity1}
\begin{split}
{\mathsf{C}}_{\mathsf{MAC}}^{\mathsf{d}}=
\log_2\left(1+\frac{p_1{\mathsf{a}}_1}{\sigma^2}+\frac{p_2{\mathsf{a}}_2}{\sigma^2}+\frac{p_1p_2}{\sigma^4}
{\mathsf{a}}_1{\mathsf{a}}_2(1-\rho_{\mathsf{u}}^2)\right).
\end{split}
\end{equation}
Consequently, problem \eqref{BC_Sum_Capacity} simplifies to the following single-variable convex form:
\begin{equation}
\begin{split}
\max_{0\leq p_1\leq P_{\mathsf{b}}}\left(\frac{p_1{\mathsf{a}}_1}{\sigma^2}+\frac{(P_{\mathsf{b}}-p_1){\mathsf{a}}_2}{\sigma^2}
+\frac{p_1(P_{\mathsf{b}}-p_1)}{\frac{\sigma^4}{{\mathsf{a}}_1{\mathsf{a}}_2(1-\rho_{\mathsf{u}}^2)}}\right).
\end{split}
\end{equation}
The optimal solution is then given by
\begin{align}
p_1^{\star}=\left\{
\begin{array}{ll}
0             & \frac{\frac{{\mathsf{a}}_1}{\sigma^2}-\frac{{\mathsf{a}}_2}{\sigma^2}}{\frac{{\mathsf{a}}_1{\mathsf{a}}_2}{\sigma^4}(1-\rho_{\mathsf{u}}^2)}\leq-P_{\mathsf{b}}\\
\frac{P_{\mathsf{b}}}{2}+\frac{\frac{{\mathsf{a}}_1}{\sigma^2}-\frac{{\mathsf{a}}_2}{\sigma^2}}{2\frac{{\mathsf{a}}_1{\mathsf{a}}_2}{\sigma^4}(1-\rho_{\mathsf{u}}^2)}           & -P_{\mathsf{b}}<\frac{\frac{{\mathsf{a}}_1}{\sigma^2}-\frac{{\mathsf{a}}_2}{\sigma^2}}{\frac{{\mathsf{a}}_1{\mathsf{a}}_2}{\sigma^4}(1-\rho_{\mathsf{u}}^2)}<P_{\mathsf{b}}\\
P_{\mathsf{b}}           & \frac{\frac{{\mathsf{a}}_1}{\sigma^2}-\frac{{\mathsf{a}}_2}{\sigma^2}}{\frac{{\mathsf{a}}_1{\mathsf{a}}_2}{\sigma^4}(1-\rho_{\mathsf{u}}^2)}\geq P_{\mathsf{b}}
\end{array} \right..
\end{align}
Once $p_1^{\star}$ is obtained, the optimized $p_2$ is set as $p_2^{\star}=P_{\mathsf{b}}-p_1^{\star}$, which yields the following sum-rate capacity:
\begin{equation}\label{BC_Sum_Capacity_Final}
\begin{split}
{\mathsf{C}}_{\mathsf{BC}}=
\log_2\left(1+\frac{p_1^{\star}{\mathsf{a}}_1}{\sigma^2}+\frac{p_2^{\star}{\mathsf{a}}_2}{\sigma^2}+\frac{p_1^{\star}p_2^{\star}}{\sigma^4}
{\mathsf{a}}_1{\mathsf{a}}_2(1-\rho_{\mathsf{u}}^2)\right).
\end{split}
\end{equation}
An upper bound for ${\mathsf{C}}_{\mathsf{BC}}$ is obtained under the assumption of no IUI, i.e., $\rho_{\mathsf{u}}=0$, which yields
\begin{align}\label{BC_Sum_Capacity_UB}
{\mathsf{C}}_{\mathsf{BC}}\leq {\mathsf{C}}_{\mathsf{BC}}^{\mathsf{ub}}=\max_{\sum_{k=1}^{2}p_k= P_{\mathsf{b}}}\sum\nolimits_{k=1}^{2}\log_2\left(1+\frac{p_k}{\sigma^2}{\mathsf{a}}_k\right),
\end{align}
where $\{p_k\}_{k=1}^{2}$ are optimized using the water-filling method on the two virtual no-IUI sub-channels. Particularly, when ${\mathsf{a}}_1={\mathsf{a}}_2$, the optimal power allocation strategy is equal power assignment. 

Utilizing the findings from \eqref{SU_NFC_Channel_Gain_Result_Asymptotic} and \eqref{Channel_Correlation_Near_Field}, we observe 
\begin{align}
\lim_{M\rightarrow\infty}({\mathsf{a}}_1-{\mathsf{a}}_2)=\frac{\xi_{\mathsf{r}}}{2}-\frac{\xi_{\mathsf{r}}}{2}=0,
\end{align}
which suggests that the optimal power allocation strategies in deriving \eqref{BC_Sum_Capacity_Final} and \eqref{BC_Sum_Capacity_UB} both tend towards equal power assignment as $M\rightarrow\infty$. Consequently, the near-field BC's sum-rate capacity and its upper bound satisfy
\begin{align}
&\lim_{M\rightarrow\infty}{\mathsf{C}}_{\mathsf{BC}}=\log_2\left(1+\frac{P_{\mathsf{b}}\xi_{\mathsf{r}}}{2\sigma^2}
+\frac{P_{\mathsf{b}}^2}{\sigma^4}\frac{\xi_{\mathsf{r}}^2}{16}(1-\rho_{\mathsf{u}}^2)\right),\label{BC_Sum_Capacity_Near_Field1}\\
&\lim_{M\rightarrow\infty}{\mathsf{C}}_{\mathsf{BC}}^{\mathsf{ub}}=2\log_2\left(1+\frac{P_{\mathsf{b}}}{\sigma^2}\frac{\xi_{\mathsf{r}}}{4}\right).
\end{align}
Both of these results are constants, which align with the law of conservation of energy. Considering $\lim_{M\rightarrow\infty}\rho_{\mathsf{u}}\ll1$ yields
\begin{subequations}\label{BC_Sum_Capacity_Near_Field2}
\begin{align}
\lim_{M\rightarrow\infty}{\mathsf{C}}_{\mathsf{BC}}&\approx \log_2\left(1+\frac{P_{\mathsf{b}}\xi_{\mathsf{r}}}{2\sigma^2}
+\frac{P_{\mathsf{b}}^2}{\sigma^4}\frac{\xi_{\mathsf{r}}^2}{16}\right)\\
&=2\log_2\left(1+\frac{P_{\mathsf{b}}}{\sigma^2}\frac{\xi_{\mathsf{r}}}{4}\right)=\lim_{M\rightarrow\infty}{\mathsf{C}}_{\mathsf{BC}}^{\mathsf{ub}}.
\end{align}
\end{subequations}
For comparison, let us consider the asymptotic BC sum-rate capacity and its upper bound under the far-field model. As discussed in Section \ref{Section: Capacity Limits for LoS Channels: Preliminaries}, we have $\lim_{M\rightarrow\infty}{\mathsf{a}}_k\simeq {\frac{M{\Psi_k A}}{{4\pi} r_k^2}}$ for $k=1,2$. This implies that $\lim_{M\rightarrow\infty}\frac{\frac{{\mathsf{a}}_1}{\sigma^2}-\frac{{\mathsf{a}}_2}{\sigma^2}}{\frac{{\mathsf{a}}_1{\mathsf{a}}_2}{\sigma^4}(1-\rho_{\mathsf{u}}^2)}
=0$ holds for ${\mathbf{a}_1}\ne {\mathbf{a}_2}$ ($\rho_{\mathsf{u}}^2\ne 1$), which leads to $p_1^{\star}=p_2^{\star}=\frac{P}{2}$. However, for ${\mathbf{a}_1}= {\mathbf{a}_2}$ (i.e., $\rho_{\mathsf{u}}^2= 1$), we have $p_1^{\star}=1$ if ${\mathsf{a}_1}\geq{\mathsf{a}_2}$ and $p_1^{\star}=0$ if ${\mathsf{a}_1}<{\mathsf{a}_2}$. In summary, the asymptotic far-field BC sum-rate capacity is given by
\begin{align}\label{BC_Sum_Capacity_Far_Field}
&\lim_{M\rightarrow\infty}{\mathsf{C}}_{\mathsf{BC}}\simeq\left\{
\begin{array}{ll}
\log_2(1+\frac{MP_{\mathsf{b}}}{\sigma^2}\max_{k=1,2}{\frac{{\Psi_k A}}{{4\pi} r_k^2}})            & {\mathbf{a}_1}= {\mathbf{a}_2}\\
\sum\nolimits_{k=1}^{2}\log_2(1+\frac{MP_{\mathsf{b}}}{2\sigma^2}{\frac{{\Psi_k A}}{{4\pi} r_k^2}})           & {\mathbf{a}_1}\ne {\mathbf{a}_2}
\end{array} \right..
\end{align}
Moreover, due to the fact that $\lim_{M\rightarrow\infty}{\mathsf{a}}_k\simeq\mathcal{O}(M)$, the optimal power allocation strategy in obtaining \eqref{BC_Sum_Capacity_UB} tends to an equal power assignment as $M\rightarrow\infty$ \cite{heath2018foundations}. Consequently, the capacity upper bound under the far-field model satisfies
\begin{equation}\label{BC_Sum_Capacity_Asymptotic_UB__Far_Field}
\lim_{M\rightarrow\infty}{\mathsf{C}}_{\mathsf{BC}}^{\mathsf{ub}}\simeq
\sum\nolimits_{k=1}^{2}\log_2\left(1+\frac{MP_{\mathsf{b}}}{2\sigma^2}{\frac{{\Psi_k A}}{{4\pi} r_k^2}}\right).
\end{equation}
\vspace{-5pt}
\begin{remark}\label{BC_Energy_Conservation}
Under the far-field model, the BC sum-rate capacity and its upper bound scale with ${\mathcal{O}}(\log_2(1+M))$, potentially approaching infinity as $M$ increases. This behavior violates the law of conservation of energy. In contrast, the BC sum-rate capacity under the near-field model converges to a finite constant. This convergence highlights the critical importance of near-field modeling in asymptotic analyses, providing a more realistic and sustainable framework for energy considerations.
\end{remark}
\vspace{-5pt}
\vspace{-5pt}
\begin{remark}\label{BC_RDMA_vs_ADMA}
As $M$ approaches infinity, the near-field BC sum-rate capacity can approach its upper bound, provided the two users are located in \textbf{different positions}. In contrast, the far-field sum-rate capacity can only approach its upper bound when the users are positioned in \textbf{different directions}. This advantageous characteristic arises from the additional range dimensions introduced by spherical-wave propagation.
\end{remark}
\vspace{-5pt}
\begin{figure}[!t]
 \centering
\includegraphics[width=0.48\textwidth]{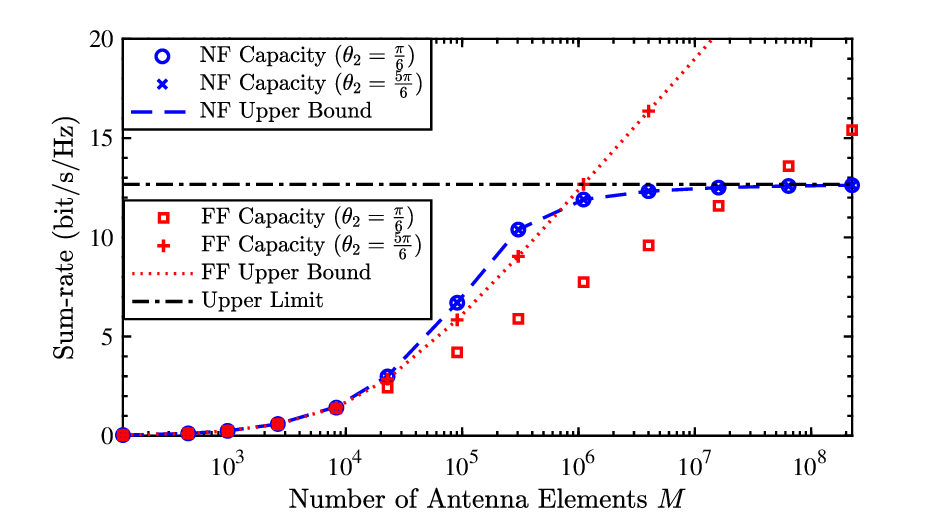}
\caption{Comparison of the BC sum-rate capacity for different channel models versus the number of antennas $M$, where ``NF'' and ``FF'' refer to ``near-field'' and ``far-field'', respectively. $d=0.0628$ m, $d=\lambda/2$, $A=\frac{\lambda^2}{4\pi}$, $r_1=15$ m, $\theta_1=\frac{\pi}{6}$, $\phi_1=\frac{\pi}{3}$, $r_2=20$ m, $\phi_2=\frac{\pi}{3}$, $M_x=M_z=\sqrt{M}$, and $\frac{P_{\mathsf{b}}}{\sigma^2}=30$ dB.}
\label{BC_Sum_Rate}
\end{figure}
To verify our results, we present the BC sum-rate capacity and its upper bound under both far-field and near-field channel models as a function of the number of antenna elements in {\figurename} {\ref{BC_Sum_Rate}}. In our simulations, the capacity upper bounds for different values of $\theta_2$ are identical. As observed, the sum-rate capacity under the far-field model exhibits unbounded growth with $M$, whereas the sum-rate capacity and its upper bound for the near-field model tend towards a finite upper limit. This behavior aligns with the law of conservation of energy, validating our assertions in Remark \ref{BC_Energy_Conservation}. Furthermore, when users are positioned in different directions, we observe a close alignment between the sum-rate capacity and its upper bound for both far-field and near-field models, attributable to the near-zero IUI in such scenarios. However, for users located in the \emph{same direction}, this alignment is only observed in the near-field channel model, as elaborated in Remark \ref{BC_RDMA_vs_ADMA}.

\begin{figure}[!t]
 \centering
\includegraphics[width=0.48\textwidth]{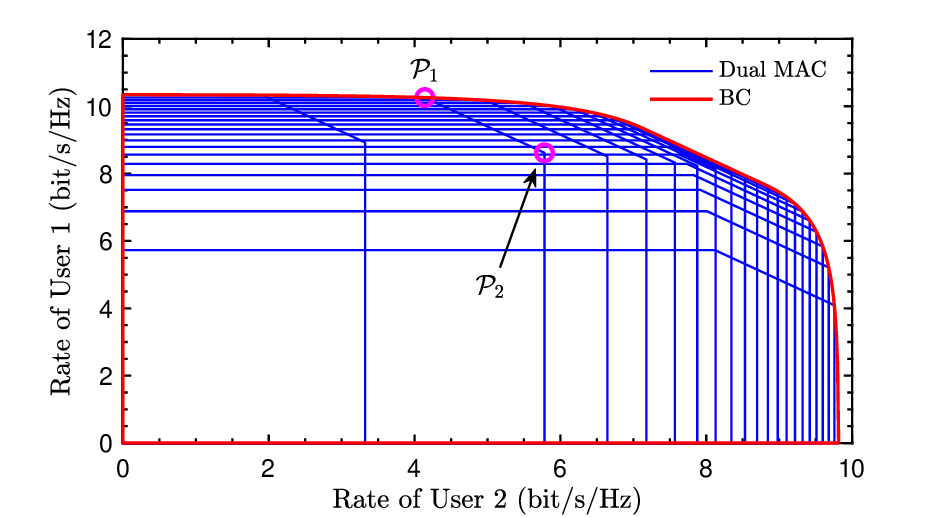}
\caption{An illustrated example of the capacity region (red line) for a two-user MISO-BC. ${\mathbf{h}}_1=[1+{\rm{j}};2+{\rm{j}}]^{\mathsf{T}}$, ${\mathbf{h}}_2=[1-{\rm{j}};2+{\rm{j}}]^{\mathsf{T}}$, and $\frac{P_{\mathsf{b}}}{\sigma^2}=20$ dB.}
\label{BC_Region}
\end{figure}
$\bullet$ \emph{Capacity Region}: We now turn our focus to the capacity region of the near-field MISO-BC. According to the MAC-BC duality, given a power allocation scheme $\{p_k\}_{k=1}^{2}$ with $\sum_{k=1}^{2}p_k= P_{\mathsf{b}}$ in \eqref{BC_Sum_Capacity}, we can obtain the capacity region of a dual MAC. \emph{The capacity region of the BC is the convex hull of all these dual MAC capacity regions} \cite{weingarten2006capacity,goldsmith2003capacity}. This implies that the boundary of the BC's capacity region is formed by the corner points of the capacity regions of the dual MACs, as depicted in {\figurename} {\ref{BC_Region}}. In the figure, the corner points of a dual MAC, i.e., $\mathcal{P}_1$ and $\mathcal{P}_2$, are labeled as circles. The blue lines denote the capacity regions of the dual MACs corresponding to a specific power allocation scheme $\{p_k\}_{k=1}^{2}$, while the red line represents the capacity region of the BC \cite{goldsmith2003capacity}.

Given a fixed $\{p_k\}_{k=1}^{2}$, the achieved rate pair $({\mathsf{R}}_{1},{\mathsf{R}}_{2})$ for users 1 and 2 can be determined. If the dirty-paper encoding order is $2\rightarrow1$, the rates are
\begin{subequations}\label{Two_User_BC_Rate_Pair_1}
\begin{align}
{\mathsf{R}}_{1}&= \log_2\left(\frac{1+\frac{p_1{\mathsf{a}}_1}{\sigma^2}+\frac{p_2{\mathsf{a}}_2}{\sigma^2}+\frac{p_1p_2}{\sigma^4}
{\mathsf{a}}_1{\mathsf{a}}_2(1-\rho_{\mathsf{u}}^2)}{1+p_2\sigma^{-2}{\mathsf{a}}_2}\right),\\
{\mathsf{R}}_{2}&= \log_2(1+p_2\sigma^{-2}{\mathsf{a}}_2).
\end{align}
\end{subequations}
Similarly, if the dirty-paper encoding order is $1\rightarrow2$, the rates are
\begin{subequations}\label{Two_User_BC_Rate_Pair_2}
\begin{align}
{\mathsf{R}}_{1}&= \log_2(1+p_1\sigma^{-2}{\mathsf{a}}_1) ,\\
{\mathsf{R}}_{2}&= \log_2\left(\frac{1+\frac{p_1{\mathsf{a}}_1}{\sigma^2}+\frac{p_2{\mathsf{a}}_2}{\sigma^2}+\frac{p_1p_2}{\sigma^4}
{\mathsf{a}}_1{\mathsf{a}}_2(1-\rho_{\mathsf{u}}^2)}{1+p_1\sigma^{-2}{\mathsf{a}}_1}\right).
\end{align}
\end{subequations}
These rate pairs correspond to the corner points of the capacity region of the dual MAC with transmission powers $\{p_k\}_{k=1}^{2}$. Following the asymptotic MAC capacity region discussed in Section \ref{Section: Capacity Limits for LoS Channels: Multiple Access Channel}, we can prove that the rate pairs at the corner points of the dual MACs satisfy 
\begin{subequations}
\begin{align}
\lim_{M\rightarrow\infty}{\mathsf{R}}_{1}&\approx \log_2(1+p_1\sigma^{-2}{\mathsf{a}}_1) ,\\
\lim_{M\rightarrow\infty}{\mathsf{R}}_{2}&\approx \log_2(1+p_2\sigma^{-2}{\mathsf{a}}_2).
\end{align}
\end{subequations}
This indicates that the capacity region of each dual MAC converges approximately from a pentagon to a rectangular shape, which resembles the upper limit region without IUI. Consequently, the boundary of the BC capacity region is approximately the convex hull of these rectangular regions. Following the methodology outlined in Section \ref{Section: Capacity Limits for LoS Channels: Multiple Access Channel}, this property of a rectangular capacity region also holds for the far-field model when users are positioned in different directions. However, if the users are located in the same direction, the capacity region of each dual MAC converges to a triangular shape. This triangular shape closely aligns with the rate region achieved by OMA techniques, as discussed in Remark \ref{MAC_Region_RDMA_vs_ADMA}.

\begin{figure}[!t]
 \centering
    \subfigure[$\theta_2=\frac{\pi}{6}$.]{
        \includegraphics[width=0.48\textwidth]{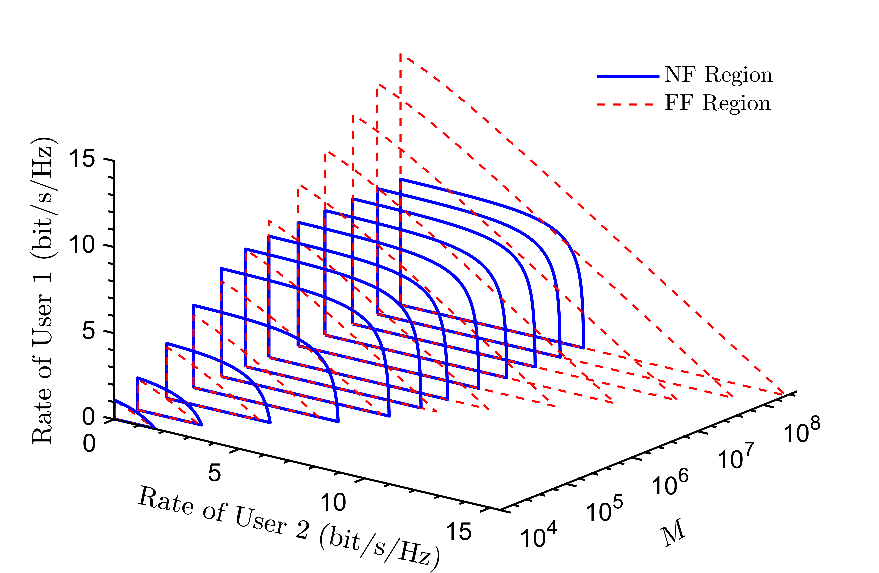}
        \label{Figure: BC_Capacity_Region_LoS1}
    }
    \subfigure[$\theta_2=\frac{5\pi}{6}$.]{
        \includegraphics[width=0.48\textwidth]{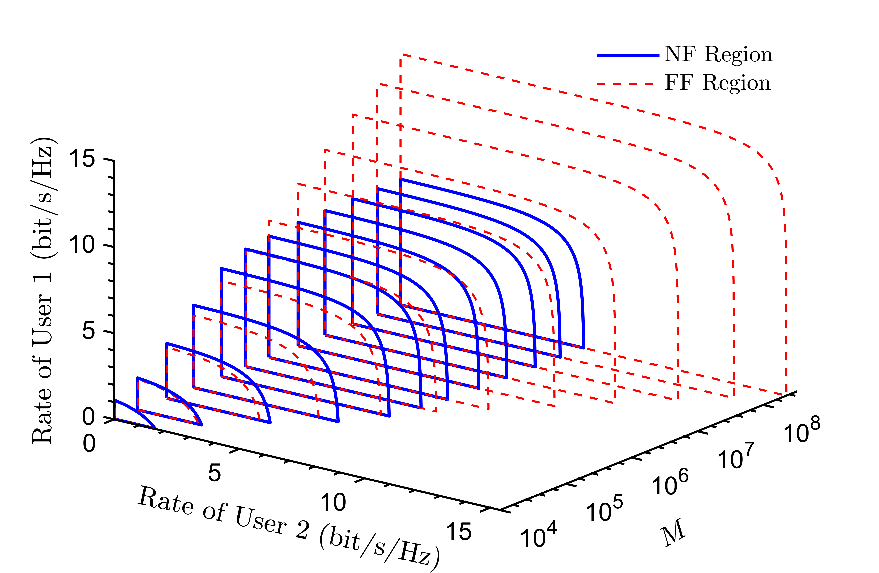}
        \label{Figure: BC_Capacity_Region_LoS2}	
    }
\caption{Comparison of the BC capacity regions for different channel models versus the number of antennas $M$, where ``NF'' and ``FF'' refer to ``near-field'' and ``far-field'', respectively. $d=0.0628$ m, $d=\lambda/2$, $A=\frac{\lambda^2}{4\pi}$, $r_1=15$ m, $\theta_1=\frac{\pi}{6}$, $\phi_1=\frac{\pi}{3}$, $r_2=20$ m, $\phi_2=\frac{\pi}{3}$, $M_x=M_z=\sqrt{M}$, and $\frac{P_{\mathsf{b}}}{\sigma^2}=30$ dB.}
\label{Figure: BC_Capacity_Region_LoS}
\end{figure}

To further validate our findings, we present the BC capacity regions under both far-field and near-field models as a function of the number of antenna elements in {\figurename} {\ref{Figure: BC_Capacity_Region_LoS}}. As observed, the capacity region under the far-field model demonstrates unbounded widening as $M$ increases, whereas the capacity region under the near-field model steadily converges towards a finite limit. Additionally, by comparing {\figurename} {\ref{Figure: BC_Capacity_Region_LoS1}} with {\figurename} {\ref{Figure: BC_Capacity_Region_LoS2}}, it can be seen that the capacity regions under the near-field model are virtually identical for scenarios where users are located in the \emph{same direction} and in \emph{different directions}. In contrast, for the far-field model, the capacity region is more constrained when users are situated in the same direction compared to when they are in different directions. This limitation arises due to the inherent challenge of far-field ADMA in mitigating IUI when users share the same angle of propagation.

$\bullet$ \emph{Extension to Cases of $K>2$}: In the following discussion, we extend the previous analysis to cases where $K>2$. The resulting channel capacity is achieved using DPC, and the capacity region can be characterized using the MAC-BC duality. The sum-rate capacity is given by
\begin{align}
{\mathsf{C}}_{\mathsf{BC}}=\max_{p_k\geq0,\sum_{k=1}^{K}p_k= P_{\mathsf{b}}}\log_2\det\left({\mathbf{I}}_M+\sum_{k=1}^{K}\frac{p_k}{\sigma^2}{\mathbf{h}}_k{\mathbf{h}}_k^{\mathsf{H}}\right).\label{BC_Sum_Capacity_General1}
\end{align}
By defining $\hat{\mathbf{H}}_{\mathsf{MAC}}\triangleq[\sqrt{p_1}{\mathbf{h}}_1,\ldots,\sqrt{p_K}{\mathbf{h}}_K]\in{\mathbbmss{C}}^{M\times K}$, we can rewrite \eqref{BC_Sum_Capacity_General1} as follows:
\begin{equation}\label{BC_Sum_Capacity_General2}
{\mathsf{C}}_{\mathsf{BC}}=\log_2\det\left({\mathbf{I}}_K+\frac{1}{\sigma^2}\hat{\mathbf{H}}_{\mathsf{MAC}}^{\mathsf{H}}\hat{\mathbf{H}}_{\mathsf{MAC}}\right),
\end{equation}
Each element of the matrix $\hat{\mathbf{H}}_{\mathsf{MAC}}^{\mathsf{H}}\hat{\mathbf{H}}_{\mathsf{MAC}}$ can be expressed as a function of the channel gain $\mathsf{a}_k=\lVert{\mathbf{h}}_k\rVert^2$ and the channel correlation factor $\frac{\lvert {\mathbf{h}}_k^{\mathsf{H}}{\mathbf{h}}_{k'}\rvert}{\lVert{\mathbf{h}}_k\rVert\lVert{\mathbf{h}}_{k'}\rVert}$. Therefore, the sum-rate capacity for $K>2$ can be analyzed using the same method as in the analysis of \eqref{BC_Sum_Capacity}. Further details are omitted here for brevity.

Regarding the capacity region, given $(p_1,\ldots,p_K)$, the capacity region of the dual MAC is defined as follows:
\begin{equation}\label{Gaussian_Vector_DMAC_Rate_Region}
\begin{split}
&\left\{({\mathsf{R}}_1,\ldots,{\mathsf{R}}_K):\sum\nolimits_{k\in{\mathscr{K}}}{\mathsf{R}}_k\right.\\
&\left.\leq\log_2\det\Big({\mathbf{I}}+\frac{1}{\sigma^2}\sum\nolimits_{k\in{\mathscr{K}}}p_k{\mathbf{h}}_k{\mathbf{h}}_k^{\mathsf{H}}\Big),\forall {\mathscr{K}}\subseteq{\mathcal{K}}\right\}.
\end{split}
\end{equation}
The capacity region of the BC is thus given by the convex hull of all the dual MAC capacity regions for all realizations of $(p_1,\ldots,p_K)$ subject to $p_k\leq0$ ($\forall k$) and $\sum_{k=1}^{K}p_k\leq P_{\mathsf{b}}$. We note that each corner point of the dual MAC capacity region is characterized by channel gains and channel correlation factors, which can be analyzed in the same way as described in the discussion of \eqref{Two_User_BC_Rate_Pair_1} or \eqref{Two_User_BC_Rate_Pair_2}.

$\bullet$ \emph{Non-orthogonal Multiple Access (NOMA)}: Although DPC is a potent capacity-achieving scheme, its practical implementation proves challenging. Consequently, non-DPC downlink non-orthogonal multiple access (NOMA) transmission schemes hold significant practical value. Unlike pre-SIC DPC, NOMA employs superposition coding and user-side SIC decoding to mitigate IUI, resulting in lower complexity compared to DPC. To date, many promising beamforming methods have been designed for NOMA, some of which can approach the performance of DPC \cite{liu2024road}. Due to its innovative approach, NOMA has recently been identified as a crucial component of the International Mobile Telecommunications (IMT)-2030 framework issued by the International Telecommunication Union (ITU) \cite{recommendation2023framework,ding2024next}. For recent advances in NOMA, please refer to \cite{sun2021new,sun2023hybrid,liu2024road} for more details.

NOMA can provide flexible resource allocation in multiuser FFC and NFC. The unique characteristics of NFC offer new design flexibility for facilitating near-field NOMA. By leveraging near-field beam focusing, signal energy can be concentrated on users located farther from the BS \cite{wu2023multiple}. Unlike far-field NOMA, where the effective channel gains of users in the same direction generally decrease with distance, near-field NOMA can achieve higher effective channel gains for users farther from the array. This implies that a `far-to-near' channel gain-based SIC order can be implemented in near-field NOMA, which is advantageous when the far user has a higher communication requirement than the near user. Early attempts to explore NOMA-based mixed near-field/far-field communications can be found in \cite{ding2023noma}. This work investigated the use of spatial beams preconfigured for legacy near-field users to serve additional far-field users.

Research on near-field NOMA is still in its infancy, and many related problems require further exploration. For example, this tutorial characterizes the sum-rate capacity of near-field BC, aiming to maximize the sum-rate under a power constraint. Additionally, it is also important to consider minimizing power consumption while satisfying the rate requirements of each user. In this context, it has been proven that there exists a scenario termed channel quasi-degradation, where NOMA achieves performance virtually equivalent to DPC \cite{chen2016application}. The works in \cite{chen2016application,zhu2020optimal} present solutions for optimal NOMA beamformers and conditions for channels being quasi-degraded. This property has garnered significant attention and is being applied in various scenarios; see \cite{sun2023application}. However, the impact of the near-field effect on channel quasi-degradation remains an open problem that requires further research efforts.

\subsection{Capacity Limits for CAP Arrays}
\begin{figure}[!t]
 \centering
\includegraphics[height=0.3\textwidth]{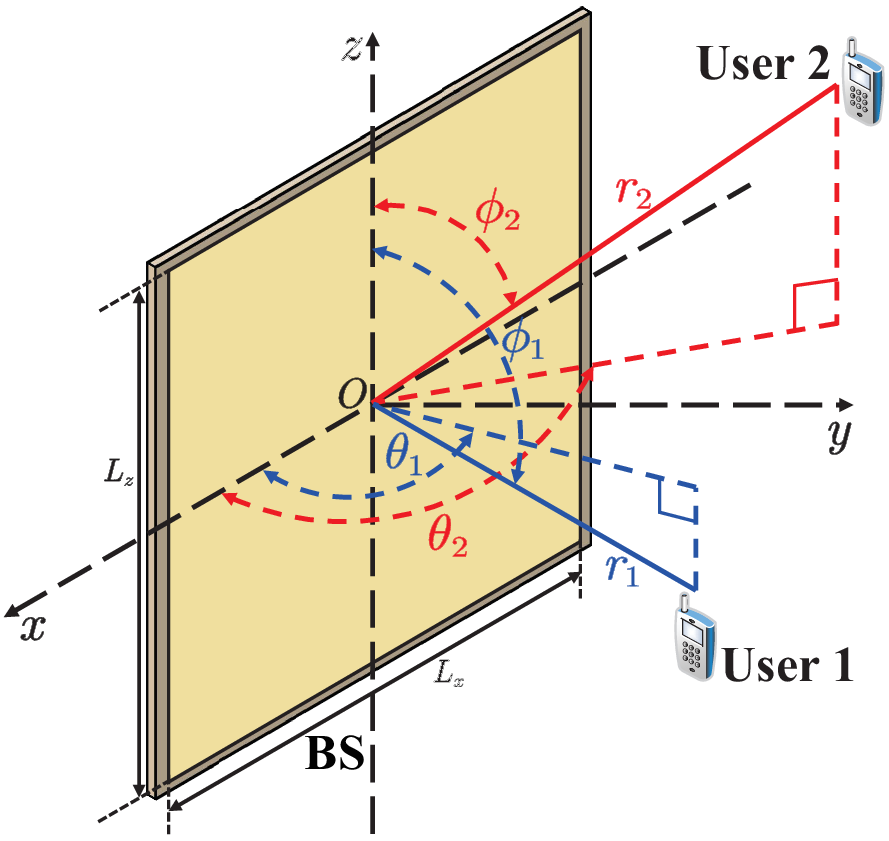}
\caption{Illustration of the array geometry of a two-user channel relaying on CAP arrays.}
\label{LoS_3D_Model_MU_CAP}
\end{figure}

Utilizing CAP arrays is considered a promising technique for improving the performance of MIMO systems with limited apertures \cite{huang2020holographic,gong2022holographic}. Unlike SPD arrays, which consist of a large number of discrete antenna elements with specific spacing, a CAP array can be conceptualized as an SPD array formed by an infinite number of antennas with infinitesimal spacing. This section investigates the capacity limits for a two-user LoS channel using CAP arrays. In this scenario, the BS is equipped with a two-dimensional CAP array with physical dimensions $L_x\times L_z$, and each user is equipped with a single hypothetical isotropic antenna element, as depicted in {\figurename} {\ref{LoS_3D_Model_MU_CAP}}. The remaining system setups are identical to those in {\figurename} {\ref{LoS_3D_Model_MU}}. Under this scenario, the channel response is characterized by Green's function. For clarity, our focus will be on the \emph{scalar Green's function}, which is given as follows:
\begin{align}
\mathsf{g}({\mathbf{r}}_k,{\mathbf{s}})=\frac{{\rm{e}}^{-{\rm{j}}\frac{2\pi}{\lambda}\lVert {\mathbf{r}}_k-{\mathbf{s}}\rVert}}{4\pi \lVert {\mathbf{r}}_k-{\mathbf{s}}\rVert},
\end{align}
where ${\mathbf{r}}_k=[r_k\Phi_k,r_k\Psi_k,r_k\Omega_k]^{\mathsf{T}}\in{\mathbbmss{R}}^{3\times1}$ and $\mathbf{s}\in{\mathbbmss{R}}^{3\times1}$ represent the coordinates of user $k$ and the point on the CAP array, respectively. Furthermore, let $\mathcal{A}_{\mathsf{S}}$ and $\mathcal{A}_{\mathsf{R}}^{k}$ denote the apertures of the BS and user $k$, respectively. In our considered case, we set $\lvert\mathcal{A}_{\mathsf{R}}^{k}\rvert=\lvert\mathcal{A}_{\mathsf{R}}\rvert\ll \lvert\mathcal{A}_{\mathsf{S}}\rvert$ for $k=1,2$. It is worth noting that the scalar Green's function accounts for spherical-wave propagation while omitting the influence of the projected aperture. Following the methodology used to derive \eqref{Near_Field_Model_Channel_Power}, we incorporate the influence of the projected aperture into $ \mathsf{g}({\mathbf{r}}_k,{\mathbf{s}})$, which yields its improved version as follows:
\begin{align}
\mathsf{h}({\mathbf{r}}_k,{\mathbf{s}})\triangleq\frac{{\rm{e}}^{-{\rm{j}}\frac{2\pi}{\lambda}\lVert {\mathbf{r}}_k-{\mathbf{s}}\rVert}}{4\pi \lVert {\mathbf{r}}_k-{\mathbf{s}}\rVert}\sqrt{\frac{\lvert{\mathbf{e}}_y^{\mathsf{T}}({\mathbf{r}}_k-{\mathbf{s}})\rvert}{\lVert{\mathbf{r}}_k-{\mathbf{s}}\rVert}}.
\end{align}
\subsubsection{Preliminaries}
Before characterizing the capacity limits, we introduce some preliminary concepts derived from uplink and downlink single-user setups.

We commence with the uplink case. Assuming that only user $k$ is transmitting signals, the observed electric field $\mathsf{y}(\mathbf{s})$ at point $\mathbf{s}\in\mathcal{A}_{\mathsf{S}} $ is the sum of the information-carrying electric field $\mathsf{e}(\mathbf{s})$ and a random noise field $\mathsf{n}(\mathbf{s})$, i.e.,
\begin{subequations}
\begin{align}
\mathsf{y}(\mathbf{s})&=\mathsf{e}(\mathbf{s})+\mathsf{n}(\mathbf{s})\\
&=\int_{\mathcal{A}_{\mathsf{R}}^k}-{\rm{j}}\eta_0\frac{2\pi}{\lambda}\mathsf{h}({\mathbf{s}},{\mathbf{r}})s_k{\mathsf{j}}_k({\mathbf{r}})
{\rm{d}}{\mathbf{r}}
+\mathsf{n}(\mathbf{s})\label{CAP_SU_Basic_Model},
\end{align}
\end{subequations}
where $s_k\in{\mathbbmss{C}}$ denotes the normalized coded data symbol, ${\mathsf{j}}_k({\mathbf{r}})\in{\mathbbmss{C}}$ denotes the source current used to convey data information with $\int_{\mathcal{A}_{\mathsf{R}}^k}\lvert{\mathsf{j}}_k({\mathbf{r}})\rvert^2{\rm{d}}{{\mathbf{r}}}$ representing the radiating power constraint, and $\mathsf{n}(\mathbf{s})$ accounts for thermal noise \cite{jensen2008capacity}. The noise field is modeled as a zero-mean complex Gaussian random process satisfying ${\mathbbmss{E}}\{{\mathsf{n}}({\mathbf{s}}){\mathsf{n}}^{\mathsf{H}}({\mathbf{s}}')\}=\overline{\sigma}^2\delta({\mathbf{s}}-{\mathbf{s}}')$, where $\delta(\cdot)$ represents the Dirac delta function. Given that $\lvert\mathcal{A}_{\mathsf{S}}\rvert\gg\lvert\mathcal{A}_{\mathsf{R}}^{k}\rvert=\lvert\mathcal{A}_{\mathsf{R}}\rvert$, \eqref{CAP_SU_Basic_Model} simplifies to
\begin{align}\label{CAP_SU_Basic_Mode2}
\mathsf{y}(\mathbf{s})=-{\rm{j}}\eta_0\frac{2\pi}{\lambda}\mathsf{h}({\mathbf{s}},\mathbf{r}_k)s_k{\mathsf{j}}_k(\mathbf{r}_k)\lvert\mathcal{A}_{\mathsf{R}}\rvert
+\mathsf{n}(\mathbf{s}).
\end{align}
In the subsequent steps, we recover $s_k$ from $\mathsf{y}(\mathbf{s})$ using a linear combiner $\mathsf{h}^{\mathsf{H}}({\mathbf{s}},\mathbf{r}_k)$ along with maximum-likelihood decoding, which leads to
\begin{equation}\label{CAP_SU_Basic_Mode3}
\begin{split}
\int_{\mathcal{A}_{\mathsf{S}}}\mathsf{y}(\mathbf{s})\mathsf{h}^{\mathsf{H}}({\mathbf{s}},\mathbf{r}_k){\rm{d}}{\mathbf{s}}&=\frac{
\int_{\mathcal{A}_{\mathsf{S}}}
\lvert\mathsf{h}({\mathbf{s}},{\mathbf{r}}_k)\rvert^2{\rm{d}}{\mathbf{s}}}{\lambda/(-{\rm{j}}\eta_0 2\pi {\mathsf{j}}_k(\mathbf{r}_k)\lvert\mathcal{A}_{\mathsf{R}}\rvert)}
s_k\\
&+\int_{\mathcal{A}_{\mathsf{S}}}\mathsf{h}^{\mathsf{H}}({\mathbf{s}},{\mathbf{r}}_k)\mathsf{n}(\mathbf{s}){\rm{d}}{\mathbf{s}}.
\end{split}
\end{equation}
Consequently, the SNR for decoding $s_k$ is given by
\begin{subequations}\label{CAP_SU_SNR_Expression}
\begin{align}
\gamma_{\mathsf{CAP}}^{(k)}&=\frac{(\eta_0\frac{2\pi}{\lambda}\lvert{\mathsf{j}}_k(\mathbf{r}_k)\rvert\lvert\mathcal{A}_{\mathsf{R}}\rvert)^2
(\int_{\mathcal{A}_{\mathsf{S}}}
\lvert\mathsf{h}({\mathbf{s}},{\mathbf{r}}_k)\rvert^2{\rm{d}}{\mathbf{s}})^2}
{\overline{\sigma}^2\int_{\mathcal{A}_{\mathsf{S}}}\lvert\mathsf{h}({\mathbf{s}},{\mathbf{r}}_k)\rvert^2{\rm{d}}{\mathbf{s}}}\\
&=\frac{(\eta_0\frac{2\pi}{\lambda}\lvert{\mathsf{j}}(\mathbf{r}_k)\rvert\lvert\mathcal{A}_{\mathsf{R}}\rvert)^2}{4\pi\overline{\sigma}^2}\overline{\mathsf{a}}_k,
\end{align}
\end{subequations}
where $\overline{\mathsf{a}}_k\triangleq4\pi\int_{\mathcal{A}_{\mathsf{S}}}
\lvert\mathsf{h}({\mathbf{s}},{\mathbf{r}}_k)\rvert^2{\rm{d}}{\mathbf{s}}=\int_{\mathcal{A}_{\mathsf{S}}}
\frac{\lvert{\mathbf{e}}_y^{\mathsf{T}}({\mathbf{r}}_k-{\mathbf{s}})\rvert}{ 4\pi\lVert {\mathbf{r}}_k-{\mathbf{s}}\rVert^3}{\rm{d}}{\mathbf{s}}$. 

We next consider the downlink case, assuming that the BS only sends signals to user $k$. The observed electric field at user $k$ can be written as follows:
{\setlength\abovedisplayskip{2pt}
\setlength\belowdisplayskip{2pt}
\begin{align}\label{Single_User_electric_radiation_field}
{\mathsf{y}}_k(\mathbf{r})=s_k\int_{{\mathcal{A}}_{\mathsf{S}}}-{\rm{j}}\eta_0\frac{2\pi}{\lambda}{\mathsf{h}}(\mathbf{r},{\mathbf{s}}){\mathsf{j}}({\mathbf{s}}){\rm{d}}{\mathbf{s}}+{\mathsf{n}}_k(\mathbf{r})
\end{align}
}for ${\mathbf{r}}\in{\mathcal{A}_{\mathsf{R}}^{k}}$, where $s_k\in{\mathbbmss{C}}$ denotes the normalized coded data symbol, ${\mathsf{j}}({\mathbf{s}})\in{\mathbbmss{C}}$ denotes the source current used to convey data information, and $\int_{\mathcal{A}_{\mathsf{S}}}\lvert{\mathsf{j}}({\mathbf{s}})\rvert^2{\rm{d}}{{\mathbf{s}}}$ represents the radiating power constraint. The term ${\mathsf{n}}_k(\mathbf{r})$ accounts for the thermal noise and is modeled as a zero-mean complex Gaussian process with ${\mathbbmss{E}}\{{\mathsf{n}}_k(\mathbf{r}){\mathsf{n}}_k^{\mathsf{H}}(\mathbf{r}')\}=\overline{\sigma}_k^2\delta(\mathbf{r}-\mathbf{r}')$.

We then design a detector ${\mathsf{v}}(\mathbf{r})$ to recover the data information contained in $s_k$. Applying this detector to ${\mathsf{y}}_k(\mathbf{r})$ yields
{\setlength\abovedisplayskip{2pt}
\setlength\belowdisplayskip{2pt}
\begin{equation}\label{Single_User_Detection}
\begin{split}
\int_{\mathcal{A}_{\mathsf{R}}^{k}}{\mathsf{v}}^{\mathsf{H}}(\mathbf{r}){\mathsf{y}}_k(\mathbf{r}){\rm{d}}{\mathbf{r}}
&=s_k\int_{\mathcal{A}_{\mathsf{R}}^{k}}{\mathsf{v}}^{\mathsf{H}}(\mathbf{r})\left(\int_{{\mathcal{A}}_{\mathsf{S}}}{\mathsf{h}}(\mathbf{r},{\mathbf{s}})
{\mathsf{j}}({\mathbf{s}}){\rm{d}}{\mathbf{s}}\right){\rm{d}}{\mathbf{r}}\\
&+\int_{\mathcal{A}_{\mathsf{R}}^{k}}{\mathsf{v}}^{\mathsf{H}}(\mathbf{r}){\mathsf{n}}_k(\mathbf{r}){\rm{d}}{\mathbf{r}}.
\end{split}
\end{equation}
}Since ${\mathsf{n}}_k(\mathbf{r})$ is a Gaussian random field, $\int_{\mathcal{A}_{\mathsf{R}}^{k}}{\mathsf{v}}^{\mathsf{H}}(\mathbf{r}){\mathsf{n}}_k(\mathbf{r}){\rm{d}}{\mathbf{r}}$ is a complex Gaussian random variable satisfying $\int_{\mathcal{A}_{\mathsf{R}}^{k}}{\mathsf{v}}^{\mathsf{H}}(\mathbf{r}){\mathsf{n}}_k(\mathbf{r}){\rm{d}}{\mathbf{r}}\sim{\mathcal{CN}}(0,\overline{\sigma}_k^2
\int_{\mathcal{A}_{\mathsf{R}}^{k}}\lvert{\mathsf{v}}(\mathbf{r})\rvert^2{\rm{d}}{\mathbf{r}})$. Consequently, the SNR for decoding $s_k$ is given by
\begin{equation}\label{Single_User_SNR}
\begin{split}
\gamma_{\mathsf{CAP}}^{(k)}=
\frac{\lvert\int_{\mathcal{A}_{\mathsf{R}}^{k}}{\mathsf{v}}^{\mathsf{H}}(\mathbf{r})(\int_{{\mathcal{A}}_{\mathsf{S}}}{\mathsf{h}}(\mathbf{r},{\mathbf{s}})
{\mathsf{j}}({\mathbf{s}}){\rm{d}}{\mathbf{s}}){\rm{d}}{\mathbf{r}}\rvert^2}
{\overline{\sigma}_k^2
\int_{\mathcal{A}_{\mathsf{R}}^{k}}\lvert{\mathsf{v}}(\mathbf{r})\rvert^2{\rm{d}}{\mathbf{r}}}.
\end{split}
\end{equation}
The next task is to find a detector that maximizes the SNR shown in \eqref{Single_User_SNR}. Observing the mathematical structure of $\gamma_{\mathsf{CAP}}^{(k)}$, we note that $\gamma_{\mathsf{CAP}}^{(k)}$ is independent of the norm of ${\mathsf{v}}(\mathbf{r})$, i.e., $\int_{\mathcal{A}_{\mathsf{R}}^{k}}\lvert{\mathsf{v}}(\mathbf{r})\rvert^2{\rm{d}}{\mathbf{r}}$. On this basis, the problem of designing ${\mathsf{v}}(\mathbf{r})$ can be equivalently transformed as follows:
\begin{equation}\label{Optimal_Detector_MISO_SU_Problem}
\argmax\nolimits_{{\mathsf{v}}(\mathbf{r})}{\left\lvert\int_{\mathcal{A}_{\mathsf{R}}^{k}}{\mathsf{v}}^{\mathsf{H}}(\mathbf{r})\left(\int_{{\mathcal{A}}_{\mathsf{S}}}{\mathsf{h}}(\mathbf{r},{\mathbf{s}})
{\mathsf{j}}({\mathbf{s}}){\rm{d}}{\mathbf{s}}\right){\rm{d}}{\mathbf{r}}
\right\rvert^2},
\end{equation}
whose optimal solution, i.e., the optimal detector, is given by
\begin{equation}\label{Optimal_Detector_MISO_SU_Answer}
{\mathsf{v}}(\mathbf{r})=\int_{{\mathcal{A}}_{\mathsf{S}}}{\mathsf{h}}(\mathbf{r},{\mathbf{s}}){\mathsf{j}}({\mathbf{s}}){\rm{d}}{\mathbf{s}}.
\end{equation}
Substituting \eqref{Optimal_Detector_MISO_SU_Answer} back into \eqref{Single_User_SNR} gives
{\setlength\abovedisplayskip{2pt}
\setlength\belowdisplayskip{2pt}
\begin{subequations}
\begin{align}
\gamma_{\mathsf{CAP}}^{(k)}&=\frac{1}{\overline{\sigma}_k^2}
\int_{\mathcal{A}_{\mathsf{R}}^{k}}\lvert{\mathsf{v}}(\mathbf{r})\rvert^2{\rm{d}}{\mathbf{r}}
=\frac{1}{\overline{\sigma}_k^2}
\int_{\mathcal{A}_{\mathsf{R}}^{k}}\left\lvert\int_{{\mathcal{A}}_{\mathsf{S}}}{\mathsf{h}}(\mathbf{r},{\mathbf{s}}){\mathsf{j}}({\mathbf{s}}){\rm{d}}{\mathbf{s}}\right\rvert^2{\rm{d}}{\mathbf{r}}\label{Single_User_Resulting_SNR1}\\
&\approx\frac{1}{{\overline{\sigma}_k^2}}
\left\lvert\int_{{\mathcal{A}}_{\mathsf{S}}}{\mathsf{h}}(\mathbf{r}_k,{\mathbf{s}}){\mathsf{j}}({\mathbf{s}}){\rm{d}}{\mathbf{s}}\right\rvert^2
\lvert\mathcal{A}_{\mathsf{R}}^{k}\rvert\label{Single_User_Resulting_SNR2}\\
&=\frac{\lvert\mathcal{A}_{\mathsf{R}}\rvert}{{\overline{\sigma}_k^2}}
\left\lvert\int_{{\mathcal{A}}_{\mathsf{S}}}{\mathsf{h}}(\mathbf{r}_k,{\mathbf{s}}){\mathsf{j}}({\mathbf{s}}){\rm{d}}{\mathbf{s}}\right\rvert^2,\label{Single_User_Resulting_SNR2_0}
\end{align}
\end{subequations}
}where the approximation in \eqref{Single_User_Resulting_SNR2} is due to the fact that $\lvert{\mathcal{A}}_{\mathsf{S}}\rvert\gg \lvert\mathcal{A}_{\mathsf{R}}^{k}\rvert$. It can be observed from \eqref{Single_User_Resulting_SNR2_0} that the SNR is maximized by setting the source current as follows:
\begin{equation}\label{Optimal_Precoder_MISO_SU_Answer}
{\mathsf{j}}({\mathbf{s}})={\sqrt{\int_{\mathcal{A}_{\mathsf{S}}}\lvert{\mathsf{j}}({\mathbf{s}})\rvert^2{\rm{d}}{\mathbf{s}}}}\frac{{\mathsf{h}}^{\mathsf{H}}({\mathbf{r}}_k,\mathbf{s})}
{\sqrt{\int_{{\mathcal{A}}_{\mathsf{S}}}\lvert{\mathsf{h}}({\mathbf{r}}_k,\mathbf{s})\rvert^2{\rm{d}}{\mathbf{s}}}}.
\end{equation}
Inserting \eqref{Optimal_Precoder_MISO_SU_Answer} into \eqref{Single_User_Resulting_SNR2_0} gives
\begin{equation}\label{Single_User_Resulting_SNR}
\begin{split}
\gamma_{\mathsf{CAP}}^{(k)}&=\frac{\int_{{\mathcal{A}}_{\mathsf{S}}}\lvert{\mathsf{j}}({\mathbf{s}})\rvert^2{\rm{d}}{\mathbf{s}}}{\overline{\sigma}_k^2}
\lvert\mathcal{A}_{\mathsf{R}}\rvert\int_{{\mathcal{A}}_{\mathsf{S}}}\lvert{\mathsf{h}}({\mathbf{r}}_k,\mathbf{s})\rvert^2{\rm{d}}{\mathbf{s}}\\
&=\frac{(\eta_0\frac{2\pi}{\lambda}\lvert\mathcal{A}_{\mathsf{R}}\rvert)^2\int_{{\mathcal{A}}_{\mathsf{S}}}\lvert{\mathsf{j}}({\mathbf{s}})\rvert^2{\rm{d}}{\mathbf{s}}}
{4\pi\overline{\sigma}_k^2}\underbrace{\int_{\mathcal{A}_{\mathsf{S}}}
\frac{\lvert{\mathbf{e}}_y^{\mathsf{T}}({\mathbf{r}}_k-{\mathbf{s}})\rvert}{ 4\pi\lVert {\mathbf{r}}_k-{\mathbf{s}}\rVert^3}{\rm{d}}{\mathbf{s}}}_{\overline{\mathsf{a}}_k}.
\end{split}
\end{equation}

Utilizing the results from \eqref{Received_SNR_Analog_SU_Near_Field_Trans_Pre} and \eqref{SU_NFC_Channel_Gain_Result_Final}, we obtain
\begin{equation}\label{SU_NFC_Channel_Gain_Result_Final_CAP}
\overline{\mathsf{a}}_k=\frac{1}{4\pi}\sum_{x\in{\mathcal{X}}_k^{'}}
\sum_{z\in{\mathcal{Z}}_k^{'}}\arctan\left(\frac{xz}{\Psi_k\sqrt{\Psi_k^2+x^2+z^2}}\right),
\end{equation}
where ${\mathcal{X}}_k^{'}=\{\frac{L_x}{2r_k}\pm\Phi_k\}$ and ${\mathcal{Z}}_k^{'}=\{\frac{L_z}{2r_k}\pm\Psi_k\}$. By comparing $\overline{\mathsf{a}}_k$ with $\mathsf{a}_k$ from \eqref{SU_NFC_Channel_Gain_Result_Final}, we find
\begin{align}
\mathsf{a}_k=\xi_{\mathsf{r}}\overline{\mathsf{a}}_k,
\end{align}
where $\xi_{\mathsf{r}}=\frac{A}{d^2}$ represents the array occupation ratio of the SPD counterpart of the CAP array illustrated in {\figurename} {\ref{LoS_3D_Model_MU_CAP}}.
\vspace{-5pt}
\begin{remark}
The relationship $\mathsf{a}_k=\xi_{\mathsf{r}}\overline{\mathsf{a}}_k$ indicates that $\overline{\mathsf{a}}_k$ represents the channel gain achieved by SPD arrays when the array occupation ratio is $1$, meaning the SPD array becomes a CAP array. This suggests that $\overline{\mathsf{a}}_k$ serves as the CAP equivalent of $\mathsf{a}_k$. Based on this observation, we term $\overline{\mathsf{a}}_k$ as the channel gain achieved by CAP arrays. In this context, the expression $\frac{(\eta_0\frac{2\pi}{\lambda}\lvert{\mathsf{j}}(\mathbf{r}_k)\rvert\lvert\mathcal{A}_{\mathsf{R}}\rvert)^2}{4\pi\overline{\sigma}^2}$ can be interpreted as the transmit SNR, equivalent to $\frac{P_{\mathsf{m}}}{\sigma^2}$ shown in \eqref{Sum_Capacity_MAC2}. It is noteworthy that when considering CAP arrays, the transmit SNR is primarily influenced by the current intensity ${\mathsf{j}}(\mathbf{r}_k)$. 
\end{remark}
\vspace{-5pt}
By letting the aperture size of the CAP array approach infinity, that is, $L_x,L_z\rightarrow\infty$, we obtain
\begin{equation}\label{CAP_SU_NFC_Channel_Gain_Result_Asymptotic}
\begin{split}
\overline{\mathsf{a}}_k\rightarrow\frac{4\lim_{x,z\rightarrow\infty}\arctan\left(\frac{xz/\Psi_k}{\sqrt{\Psi_k^2+x^2+z^2}}\right)}{4\pi}
=\frac{4\frac{\pi}{2}}{4\pi }=\frac{1}{2}.
\end{split}
\end{equation}
This result suggests that when using an infinitely large CAP array for reception, only half of the power transmitted by an isotropic source is captured, while the other half does not reach the array. This finding is consistent with the law of energy conservation and aligns with the previous discussion of ${\mathsf{a}}_k$. Next, we analyze the channel gain under the far-field model as described in \eqref{CAP_UPA_FFC_Model}, where the channel gain of user $k$ is given by ${\frac{{\Psi_k }}{{4\pi} r_k^2}}L_xL_z$, which scales linearly with $L_xL_z$ and thereby violates the law of energy conservation. 
\vspace{-5pt}
\begin{remark}
The above asymptotic analyses do not take into account the influence of the reactive region. This omission is justified by the fact that when we use \eqref{T_Green_Standard_Scalar} to model the channel response, we obtain the same result as presented in \eqref{SU_NFC_Channel_Gain_Result_Evanescent2} regarding the channel gain. Consequently, it can be concluded that the influence of evanescent waves can be disregarded when conducting asymptotic analyses in the context of NFC supported by CAP arrays.
\end{remark}
\vspace{-5pt}
Following the methodology in defining the channel gain for CAP arrays, we can also define the channel correlation factor akin to $\rho_{\mathsf{u}}$ in \eqref{Channel_Correlation_SPD}, which is given by
\begin{align}
\overline{\rho}_{\mathsf{u}}\triangleq \frac{\left\lvert {\int_{\mathcal{A}_{\mathsf{S}}}
\mathsf{h}^{\mathsf{H}}({\mathbf{s}},{\mathbf{r}}_1)\mathsf{h}({\mathbf{s}},{\mathbf{r}}_2){\rm{d}}{\mathbf{s}}}\right\rvert}
{\sqrt{\int_{\mathcal{A}_{\mathsf{S}}}
\lvert\mathsf{h}({\mathbf{s}},{\mathbf{r}}_1)\rvert^2{\rm{d}}{\mathbf{s}}}
\sqrt{\int_{\mathcal{A}_{\mathsf{S}}}
\lvert\mathsf{h}({\mathbf{s}},{\mathbf{r}}_2)\rvert^2{\rm{d}}{\mathbf{s}}}}\in[0,1].
\end{align}
For far-field channels, we have \cite{liu2023near,liu2023nearfield}
\begin{align}\label{CAP_Channel_Correlation_Far_Field}
\lim_{L_xL_z\rightarrow\infty}\overline{\rho}_{\mathsf{u}}=\left\{
\begin{array}{ll}
1             & {\mathbf{a}_1}= {\mathbf{a}_2}\\
0           & {\mathbf{a}_1}\ne {\mathbf{a}_2}
\end{array} \right.,
\end{align}
where $\mathbf{a}_k=[\Phi_k,\Psi_k,\Omega_k]^{\mathsf{T}}$ represents the directional vector of user $k=1,2$. For near-field channels, we have \cite{liu2023near,liu2023nearfield}
\begin{align}\label{CAP_Channel_Correlation_Near_Field}
\lim_{L_xL_z\rightarrow\infty}\overline{\rho}_{\mathsf{u}}\left\{
\begin{array}{ll}
=1             & {\mathbf{r}_1}= {\mathbf{r}_2}\\
\ll 1           & {\mathbf{r}_1}\ne {\mathbf{r}_2}
\end{array} \right..
\end{align}
\vspace{-5pt}
\begin{remark}
The results in \eqref{CAP_Channel_Correlation_Far_Field} and \eqref{CAP_Channel_Correlation_Near_Field} suggest that far-field ADMA cannot effectively handle IUI when users are located in the same direction. However, this limitation can be mitigated by near-field RDMA, which can distinguish users in the same direction due to the additional range information introduced by spherical-wave propagation. These results are consistent with their SPD counterparts shown in Section \ref{Section: Capacity Limits for LoS Channels: Preliminaries}.
\end{remark}
\vspace{-5pt}
\subsubsection{Multiple Access Channel}
Based on the above results, we now study the near-field capacity limits achieved by CAP arrays. We commence with the MAC, where the two users simultaneously transmit their respective messages to the BS. The observed electric field $\mathsf{y}(\mathbf{s})$ at point $\mathbf{s}\in\mathcal{A}_{\mathsf{S}} $ is the sum of the information-carrying electric fields $\mathsf{e}_1(\mathbf{s})$ and $\mathsf{e}_2(\mathbf{s})$, along with a random noise field $\mathsf{n}(\mathbf{s})$, i.e.,
\begin{subequations}\label{CAP_MAC_Basic_Model}
\begin{align}
\mathsf{y}(\mathbf{s})&=\mathsf{e}_1(\mathbf{s})+\mathsf{e}_2(\mathbf{s})+\mathsf{n}(\mathbf{s})\\
&=\int_{\mathcal{A}_{\mathsf{R}}^{1}}-{\rm{j}}\eta_0\frac{2\pi}{\lambda}\mathsf{h}({\mathbf{s}},{\mathbf{r}}_1')s_1{\mathsf{j}}_1({\mathbf{r}}_1')
{\rm{d}}{\mathbf{r}}_1'\\
&+\int_{\mathcal{A}_{\mathsf{R}}^{2}}-{\rm{j}}\eta_0\frac{2\pi}{\lambda}\mathsf{h}({\mathbf{s}},{\mathbf{r}}_2')s_2{\mathsf{j}}_2({\mathbf{r}}_2')
{\rm{d}}{\mathbf{r}}_2'
+\mathsf{n}(\mathbf{s}).
\end{align}
\end{subequations}
Here, $s_k\in{\mathbbmss{C}}$ represents the normalized data symbol sent by user $k=1,2$, ${\mathsf{j}}_k({\mathbf{r}}_k')$ is the associated source current conveying data information with $\int_{\mathcal{A}_{\mathsf{R}}^{k}}\lvert{\mathsf{j}}_k({\mathbf{r}}_k')\rvert^2{\rm{d}}{{\mathbf{r}}_k'}$ being the radiating power constraint, and $\mathsf{n}(\mathbf{s})$ accounts for thermal noise. The noise field is modeled as a zero-mean complex Gaussian random process satisfying ${\mathbbmss{E}}\{{\mathsf{n}}({\mathbf{s}}){\mathsf{n}}^{\mathsf{H}}({\mathbf{s}}')\}=\overline{\sigma}^2\delta({\mathbf{s}}-{\mathbf{s}}')$. Given that $\lvert\mathcal{A}_{\mathsf{S}}\rvert\gg\lvert\mathcal{A}_{\mathsf{R}}^{k}\rvert=\lvert\mathcal{A}_{\mathsf{R}}\rvert$, \eqref{CAP_MAC_Basic_Model} simplifies to the following form:
\begin{align}\label{CAP_MAC_Basic_Mode2}
\mathsf{y}(\mathbf{s})=\sum_{k=1}^{2}-{\rm{j}}\eta_0\frac{2\pi}{\lambda}\mathsf{h}({\mathbf{s}},\mathbf{r}_k)s_k{\mathsf{j}}_k(\mathbf{r}_k)\lvert\mathcal{A}_{\mathsf{R}}\rvert
+\mathsf{n}(\mathbf{s}).
\end{align}

As illustrated in {\figurename} {\ref{MAC_Region}}, the MAC capacity region is achieved through SIC decoding along with time sharing. Notably, the corner points ${\mathcal{P}}_1$ and ${\mathcal{P}}_2$ are attained by the SIC decoding orders $2\rightarrow1$ and $1\rightarrow2$, respectively. The line segment connecting ${\mathcal{P}}_1$ and ${\mathcal{P}}_2$ is achieved via a time sharing between ${\mathcal{P}}_1$ and ${\mathcal{P}}_2$. Besides, the sum-rate capacity is always the same regardless of the decoding order. 

Let us begin our discussion with the corner point achieved by the decoding order $2\rightarrow1$. In this case, $s_1$ is decoded without IUI and thus the resulting rate is determined by the received SNR as shown in \eqref{CAP_SU_SNR_Expression}. To keep consistence with the SPD arrays, we define $\frac{(\eta_0\frac{2\pi}{\lambda}\lvert{\mathsf{j}}_k(\mathbf{r}_k)\rvert\lvert\mathcal{A}_{\mathsf{R}}\rvert)^2}{4\pi\overline{\sigma}^2}\triangleq\frac{P_{\mathsf{m}}}{\sigma^2}$ as the transmit SNR for $k=1,2$, and the achieved rate for user 1 is expressed as follows:
\begin{align}
{\mathsf{R}}_{1}=\log_2(1+P_{\mathsf{m}}/\sigma^{-2}\overline{\mathsf{a}}_1).
\end{align}
We then analyze the achieved rate for user 2. Specifically, when decoding $s_2$, we must consider the influence of IUI. Following the principle of multiuser detection \cite{tse2005fundamentals}, we initially using a linear transform ${\mathsf{W}}({\mathbf{s}}',{\mathbf{s}})$ to whiten the interference-plus-noise term $\mathsf{z}(\mathbf{s})\triangleq-{\rm{j}}\eta_0\frac{2\pi}{\lambda}\mathsf{h}({\mathbf{s}},\mathbf{r}_1)s_1{\mathsf{j}}_1(\mathbf{r}_1)\lvert\mathcal{A}_{\mathsf{R}}\rvert
+\mathsf{n}(\mathbf{s})$ into an AWGN field: 
\begin{equation}\label{CAP_MMSE_SIC}
\begin{split}
\int_{\mathcal{A}_{\mathsf{S}}}{\mathsf{W}}({\mathbf{s}}',{\mathbf{s}})\mathsf{z}(\mathbf{s}){\rm{d}}{\mathbf{s}}\triangleq {\mathsf{n}}_{\mathsf{eq}}({\mathbf{s}}').
\end{split}
\end{equation}
Subsequently, we employ maximal-ratio combining (MRC) on the resulting channel: 
\begin{equation}\label{CAP_MAC_AWGN_Channel}
\begin{split}
&\int_{\mathcal{A}_{\mathsf{S}}}{\mathsf{W}}({\mathbf{s}}',{\mathbf{s}})\mathsf{y}(\mathbf{s}){\rm{d}}{\mathbf{s}}\\
&=s_2\int_{\mathcal{A}_{\mathsf{S}}}{\mathsf{W}}({\mathbf{s}}',{\mathbf{s}})\frac{-{\rm{j}}\eta_02\pi}{\lambda}\mathsf{h}({\mathbf{s}},\mathbf{r}_2)
{\mathsf{j}}_2(\mathbf{r}_2)\lvert\mathcal{A}_{\mathsf{R}}\rvert{\rm{d}}{\mathbf{s}}
+{\mathsf{n}}_{\mathsf{eq}}({\mathbf{s}}').
\end{split}
\end{equation}
As proved in \cite[Lemma 3]{ouyang2024performance}, ${\mathsf{W}}({\mathbf{s}}',{\mathbf{s}})$ can be designed as follows:
\begin{equation}\label{CAP_LMMSE_Combiner}
\begin{split}
{\mathsf{W}}({\mathbf{s}}',{\mathbf{s}})={\delta({\mathbf{s}}'-{\mathbf{s}})}
+{\lambda_{\star}}\mathsf{h}({\mathbf{s}}',\mathbf{r}_1)
\mathsf{h}^{\mathsf{H}}({\mathbf{s}},\mathbf{r}_1),
\end{split}
\end{equation}
where $\delta(\cdot)$ denotes the Dirac delta function, and the regularizer ${\lambda_{\star}}$ is set as follows: 
\begin{equation}\label{CAP_LMMSE_Regularizer}
\begin{split}
{\lambda_{\star}}=-\frac{1}{\overline{\mathsf{a}}_1}\pm\frac{1}{\overline{\mathsf{a}}_1\sqrt{1+\overline{\mathsf{a}}_1P_{\mathsf{m}}/\sigma^2}}.
\end{split}
\end{equation}
In fact, \cite{ouyang2024performance} demonstrates that ${\mathsf{W}}({\mathbf{s}}',{\mathbf{s}})$ not only whitens the interference-plus-noise term but also remains invertible, or information lossless. This property is the fundamental reason for using this detector to aid in data recovery. By substituting equations \eqref{CAP_LMMSE_Combiner} and \eqref{CAP_LMMSE_Regularizer} into \eqref{CAP_MMSE_SIC} and performing basic mathematical manipulations, we obtain the correlation function of ${\mathsf{n}}_{\mathsf{eq}}({\mathbf{s}}')$ as follows:
\begin{align}\label{MMSE_Equal_Noise_Correlation}
{\mathbbmss{E}}\{{\mathsf{n}}_{\mathsf{eq}}^{\mathsf{H}}({\mathbf{s}}'){\mathsf{n}}_{\mathsf{eq}}({\mathbf{s}}'')\}=\overline{\sigma}^2\delta({\mathbf{s}}'-{\mathbf{s}}'').
\end{align}
The result in \eqref{MMSE_Equal_Noise_Correlation} implies that the following channel:
\begin{equation}
\begin{split}
\int_{\mathcal{A}_{\mathsf{S}}}({\mathsf{W}}({\mathbf{s}}',{\mathbf{s}})(-{\rm{j}}\eta_0\frac{2\pi}{\lambda}
\mathsf{h}({\mathbf{s}},\mathbf{r}_2)s_2{\mathsf{j}}_2(\mathbf{r}_2)\lvert\mathcal{A}_{\mathsf{R}}\rvert))
{\rm{d}}{\mathbf{s}}+{\mathsf{n}}_{\mathsf{eq}}({\mathbf{s}}')\nonumber
\end{split}
\end{equation}
forms \emph{a single-stream AWGN channel}. Hence, the corresponding channel capacity is achieved by employing MRC combining, and the resulting SNR is computed as follows:
\begin{equation}\label{MAC_MMSE_SNR_General}
\begin{split}
\gamma_{\mathsf{CAP}}^{\mathsf{MAC},2}
&=\frac{(\eta_0\frac{2\pi}{\lambda}\lvert{\mathsf{j}}_2(\mathbf{r}_2)\rvert\lvert\mathcal{A}_{\mathsf{R}}\rvert)^2}{\overline{\sigma}^2}\\
&\times{\int_{\mathcal{A}_{\mathsf{S}}}\left\vert
\int_{\mathcal{A}_{\mathsf{S}}}{\mathsf{W}}({\mathbf{s}}',{\mathbf{s}})\mathsf{h}({\mathbf{s}},\mathbf{r}_2)
{\rm{d}}{\mathbf{s}}\right\rvert^2{\rm{d}}{\mathbf{s}}'}.
\end{split}
\end{equation}
Inserting \eqref{CAP_LMMSE_Combiner} into \eqref{MAC_MMSE_SNR_General} gives
\begin{equation}\label{MAC_MMSE_SNR_General1}
\begin{split}
\gamma_{\mathsf{CAP}}^{\mathsf{MAC},2}
=\frac{(\eta_0\frac{2\pi}{\lambda}\lvert{\mathsf{j}}_2(\mathbf{r}_2)\rvert\lvert\mathcal{A}_{\mathsf{R}}\rvert)^2}{\overline{\sigma}^24\pi}
(\overline{\mathsf{a}}_2+(\lambda_{\star}^2\overline{\mathsf{a}}_1+2\lambda_{\star})\overline{\mathsf{a}}_1
\overline{\mathsf{a}}_2\overline{\rho}_{\mathsf{u}}^2),
\end{split}
\end{equation}
which, together with the definition of $\frac{(\eta_0\frac{2\pi}{\lambda}\lvert{\mathsf{j}}_k(\mathbf{r}_k)\rvert\lvert\mathcal{A}_{\mathsf{R}}\rvert)^2}{4\pi\overline{\sigma}^2}\triangleq\frac{P_{\mathsf{m}}}{\sigma^2}$, yields
\begin{align}
{\mathsf{R}}_{2}
=\log_2\left(1+\frac{P_{\mathsf{m}}(\overline{\mathsf{a}}_2+(\lambda_{\star}^2\overline{\mathsf{a}}_1+2\lambda_{\star})\overline{\mathsf{a}}_1
\overline{\mathsf{a}}_2\overline{\rho}_{\mathsf{u}}^2)}{\sigma^2}\right).\nonumber
\end{align}
Moreover, according to \eqref{CAP_LMMSE_Regularizer}, we can obtain
\begin{align}
\lambda_{\star}^2\overline{\mathsf{a}}_1+2\lambda_{\star}=-\frac{1}{\overline{\mathsf{a}}_1}\frac{\overline{\mathsf{a}}_1P_{\mathsf{m}}/\sigma^2}{1+\overline{\mathsf{a}}_1P_{\mathsf{m}}/\sigma^2}.
\end{align}
Taken together, the achieved rate for user 2 is given by
\begin{align}
{\mathsf{R}}_{2}
=\log_2\left(1+\frac{P_{\mathsf{m}}}{\sigma^2}{\overline{\mathsf{a}}}_2\left(1-\overline{\rho}_{\mathsf{u}}^2\left(1-\frac{1}{1+\frac{P_{\mathsf{m}}}{\sigma^2}{\overline{\mathsf{a}}}_1}\right)\right)\right),
\end{align}
which can rewritten as follows:
\begin{align}
{\mathsf{R}}_{2}
=\log_2\left(\frac{1+\frac{P_{\mathsf{m}}}{\sigma^2}(\overline{\mathsf{a}}_1+\overline{\mathsf{a}}_2)+\frac{P_{\mathsf{m}}^2}{\sigma^4}\overline{\mathsf{a}}_1
\overline{\mathsf{a}}_2(1-\overline{\rho}_{\mathsf{u}}^2)}
{1+\frac{P_{\mathsf{m}}}{\sigma^2}\overline{\mathsf{a}}_1}\right).
\end{align}

By following the same steps, we can derive the rate pair at the corner point achieved through the decoding order $1\rightarrow2$, which is given by
\begin{subequations}\label{Two_User_MAC_Corner_Point_Rate}
\begin{align}
{\mathsf{R}}_{1}&= \log_2\left(\frac{1+\frac{P_{\mathsf{m}}}{\sigma^2}(\overline{\mathsf{a}}_1+\overline{\mathsf{a}}_2)+\frac{P_{\mathsf{m}}^2}{\sigma^4}\overline{\mathsf{a}}_1
\overline{\mathsf{a}}_2(1-\overline{\rho}_{\mathsf{u}}^2)}
{1+\frac{P_{\mathsf{m}}}{\sigma^2}\overline{\mathsf{a}}_2}\right),\label{Two_User_MAC_Corner_Point_Rate1}\\
{\mathsf{R}}_{2}&= \log_2(1+P_{\mathsf{m}}/\sigma^{-2}\overline{\mathsf{a}}_2).\label{Two_User_MAC_Corner_Point_Rate2}
\end{align}
\end{subequations}
The sum-rate capacity is thus given by
\begin{align}
{\mathsf{C}}_{\mathsf{MAC}}=
\log_2\left(1+\frac{P_{\mathsf{m}}}{\sigma^2}(\overline{\mathsf{a}}_1+\overline{\mathsf{a}}_2)+\frac{P_{\mathsf{m}}^2}{\sigma^4}\overline{\mathsf{a}}_1
\overline{\mathsf{a}}_2(1-\overline{\rho}_{\mathsf{u}}^2)\right).
\end{align}
We note that the expressions for the sum-rate capacity and the corner points closely resemble those for SPD arrays. Therefore, the insights derived from the MAC relying on SPD arrays can be straightforwardly extended to the case of CAP arrays. Consequently, we shall omit further discussions on this matter here.

\begin{figure}[!t]
    \centering
    \subfigbottomskip=0pt
	\subfigcapskip=-5pt
\setlength{\abovecaptionskip}{0pt}
    \subfigure[Capacity region vs. $\xi_{\mathsf{r}}$. $d=\frac{\lambda}{2}$]
    {
        \includegraphics[width=0.48\textwidth]{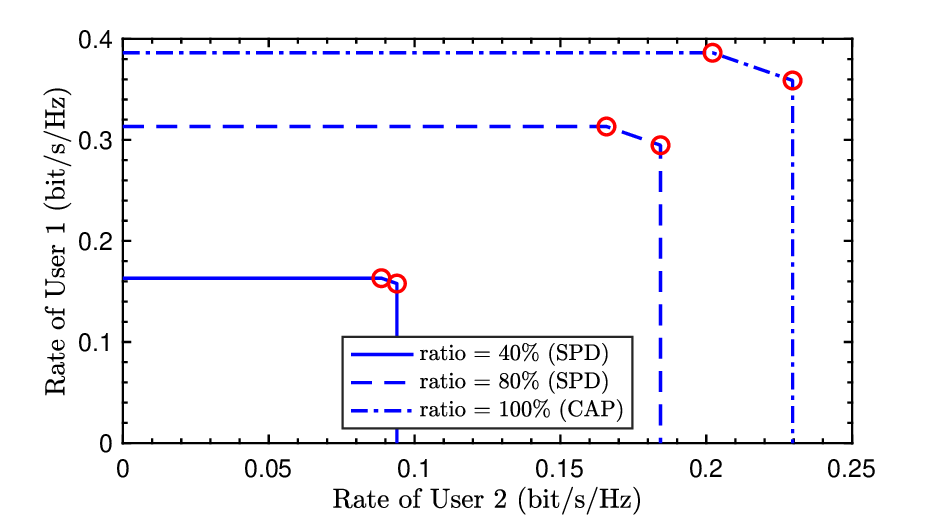}
	   \label{fig4a}	
    }
   \subfigure[Capacity region vs. $\lvert{\mathcal{A}}_{\mathsf{S}}\rvert$.]
    {
        \includegraphics[width=0.48\textwidth]{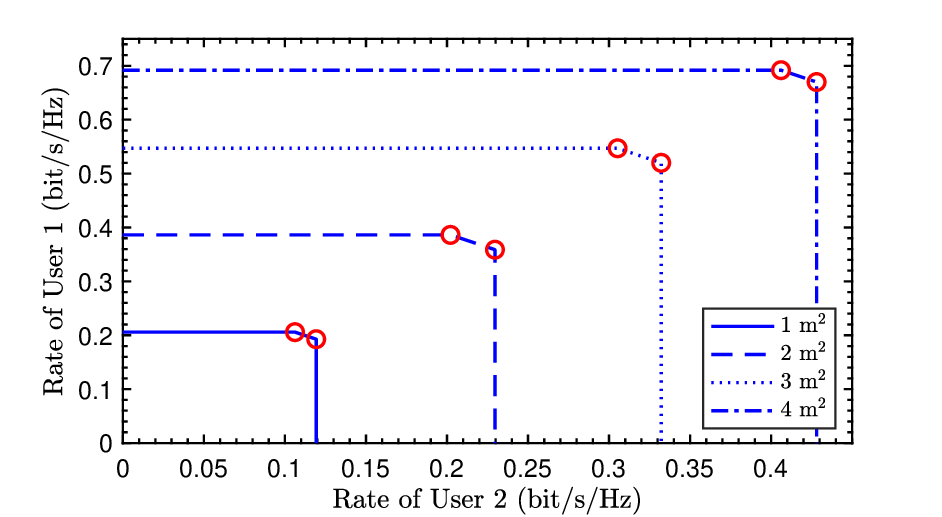}
	   \label{fig4b}	
    }
\caption{MAC capacity regions of CAP arrays. $\theta_1=\theta_2=\frac{\pi}{6}$, $\phi_1=\phi_2=\frac{\pi}{3}$, $r_1=10$ m, $r_2=20$ m, $\lambda = 0.0107$ m, $d=\frac{\lambda}{2}$, $A=\frac{\lambda^2}{4\pi}$, $M_x=M_z$, and $L_x=L_z$, and $\frac{P_{\mathsf{m}}}{\sigma^2}=30$ dB.}
    \label{Figure4}
    \vspace{-10pt}
\end{figure}
{\figurename} {\ref{Figure4}} illustrates the capacity regions achieved by the CAP and SPD arrays, where the red points represent the achieved rates by \emph{SIC decoding}. The rate tuple on the line segment connecting these two points is achieved by the \emph{time-sharing strategy}. As depicted in {\figurename} {\ref{fig4a}}, the CAP array yields a broader capacity region than the SPD array, and their gap decreases as the array occupation ratio $\xi_{\mathsf{r}}$ increases, which is as expected. In {\figurename} {\ref{fig4b}}, the capacity regions achieved by the CAP array for various values of $\lvert{\mathcal{A}}_{\mathsf{S}}\rvert$ are illustrated. It is observed that the capacity region gradually extends as $\lvert{\mathcal{A}}_{\mathsf{S}}\rvert$ increases, which transitions from a pentagon to a rectangular shape. This phenomenon occurs because the IUI gradually decreases as the aperture size increases, which aligns with the results shown in {\figurename} {\ref{Figure: MAC_Capacity_Region_LoS}}. These findings underscore the superiority of CAP arrays over traditional SPD arrays in terms of channel capacity.

\subsubsection*{Extension to Cases of $K>2$} We next consider the scenario involving more than two users. In this context, the capacity region can still be achieved using point-to-point Gaussian codes and SIC decoding in conjunction with time sharing. However, unlike the two-user case, deriving a closed-form solution for the capacity-achieving detector when $K>2$ presents a significant challenge for CAP arrays. This complexity arises due to the intricate operator calculations within Hilbert space. Consequently, obtaining a closed-form expression for the sum-rate capacity remains a formidable task, warranting further research attention.
\subsubsection{Broadcast Channel}
We next analyze the channel capacity achieved by CAP arrays in the BC. The observed electric field at user $k$ can be expressed:
\begin{align}\label{CAP_BC_Basic_Model}
\mathsf{y}_k(\mathbf{r})
&=\int_{\mathcal{A}_{\mathsf{S}}}-{\rm{j}}\eta_0\frac{2\pi}{\lambda}\mathsf{h}({\mathbf{r}},{\mathbf{s}}){\mathsf{j}}({\mathbf{s}})
{\rm{d}}{\mathbf{s}}+{\mathsf{n}}_k(\mathbf{r}),
\end{align}
for ${\mathbf{r}}\in{\mathcal{A}_{\mathsf{R}}^{k}}$, where ${\mathsf{j}}({\mathbf{s}})\in{\mathbbmss{C}}$ denotes the source current used to convey data information, and $\int_{\mathcal{A}_{\mathsf{S}}}\lvert{\mathsf{j}}({\mathbf{s}})\rvert^2{\rm{d}}{{\mathbf{s}}}$ represents the radiating power constraint. The term ${\mathsf{n}}_k(\mathbf{r})$ accounts for the thermal noise and is modeled as a zero-mean complex Gaussian process with ${\mathbbmss{E}}\{{\mathsf{n}}_k(\mathbf{r}){\mathsf{n}}_k^{\mathsf{H}}(\mathbf{r}')\}=\overline{\sigma}_k^2\delta(\mathbf{r}-\mathbf{r}')$. For brevity, we assume that $\overline{\sigma}_k^2=\overline{\sigma}^2$ for $k=1,2$. The source current is constructed as follows:
\begin{align}
{\mathsf{j}}({\mathbf{s}})=\sum\nolimits_{k=1}^{2}s_k {\mathsf{j}}_k(\mathbf{s}),
\end{align}
where $s_k\in{\mathbbmss{C}}$ denotes the normalized coded data symbol for user $k$, and ${\mathsf{j}}_k(\mathbf{s})\in{\mathbbmss{C}}$ is the associated source current.

Let ${\mathsf{v}}_k(\mathbf{r})\in{\mathbbmss{C}}$ denote the detector used by user $k$ to recover $s_k$ from \eqref{CAP_BC_Basic_Model}. The resulting signal-to-noise-plus-interference ratio (SINR) can be written as follows:
\begin{equation}\label{Multiple_User_SINR_Step1}
\begin{split}
&
\frac{\lvert\int_{{\mathcal{A}_{\mathsf{R}}^{k}}}{\mathsf{v}}_k^{\mathsf{H}}(\mathbf{r}){\mathsf{h}}_{k}(\mathbf{r}){\rm{d}}{\mathbf{r}}\rvert^2}
{\frac{\overline{\sigma}^2}{(\eta_0\frac{2\pi}{\lambda})^2}\int_{{\mathcal{A}_{\mathsf{R}}^{k}}}\lvert{\mathsf{v}}_k(\mathbf{r})\rvert^2{\rm{d}}{\mathbf{r}}+\sum_{k'\ne k}\lvert\int_{{\mathcal{A}_{\mathsf{R}}^{k}}}{\mathsf{v}}_k^{\mathsf{H}}(\mathbf{r})
{\mathsf{h}}_{k'}(\mathbf{r}){\rm{d}}{\mathbf{r}}\rvert^2}\\
&\approx \frac{\lvert\int_{{\mathcal{A}_{\mathsf{R}}^{k}}}{\mathsf{v}}_k(\mathbf{r}){\rm{d}}{\mathbf{r}}\rvert^2
\lvert{\mathsf{h}}_{k}(\mathbf{r}_k)\rvert^2}
{\frac{\overline{\sigma}^2}{(\eta_0\frac{2\pi}{\lambda})^2}\int_{{\mathcal{A}_{\mathsf{R}}^{k}}}\lvert{\mathsf{v}}_k(\mathbf{r})\rvert^2{\rm{d}}{\mathbf{r}}+\sum_{k'\ne k}\lvert\int_{{\mathcal{A}_{\mathsf{R}}^{k}}}{\mathsf{v}}_k(\mathbf{r})
{\rm{d}}{\mathbf{r}}\rvert^2\lvert{\mathsf{h}}_{k'}(\mathbf{r}_k)\rvert^2},
\end{split}
\end{equation}
where ${\mathsf{h}}_{k'}(\mathbf{r})\triangleq\int_{{\mathcal{A}}_{\mathsf{S}}}{\mathsf{h}}(\mathbf{r},{\mathbf{s}}){\mathsf{j}}_{k'}({\mathbf{s}}){\rm{d}}{\mathbf{s}}$, and the last approximation is due to the fact that $\lvert\mathcal{A}_{\mathsf{S}}\rvert\gg\lvert\mathcal{A}_{\mathsf{R}}^{k}\rvert=\lvert\mathcal{A}_{\mathsf{R}}\rvert$. By continuously using this property, we obtain
\begin{subequations}
\begin{align}
&\left\lvert\int_{\mathcal{A}_{\mathsf{R}}^{k}}{\mathsf{v}}_k(\mathbf{r}){\rm{d}}{\mathbf{r}}\right\rvert^2\approx \lvert{\mathsf{v}}_k(\mathbf{r}_k)\rvert^2\lvert\mathcal{A}_{\mathsf{R}}\rvert^2,\\
&\int_{\mathcal{A}_{\mathsf{R}}^{k}}\lvert{\mathsf{v}}_k(\mathbf{r})\rvert^2{\rm{d}}{\mathbf{r}}\approx \lvert{\mathsf{v}}_k(\mathbf{r}_k)\rvert^2\lvert\mathcal{A}_{\mathsf{R}}\rvert,
\end{align}
\end{subequations}
which simplifies \eqref{Multiple_User_SINR_Step1} as follows:
\begin{equation}\label{Multiple_User_SINR_Step2}
\begin{split}
{\mathsf{SINR}}_k&=\frac{\lvert{\mathsf{h}}_{k}(\mathbf{r}_k)\rvert^2}
{\frac{\overline{\sigma}^2}{(\eta_0\frac{2\pi}{\lambda})^2\lvert\mathcal{A}_{\mathsf{R}}\rvert}+\sum_{k'\ne k}\lvert{\mathsf{h}}_{k'}(\mathbf{r}_k)\rvert^2}\\
&=\frac{\left\lvert\int_{{\mathcal{A}}_{\mathsf{S}}}{\mathsf{h}}(\mathbf{r}_k,{\mathbf{s}}){\mathsf{j}}_{k}({\mathbf{s}}){\rm{d}}{\mathbf{s}}\right\rvert^2}
{\frac{\overline{\sigma}^2}{(\eta_0\frac{2\pi}{\lambda})^2\lvert\mathcal{A}_{\mathsf{R}}\rvert}+\sum_{k'\ne k}\left\lvert\int_{{\mathcal{A}}_{\mathsf{S}}}{\mathsf{h}}(\mathbf{r}_k,{\mathbf{s}}){\mathsf{j}}_{k'}({\mathbf{s}}){\rm{d}}{\mathbf{s}}\right\rvert^2}.
\end{split}
\end{equation}
The sum-rate is thus given by
\begin{equation}
{\mathsf{R}}_{\mathsf{sum}}=\sum\nolimits_{k=1}^{2}\log_2(1+{\mathsf{SINR}}_k).
\end{equation}

Given a downlink Gaussian channel, it is widely established that the capacity region is achieved by DPC. For the DPC order $\varepsilon(2)\rightarrow\varepsilon(1)$ with $\{\varepsilon(k)\}_{k=1}^{2}=\{k\}_{k=1}^{2}$, the rate of user $\varepsilon(k)$ can be written as follows:
\begin{equation}
{\mathsf{R}}_{\varepsilon(k)}=\log_2\left(1+\frac{\lvert{\mathsf{h}}_{\varepsilon(k)}(\mathbf{r}_{\varepsilon(k)})\rvert^2}
{\frac{\overline{\sigma}^2}{(\eta_0\frac{2\pi}{\lambda})^2\lvert\mathcal{A}_{\mathsf{R}}\rvert}+\sum_{k'< k}\lvert{\mathsf{h}}_{\varepsilon(k')}(\mathbf{r}_{\varepsilon(k)})\rvert^2}\right).
\end{equation}
Consider the DPC order $2\rightarrow1$, the achieved rate for user $1$ and user $2$ can be written as follows:
{\setlength\abovedisplayskip{2pt}
\setlength\belowdisplayskip{2pt}
\begin{subequations}
\begin{align}
&{\mathsf{R}}_1=\log_2\left(1+\frac{{\left\lvert\int_{{\mathcal{A}}_{\mathsf{S}}}{\mathsf{h}}(\mathbf{r}_1,{\mathbf{s}}){\mathsf{j}}_{1}({\mathbf{s}}){\rm{d}}{\mathbf{s}}\right\rvert^2}
}{\frac{\overline{\sigma}^2}{(\eta_0\frac{2\pi}{\lambda})^2\lvert\mathcal{A}_{\mathsf{R}}\rvert}}\right),\label{Duality_Rate1}\\
&{\mathsf{R}}_2=\log_2\left(1+\frac{\left\lvert\int_{{\mathcal{A}}_{\mathsf{S}}}{\mathsf{h}}(\mathbf{r}_2,{\mathbf{s}}){\mathsf{j}}_{2}({\mathbf{s}}){\rm{d}}{\mathbf{s}}\right\rvert^2}
{\frac{\overline{\sigma}^2}{(\eta_0\frac{2\pi}{\lambda})^2\lvert\mathcal{A}_{\mathsf{R}}\rvert}+
\left\lvert\int_{{\mathcal{A}}_{\mathsf{S}}}{\mathsf{h}}(\mathbf{r}_2,{\mathbf{s}}){\mathsf{j}}_{1}({\mathbf{s}}){\rm{d}}{\mathbf{s}}\right\rvert^2}\right).\label{Duality_Rate2}
\end{align}
\end{subequations}
}For clarity, we define $\frac{(\eta_0\frac{2\pi}{\lambda})^2\lvert\mathcal{A}_{\mathsf{R}}\rvert}{\overline{\sigma}^2}\sum_{k=1}^{2}
\int_{{\mathcal{A}}_{\mathsf{S}}}\lvert{\mathsf{j}}_k(\mathbf{s})\rvert^2\triangleq\frac{P_{\mathsf{b}}}{\sigma^2}$ as the total transmit SNR.

Determining the optimal $\{\varepsilon(k)\}$ and the associated $\{{\mathsf{j}}_{\varepsilon(k)}({\mathbf{s}})\}$ is a challenging task, and the MAC-BC duality can be employed to characterize the BC capacity region. Specifically, let ${\mathsf{p}}_k$ denote the transmit SNR for user $k$ in the dual MAC described in \eqref{CAP_MAC_Basic_Model}, the sum-rate capacity of the BC can be written as follows:
\begin{equation}\label{BC_Sum_Capacity1_CAP}
{\mathsf{C}}_{\mathsf{BC}}=\max\nolimits_{{\mathsf{p}}_k\geq0,{\mathsf{p}}_1+{\mathsf{p}}_2\leq P_{\mathsf{b}}/\sigma^2}{\mathsf{C}}_{\mathsf{MAC}}^{\mathsf{d}},
\end{equation}
where
\begin{equation}\label{BC_Sum_Capacity2_CAP}
\begin{split}
{\mathsf{C}}_{\mathsf{MAC}}^{\mathsf{d}}=
\log_2\left(1+{{\mathsf{p}}_1\overline{\mathsf{a}}_1}+{{\mathsf{p}}_2\overline{\mathsf{a}}_2}+{{\mathsf{p}}_1{\mathsf{p}}_2}
\overline{\mathsf{a}}_1\overline{\mathsf{a}}_2(1-\overline{\rho}_{\mathsf{u}}^2)\right)
\end{split}
\end{equation}
represents the sum-rate capacity of the dual MAC. Note that the dual-MAC sum-rate capacity is maximized when ${\mathsf{p}}_1+{\mathsf{p}}_2=\frac{ P_{\mathsf{b}}}{\sigma^2}$. Furthermore, given $({\mathsf{p}}_1,{\mathsf{p}}_2)$, we can utilize the MAC-BC duality to determine the corresponding source currents $\{{\mathsf{j}}_k(\mathbf{s})\}_{k=1}^{2}$ that achieve the BC sum-rate capacity. More specifically, the source currents corresponding to the DPC encoding order $2\rightarrow1$ are given as follows:
\begin{subequations}
\begin{align}
{\mathsf{j}}_{1}({\mathbf{s}})&=\sqrt{{\mathsf{p}}_1}\frac{{\mathsf{h}}^{\mathsf{H}}(\mathbf{r}_1,{\mathbf{s}})-\frac{{\mathsf{p}}_2}{1+{\mathsf{p}}_2\overline{\mathsf{a}}_2}\rho
{\mathsf{h}}^{\mathsf{H}}(\mathbf{r}_2,{\mathbf{s}})}{\sqrt{\frac{(\eta_0\frac{2\pi}{\lambda})^2\lvert\mathcal{A}_{\mathsf{R}}\rvert}{\overline{\sigma}^2}}
\sqrt{\overline{\mathsf{a}}_1-\frac{{\mathsf{p}}_2}{1+{\mathsf{p}}_2\overline{\mathsf{a}}_2}|\rho|^2}},\label{Duality_Current1}\\
{\mathsf{j}}_{2}({\mathbf{s}})&=\sqrt{{\mathsf{p}}_2}\frac{\sqrt{1+\frac{(\eta_0\frac{2\pi}{\lambda})^2\lvert\mathcal{A}_{\mathsf{R}}\rvert}{\overline{\sigma}^2}
|\int_{{\mathcal{A}}_{\mathsf{S}}}{\mathsf{h}}(\mathbf{r}_2,{\mathbf{s}}){\mathsf{j}}_{1}({\mathbf{s}})
{\rm{d}}{\mathbf{s}}|^2}}
{\sqrt{\frac{(\eta_0\frac{2\pi}{\lambda})^2\lvert\mathcal{A}_{\mathsf{R}}\rvert}{\overline{\sigma}^2}}\sqrt{\overline{\mathsf{a}}_2}}{\mathsf{h}}^{\mathsf{H}}(\mathbf{r}_2,{\mathbf{s}}),
\label{Duality_Current2}
\end{align}
\end{subequations}
where $\rho=\int_{{\mathcal{A}}_{\mathsf{S}}}{\mathsf{h}}^{\mathsf{H}}(\mathbf{r}_1,{\mathbf{s}}){\mathsf{h}}(\mathbf{r}_2,{\mathbf{s}}){\rm{d}}{\mathbf{s}}$. It the sequel, we proceed to verify whether $\{{\mathsf{j}}_k(\mathbf{s})\}_{k=1}^{2}$ satisfy the power budget. For clarity, we define $\hat{\mathsf{j}}_{k}({\mathbf{s}})\triangleq\sqrt{\frac{(\eta_0\frac{2\pi}{\lambda})^2\lvert\mathcal{A}_{\mathsf{R}}\rvert}{\overline{\sigma}^2}}{\mathsf{j}}_{k}({\mathbf{s}})$ for $k=1,2$, which yields
\begin{align}\label{Two_User_BC_Power_Constraint}
\frac{(\eta_0\frac{2\pi}{\lambda})^2\lvert\mathcal{A}_{\mathsf{R}}\rvert}{\overline{\sigma}^2}\sum_{k=1}^{2}
\int_{{\mathcal{A}}_{\mathsf{S}}}\lvert{\mathsf{j}}_k(\mathbf{s})\rvert^2=\sum_{k=1}^{2}
\int_{{\mathcal{A}}_{\mathsf{S}}}\lvert\hat{\mathsf{j}}_k(\mathbf{s})\rvert^2.
\end{align}
Equation \eqref{Two_User_BC_Power_Constraint} can be rewritten as follows:
\begin{align}
\eqref{Two_User_BC_Power_Constraint}&={\mathsf{p}}_2+{\mathsf{p}}_2\left\lvert\int_{{\mathcal{A}}_{\mathsf{S}}}{\mathsf{h}}(\mathbf{r}_2,{\mathbf{s}})\hat{\mathsf{j}}_{1}({\mathbf{s}})
{\rm{d}}{\mathbf{s}}\right\rvert^2+\int_{{\mathcal{A}}_{\mathsf{S}}}\lvert\hat{\mathsf{j}}_{1}({\mathbf{s}})\rvert^2{\rm{d}}{\mathbf{s}},
\end{align}
which, together with the fact that $\int_{{\mathcal{A}}_{\mathsf{S}}}\lvert\hat{\mathsf{j}}_{1}({\mathbf{s}})\rvert^2{\rm{d}}{\mathbf{s}}=
\int_{{\mathcal{A}}_{\mathsf{S}}}\int_{{\mathcal{A}}_{\mathsf{S}}}\hat{\mathsf{j}}_{1}({\mathbf{s}})\delta({\mathbf{s}},{\mathbf{s}}')
\hat{\mathsf{j}}_{1}^{\mathsf{H}}({\mathbf{s}}'){\rm{d}}{\mathbf{s}}{\rm{d}}{\mathbf{s}}'$ and $\lvert\int_{{\mathcal{A}}_{\mathsf{S}}}{\mathsf{h}}(\mathbf{r}_2,{\mathbf{s}})\hat{\mathsf{j}}_{1}({\mathbf{s}})
{\rm{d}}{\mathbf{s}}\rvert^2=\int_{{\mathcal{A}}_{\mathsf{S}}}\int_{{\mathcal{A}}_{\mathsf{S}}}{\mathsf{h}}(\mathbf{r}_2,{\mathbf{s}})\hat{\mathsf{j}}_{1}({\mathbf{s}})
{\mathsf{h}}^{\mathsf{H}}(\mathbf{r}_2,{\mathbf{s}}')\hat{\mathsf{j}}^{\mathsf{H}}_{1}({\mathbf{s}}'){\rm{d}}{\mathbf{s}}{\rm{d}}{\mathbf{s}}'$, yields 
\begin{equation}
\begin{split}
\eqref{Two_User_BC_Power_Constraint}&={\mathsf{p}}_2+\int_{{\mathcal{A}}_{\mathsf{S}}}\int_{{\mathcal{A}}_{\mathsf{S}}}\hat{\mathsf{j}}_{1}({\mathbf{s}})
({\mathsf{p}}_2{\mathsf{h}}(\mathbf{r}_2,{\mathbf{s}}){\mathsf{h}}^{\mathsf{H}}(\mathbf{r}_2,{\mathbf{s}}')\\
&+\delta({\mathbf{s}},{\mathbf{s}}'))
\hat{\mathsf{j}}^{\mathsf{H}}_{1}({\mathbf{s}}'){\rm{d}}{\mathbf{s}}{\rm{d}}{\mathbf{s}}'.
\end{split}
\end{equation}
By noting that 
\begin{equation}
\begin{split}
&\int_{{\mathcal{A}}_{\mathsf{S}}}\left({\mathsf{h}}^{\mathsf{H}}(\mathbf{r}_1,{\mathbf{s}})-\frac{{\mathsf{p}}_2}{1+{\mathsf{p}}_2\overline{\mathsf{a}}_2}\rho
{\mathsf{h}}^{\mathsf{H}}(\mathbf{r}_2,{\mathbf{s}})\right)\\
&\times({\mathsf{p}}_2{\mathsf{h}}(\mathbf{r}_2,{\mathbf{s}}){\mathsf{h}}^{\mathsf{H}}(\mathbf{r}_2,{\mathbf{s}}')
+\delta({\mathbf{s}},{\mathbf{s}}'))
{\rm{d}}{\mathbf{s}}\\
&={\mathsf{h}}^{\mathsf{H}}(\mathbf{r}_2,{\mathbf{s}}')\left({\mathsf{p}}_2\rho-\frac{{\mathsf{p}}_2\rho{\mathsf{p}}_2\overline{\mathsf{a}}_2}{1+{\mathsf{p}}_2\overline{\mathsf{a}}_2}
-\frac{{\mathsf{p}}_2\rho}{1+{\mathsf{p}}_2\overline{\mathsf{a}}_2}\right)+{\mathsf{h}}^{\mathsf{H}}(\mathbf{r}_1,{\mathbf{s}}')\\
&={\mathsf{h}}^{\mathsf{H}}(\mathbf{r}_1,{\mathbf{s}}'),
\end{split}
\end{equation}
we obtain
\begin{align}
\eqref{Two_User_BC_Power_Constraint}&={\mathsf{p}}_2+{\mathsf{p}}_1\frac{\int_{{\mathcal{A}}_{\mathsf{S}}}{\mathsf{h}}^{\mathsf{H}}(\mathbf{r}_1,{\mathbf{s}}')
({\mathsf{h}}(\mathbf{r}_1,{\mathbf{s}}')-\frac{{\mathsf{p}}_2\rho^{\mathsf{H}}
{\mathsf{h}}(\mathbf{r}_2,{\mathbf{s}}')}{1+{\mathsf{p}}_2\overline{\mathsf{a}}_2}){\rm{d}}{\mathbf{s}}'}
{\overline{\mathsf{a}}_1-\frac{{\mathsf{p}}_2}{1+{\mathsf{p}}_2\overline{\mathsf{a}}_2}|\rho|^2}\nonumber\\
&={\mathsf{p}}_2+{\mathsf{p}}_1\frac{\overline{\mathsf{a}}_1-\frac{{\mathsf{p}}_2}{1+{\mathsf{p}}_2\overline{\mathsf{a}}_2}|\rho|^2}{\overline{\mathsf{a}}_1-\frac{{\mathsf{p}}_2}{1+{\mathsf{p}}_2\overline{\mathsf{a}}_2}|\rho|^2}
={\mathsf{p}}_2+{\mathsf{p}}_1=\frac{P_{\mathsf{b}}}{\sigma^2},
\end{align}
which satisfies the power budget. Substituting \eqref{Duality_Current1} and \eqref{Duality_Current2} into \eqref{Duality_Rate1} and \eqref{Duality_Rate2} gives
\begin{subequations}
\begin{align}
{\mathsf{R}}_1&=\log_2\left(1+{\mathsf{p}}_1{\overline{\mathsf{a}}}_1\left(1-\overline{\rho}_{\mathsf{u}}^2\left(1-\frac{1}{1+{\mathsf{p}}_2{\overline{\mathsf{a}}}_2}\right)\right)\right)\\
&=\log_2\left(\frac{1+{\mathsf{p}}_1\overline{\mathsf{a}}_1+{\mathsf{p}}_2\overline{\mathsf{a}}_2+{\mathsf{p}}_1{\mathsf{p}}_2
\overline{\mathsf{a}}_1\overline{\mathsf{a}}_2(1-\overline{\rho}_{\mathsf{u}}^2)}
{1+{\mathsf{p}}_2\overline{\mathsf{a}}_2}\right)
\end{align}
\end{subequations}
and
\begin{equation}
{\mathsf{R}}_2=\log_2\left(1+{\mathsf{p}}_2{\overline{\mathsf{a}}}_2\right),
\end{equation}
respectively, which are akin to the results in \eqref{Two_User_MAC_Corner_Point_Rate}. The above results thus verify the MAC-BC duality under the DPC encoding order $2\rightarrow1$.

Following similar steps, we can also obtain the capacity-achieving currents and the achieved rates under the DPC order $1\rightarrow2$. Specifically, we have 
\begin{subequations}
\begin{align}
{\mathsf{j}}_{1}({\mathbf{s}})&=\sqrt{{\mathsf{p}}_1}\frac{\sqrt{1+\frac{(\eta_0\frac{2\pi}{\lambda})^2\lvert\mathcal{A}_{\mathsf{R}}\rvert}{\overline{\sigma}^2}
|\int_{{\mathcal{A}}_{\mathsf{S}}}{\mathsf{h}}(\mathbf{r}_1,{\mathbf{s}}){\mathsf{j}}_{2}({\mathbf{s}})
{\rm{d}}{\mathbf{s}}|^2}}
{\sqrt{\frac{(\eta_0\frac{2\pi}{\lambda})^2\lvert\mathcal{A}_{\mathsf{R}}\rvert}{\overline{\sigma}^2}}\sqrt{\overline{\mathsf{a}}_1}}
{\mathsf{h}}^{\mathsf{H}}(\mathbf{r}_1,{\mathbf{s}}),\label{Duality_Current1_Another}\\
{\mathsf{j}}_{2}({\mathbf{s}})&=\sqrt{{\mathsf{p}}_2}\frac{{\mathsf{h}}^{\mathsf{H}}(\mathbf{r}_2,{\mathbf{s}})-\frac{{\mathsf{p}}_1}{1+{\mathsf{p}}_1\overline{\mathsf{a}}_1}\rho'
{\mathsf{h}}^{\mathsf{H}}(\mathbf{r}_1,{\mathbf{s}})}{\sqrt{\frac{(\eta_0\frac{2\pi}{\lambda})^2\lvert\mathcal{A}_{\mathsf{R}}\rvert}{\overline{\sigma}^2}}
\sqrt{\overline{\mathsf{a}}_2-\frac{{\mathsf{p}}_1}{1+{\mathsf{p}}_1\overline{\mathsf{a}}_1}|\rho|^2}},\label{Duality_Current2_Another}
\end{align}
\end{subequations}
where $\rho'=\int_{{\mathcal{A}}_{\mathsf{S}}}{\mathsf{h}}^{\mathsf{H}}(\mathbf{r}_2,{\mathbf{s}}){\mathsf{h}}(\mathbf{r}_1,{\mathbf{s}}){\rm{d}}{\mathbf{s}}$. The corresponding rates are given by
\begin{subequations}
\begin{align}
{\mathsf{R}}_1&=\log_2\left(1+{\mathsf{p}}_1{\overline{\mathsf{a}}}_1\right),\\
{\mathsf{R}}_2&=\log_2\left(1+{\mathsf{p}}_2{\overline{\mathsf{a}}}_2\left(1-\overline{\rho}_{\mathsf{u}}^2\left(1-\frac{1}{1+{\mathsf{p}}_1{\overline{\mathsf{a}}}_1}\right)\right)\right),
\end{align}
\end{subequations}
which are essentially the rates achieved in the dual MAC under SIC order $2\rightarrow1$. 

\begin{figure}[!t]
 \centering
\includegraphics[width=0.48\textwidth]{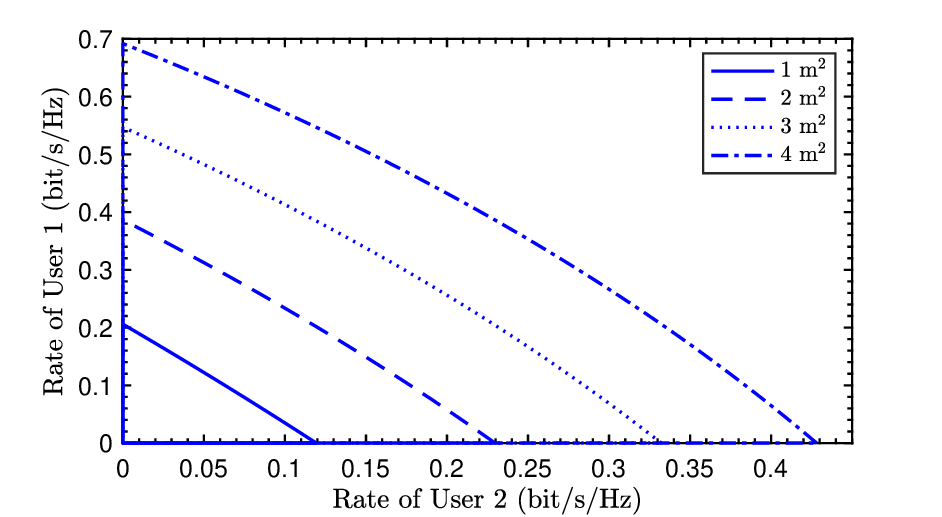}
\caption{BC capacity regions of CAP arrays under different aperture sizes. $\theta_1=\theta_2=\frac{\pi}{6}$, $\phi_1=\phi_2=\frac{\pi}{3}$, $r_1=10$ m, $r_2=20$ m, $\lambda = 0.0107$ m, $d=\frac{\lambda}{2}$, $A=\frac{\lambda^2}{4\pi}$, and $L_x=L_z$, and $\frac{P_{\mathsf{b}}}{\sigma^2}=30$ dB.}
\label{Figure: BC capacity}
\end{figure}

In {\figurename} {\ref{Figure: BC capacity}}, the BC capacity regions achieved by the CAP array for various values of $\lvert{\mathcal{A}}_{\mathsf{S}}\rvert$ are illustrated. It is observed that the capacity region gradually extends as $\lvert{\mathcal{A}}_{\mathsf{S}}\rvert$ increases.

We note that the expressions for the sum-rate capacity and per-user rate are analogous to those for SPD arrays. Consequently, the insights derived from channel capacity limits in SPD array-based BC readily extend to CAP arrays. Therefore, we will not delve further into this aspect here. It is also worth noting that the capacity region of a BC is closely related to its dual MAC. Since the capacity region of a CAP MAC is only available for the two-user case, we can only provide the channel capacity of a CAP BC for $K=2$. The channel capacity for $K>2$ remains an open problem, necessitating further research endeavors. While DPC is known to be capacity-achieving, its implementation is complex. A comprehensive exploration of the design of $\{{\mathsf{j}}_{k}(\mathbf{s})\}$ for CAP arrays is provided in Section \ref{sec4}.
\subsection{Summary and Discussions}
The above arguments suggest that the near-field information-theoretic capacity limits of CAP arrays closely resemble their SPD counterparts. From this careful analysis, several key conclusions emerge.
\begin{enumerate}
  \item Utilizing the far-field channel model to describe the asymptotic behavior of channel capacity with a large aperture size may lead to violations of the fundamental law of energy conservation. This discrepancy arises from the expansive nature of the near-field region as the array aperture size approaches infinity. Therefore, it is essential to adopt channel models based on spherical-wave propagation to accurately depict the physical reality. 
  \item Both SPD and CAP arrays exhibit no need to account for the influence of the reactive near-field region in asymptotic analyses when the number of antennas tends towards infinity.
  \item Both far-field ADMA and near-field RDMA effectively manage interference among users located in different spatial locations. However, ADMA faces significant challenges when users are aligned in the same direction, resulting in pronounced interference that is difficult to mitigate. Conversely, near-field RDMA, benefiting from additional range dimensions introduced by spherical-wave propagation, continues to effectively handle interference even in such challenging scenarios. This suggests that the near-field effect can provide more flexibility for MA in terms of interference management.
\end{enumerate}
\section{Beamforming Design for NFC-NGMA}
\label{sec4}
Beamforming is an important enabling technique for multi-antenna multi-user systems. In this section, we will introduce the beamforming design for NFC-NGMA in the spatial domain and wavenumber domain, respectively.

\subsection{Spatial-domain Beamforming}
SDMA is a crucial multiple-access technique for multi-antenna systems. It enables multi-user multiplexing in the spatial domain through precoding techniques. Therefore, SDMA can support multiple users in a single time-frequency resource block. Precoding is critical to mitigate IUI in the spatial domain for SDMA, which can be either non-linear or linear. Some non-linear precoding techniques have been proven optimal to attain the capacity region of SDMA, such as the aforementioned DPC \cite{costa1983writing}. However, the extremely high complexity of non-linear precoding makes it difficult to implement in practice. In this case, the more practical linear precoding, which is also frequently named beamforming, has attracted more attention. For semi-orthogonal multi-user channels, beamforming is able to achieve a capacity region comparable to that of DPC. Let us consider a multi-user-MISO system where $K$ users are served by an $M$-antenna SPD array. In a rich scattering environment, the multi-user channel matrix can be expressed as 
\begin{equation}
  \mathbf{H} = \tilde{\mathbf{H}} \mathsf{diag}(\boldsymbol{\beta})^{1/2},
\end{equation}    
where $\tilde{\mathbf{H}} \in \mathbbmss{C}^{M \times K}$ and $\boldsymbol{\beta} \in \mathbbmss{C}^{K \times 1}$ account for the small-scale and large-scale fading coefficients, respectively. The $k$th column of $\mathbf{H}$ is the channel vector for user $k$. It has been widely shown that as the number of antennas scales up to be much larger than the number of users, i.e., $M \gg K$, the multi-user channels are asymptotically orthogonal \cite{marzetta2010noncooperative, rusek2012scaling}:
\begin{equation} \label{zw_ortho_mMIMO}
  \frac{1}{M} \mathbf{H}^{\mathsf{H}} \mathbf{H} \rightarrow \mathsf{diag}(\boldsymbol{\beta}).
\end{equation}
The above result implies that IUI can be automatically eliminated through matched filtering. This was the initial motivation of massive MIMO \cite{marzetta2010noncooperative}. However, for typical NFC scenarios over high-frequency bands (e.g. mmWave and THz bands), the assumption of rich scattering no longer holds in general, in which case the communication channel is mainly contributed by a small number of resolvable paths \cite{rangan2014millimeter}. Furthermore, according to the measurements in \cite{rangan2014millimeter}, the path loss of the NLoS path is more than 10 dB larger than that of the LoS path. Therefore, the conclusions in \eqref{zw_ortho_mMIMO} need to be revisited in NFC systems when there is no significant scattering or even only LoS paths.

\subsubsection{Near-Field Beamfocusing for RDMA}\label{Section: Near-Field Beamfocusing for RDMA}
By taking into account only the LoS path, the near-field channel vector for user $k$ (i.e., the $k$th column of $\mathbf{H}$) can be expressed as, c.f. \eqref{SPD_UPA_NFC_Model_USW}
\begin{equation} \label{zw_NF_MISO_LOS}
  \mathbf{h}_k = \beta_k \left[ {\rm{e}}^{-{\rm{j}} \frac{2\pi}{\lambda} \|\mathbf{r}_k - \mathbf{s}_1 \| }, ...., {\rm{e}}^{-{\rm{j}} \frac{2\pi}{\lambda} \|\mathbf{r}_k - \mathbf{s}_M \| }   \right]^{\mathsf{T}},
\end{equation} 
where $\mathbf{r}_k$ and $\mathbf{s}_m$ denote the coordinate vectors of the $k$th user and the $m$th antenna at the SPD array and $\beta_k$ denote the uniform array gain. Here, we assume that the distance between the user and the SPD array is larger than the uniform-power distance for brevity. More specifically, the coordinate vector of user $k$ can be expressed as 
\begin{equation}
  \mathbf{r}_k = r_k \mathbf{k}(\theta_k, \phi_k),
\end{equation} 
where $r_k$ denotes the distance between user $k$ and the center of the SPD array and $\mathbf{k}(\theta_k, \phi_k) = [\cos \theta_k \sin \phi_k, \sin \theta_k \sin \phi_k, \cos \phi_k]^{\mathsf{T}}$ is the direction vector with $\theta_k$ and $\phi_k$ being the azimuth and elevation directions, respectively. 
For comparison, the channel vector based on the far-field channel model is given by 
\begin{equation}
  \mathbf{h}_k^{\text{far}} = \beta_k \left[ {\rm{e}}^{{\rm{j}} \frac{2\pi}{\lambda} \mathbf{k}^{\mathsf{T}}(\theta_k, \phi_k) \mathbf{s}_1 }, ...., {\rm{e}}^{-{\rm{j}} \frac{2\pi}{\lambda} \mathbf{k}^{\mathsf{T}}(\theta_k, \phi_k) \mathbf{s}_M }   \right]^{\mathsf{T}},
\end{equation} 
where $\theta_k$ and $\phi_k$ denote the azimuth and elevation directions of user $k$ with respect to the SPD array. It can be observed that the near-field LoS-channel vector depends on the location of the user. Hence, when the matched filtering is used for beamforming, the beam that focuses on the location of this user can be generated, which is referred to as \emph{beamfocusing}. This is fundamentally different from utilizing matched filtering in the far field, where the channel vector is merely related to the directions $\theta_k$ and $\phi_k$. Hence, the far-field matched filtering can only steer a beam toward a specific direction. In the following, we will unveil how the beamfocusing enables a new paradigm of SDMA, namely RDMA.

\begin{figure}[!t]
  \centering
 \setlength{\abovecaptionskip}{0pt}
 \includegraphics[width=0.48\textwidth]{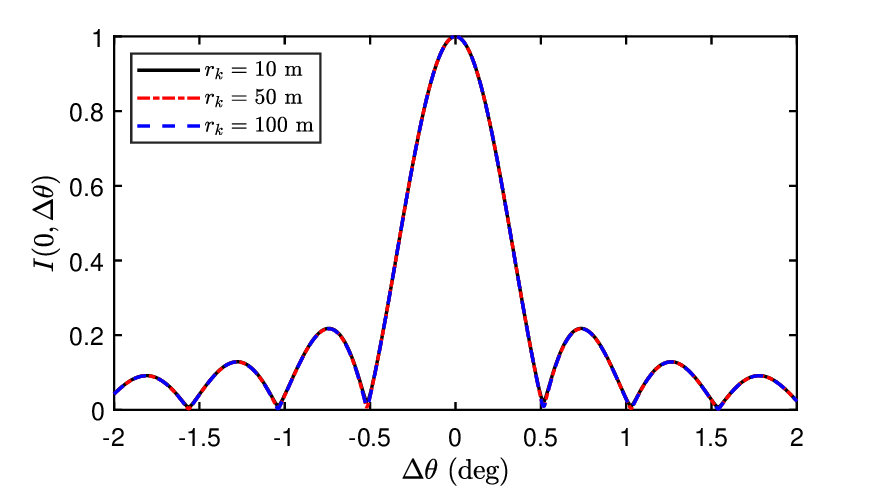}
 \caption{Illustration of interference function $I(0, \Delta \theta)$ in the angular domain with $\theta_k = \pi / 3$.}
 \label{IUI_angle}
 \vspace{-10pt}
\end{figure}

The essence of beamforming in SDMA lies in mitigating IUI while maximizing beamforming gain. Notably, beamfocusing based on matched filtering maximizes this gain. Consequently, our analysis shifts towards quantifying the extent of interference this approach imposes on other users. For simplicity, we assume the SPD array is a ULA, which exhibits the same array response for all elevation directions. Thus, it is safe to set $\phi_k = \frac{\pi}{2}$ and assume that the ULA is deployed on the $x$-axis. In this case, the near-field channel vector for user $k$ is only related to $r_k$ and $\theta_k$. The distance $\|\mathbf{r}_k - \mathbf{s}_m\|$ in \eqref{zw_NF_MISO_LOS} can be simplified as 
\begin{equation}
  \|\mathbf{r}_k - \mathbf{s}_m\| = \sqrt{ r_k^2 + \delta_m^2 d^2 - 2 r_k \delta_m d \cos \theta_k },
\end{equation}
where $d$ denotes the antenna spacing and $\delta_m = m - 1 - \frac{M-1}{2}$. Let us now consider another user $i$ with distance $r_i = r_k + \Delta r$ and direction $\theta_i = \theta_k + \Delta \theta$. Then, the interference caused by beamfocusing for user $k$ to user $i$ can be reflected by the cross-correlation between their channel vectors, which is given by 
\begin{align}
  &I(\Delta r, \Delta \theta) = \frac{1}{\beta_k \beta_i M} |\mathbf{h}_k^{\mathsf{H}} \mathbf{h}_i| \nonumber \\
  = &\frac{1}{M} \left| \sum_{m=1}^M {\rm{e}}^{-{\rm{j}} \frac{2 \pi}{\lambda} \left( \|\mathbf{r}_i - \mathbf{s}_m\| - \|\mathbf{r}_k - \mathbf{s}_m\| \right) } \right| \nonumber \\
  \overset{(a)}{\approx} &\frac{1}{M} \left| \sum_{m=1}^M {\rm{e}}^{{\rm{j}} \left( a \delta_m  + b \delta_m^2 \right) } \right|
  \approx \left| \int_{-\frac{1}{2}}^{\frac{1}{2}} {\rm{e}}^{{\rm{j}} \left(M a x + M^2 b x^2 \right) } {\rm{d}}x \right| \nonumber \\
  = & \frac{1}{2M} \sqrt{\frac{\pi}{ |b|}} \Bigg| \left( {\rm{e}}^{{\rm{j}} \frac{3\pi}{4}} \sin\left(\frac{a^2}{4b}\right) -  {\rm{e}}^{{\rm{j}} \frac{\pi}{4}} \cos\left(\frac{a^2}{4b}\right)  \right) \nonumber\\
  &\times \left( \mathrm{erf}\left( \frac{{\rm{j}} (M b + a)}{2\sqrt{-{\rm{j}}b}} \right) + \mathrm{erf}\left( \frac{{\rm{j}} (M b - a)}{2\sqrt{-{\rm{j}}b}} \right)\right) \Bigg|,\label{Correlation_Faactor_Approximate_Exp}
\end{align}     
where $a = \frac{2\pi}{\lambda} d \left( \cos (\theta_k + \Delta \theta) - \cos \theta_k \right)$, $b = \frac{2 \pi}{\lambda} d^2  \left( \frac{\sin^2 \theta_k}{2 r_k} - \frac{\sin^2 (\theta_k + \Delta \theta)}{2 (r_k + \Delta r)} \right) $, and $\mathrm{erf}(x) = \frac{2}{\sqrt{\pi}} \int_{0}^{x} {\rm{e}}^{-t^2} dt$ is the error function.   
In step $(a)$ of the above derivation, we exploit the Fresnel approximation \cite{liu2023nearfield}:
\begin{equation}
  \|\mathbf{r}_k - \mathbf{s}_m\| \approx r_k - \delta_m d \cos \theta_k + \frac{\delta_n^2 d^2 \sin^2 \theta_k}{2 r_k},
\end{equation}  
which is accurate for $r_k$ and $r_i$  greater than the uniform power distance. We first exam the interference function $I(\Delta r, \Delta \theta)$ in the angular domain, keeping $\Delta r = 0$ as shown in Fig. \ref{IUI_angle}. It is noted that the distance from the BS to user $k$ barely impacts the interference region. Thus, akin to far-field beamforming, near-field beamfocusing also shows a limited interference region in the angular domain

\begin{figure}[!t]
  \centering
 \setlength{\abovecaptionskip}{0pt}
 \includegraphics[width=0.48\textwidth]{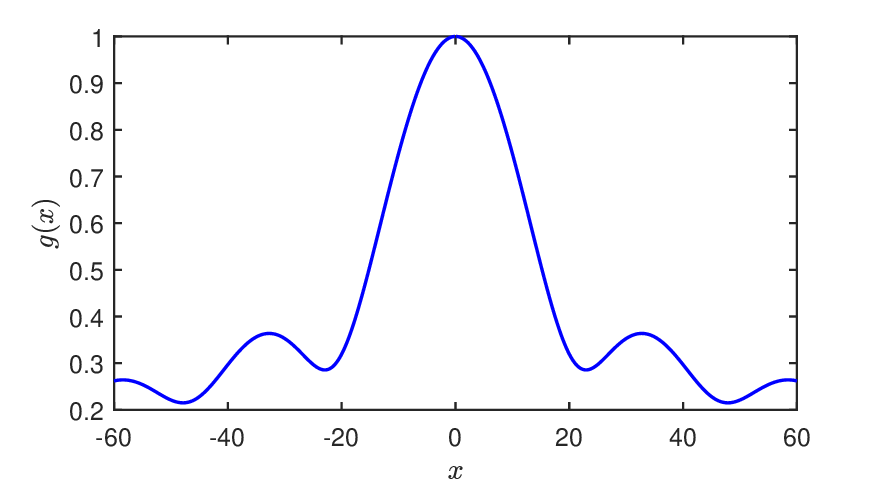}
 \caption{Illustration of function $g(x) = \sqrt{\frac{\pi}{|x|}} \left| \mathrm{erf} \left( \frac{{\rm{e}}^{-{\rm{j}} \frac{\pi}{4} }\sqrt{x}}{2} \right) \right|$.}
 \label{IUI_distance}
 \vspace{-10pt}
\end{figure}

\begin{figure}[!t]
  \centering
 \setlength{\abovecaptionskip}{0pt}
 \includegraphics[width=0.48\textwidth]{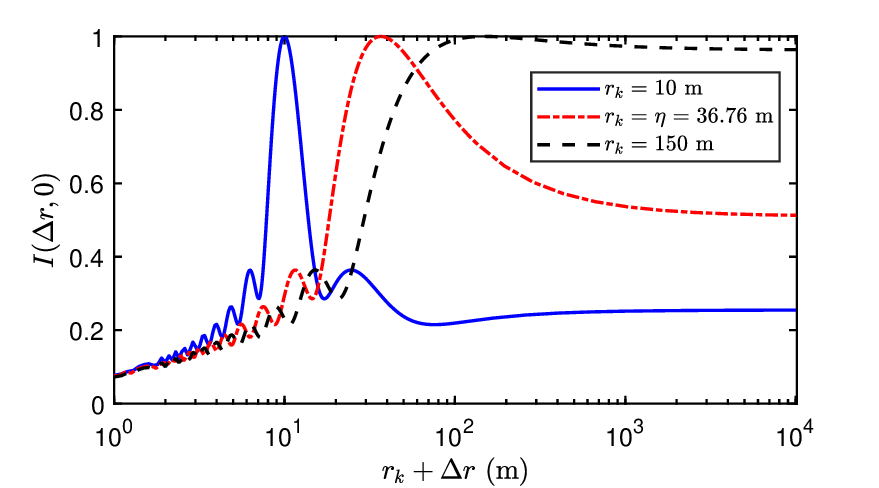}
 \caption{Illustration of interference function $I(\Delta r, 0)$ in the range domain with $\theta_k = \pi / 2$.}
 \label{IUI_distance_2}
 \vspace{-10pt}
\end{figure}

Now, we shift our focus to the interference region of near-field beamfocusing in the range domain. When $\Delta \theta = 0$, we have $a = 0$. In this case, $I(\Delta r, \Delta \theta)$ can be simplified as 
\begin{align}
  I(\Delta r, 0) =  &\sqrt{\frac{\pi}{|M^2 b|}} \Bigg| \mathrm{erf} \left(\frac{{\rm{e}}^{-{\rm{j}} \frac{\pi}{4}} \sqrt{M^2 b}}{2}  \right) \Bigg| \nonumber \\
  = &g( M^2b ), 
\end{align}  
where $g(x) \triangleq \sqrt{\frac{\pi}{|x|}} \left| \mathrm{erf} \left( \frac{{\rm{e}}^{-{\rm{j}} \frac{\pi}{4}} \sqrt{x}}{2} \right) \right| $. The value of function $g(x)$ is illuminated in Fig. \ref{IUI_distance}. According to the numerical results, we have $g(x) \le 0.5$ when $|x| \le 15$. Therefore, the 3-dB interference region, where $I(\Delta r, 0) \le 0.5$, of user $k$ in the range domain can be obtained by 
\begin{equation}
  |M^2 b| \le 15 \quad \Leftrightarrow \quad \left| \frac{1}{r_k} - \frac{1}{r_k + \Delta r} \right| \le \frac{15 \lambda}{\pi M^2 d^2 \sin^2\theta_k}.
\end{equation}   
By defining $\eta = \frac{\pi M^2 d^2 \sin^2\theta_k}{15 \lambda}$, the above formula can be further simplified as 
\begin{align} \label{interference_region}
  \begin{cases}
    - \frac{r_k^2}{\eta + r_k} \le \Delta r \le \frac{r_k^2}{\eta - r_k}, & \text{if } r_k < \eta, \\
    - \frac{r_k^2}{\eta + r_k} \le \Delta r \le + \infty, & \text{if } r_k \ge \eta.
  \end{cases}
\end{align}
\vspace{-5pt}
\begin{remark}
    \emph{(RDMA)} In \eqref{interference_region}, we can observe that when the distance of user $k$ is smaller than a threshold $\eta$, i.e., $r_k < \eta$, the matched-filtering-based beamfocusing for user $k$ will generate a finite 3-dB interference region in the range domain, as shown in Fig. \ref{IUI_distance_2}. Hence, multiplexing signals of multiple users located in similar directions but at different distances becomes possible, which is referred to as RDMA. However, as the distance $r_k$ becomes larger than $\eta$, the 3-dB interference region becomes infinite. In other words, the signal for user $k$ will interfere with all users in similar directions, resulting in the low effectiveness of SDMA.        
\end{remark}
\vspace{-5pt}
\vspace{-5pt}
\begin{remark}
    \emph{(RDMA Region)} We can also observe from \eqref{interference_region} that the threshold $\eta$ has a different value as the Rayleigh distance. Recall that the Rayleigh distance of ULA for direction $\theta_k$ is approximated to be
    \begin{equation}
      r_R = \frac{2 D^2 \sin^2\theta_k}{\lambda},
    \end{equation}  
    where $D \approx Md$. The threshold $\eta$ can be represented in a similar way as 
    \begin{equation}
      \eta = \frac{\pi D^2 \sin^2\theta_k}{15 \lambda} \approx \frac{1}{10} r_R.
    \end{equation}  
    This implies that the finite 3-dB interference region can be achieved within $1/10$ of the classical near-field region.
\end{remark}
\vspace{-5pt}

It is worth noting that channel correlation is an important metric that measures the resolution of near-field beamforming, i.e., how effectively users at different locations can be distinguished. For simplicity, we denote the channel correlation factor as $\rho_{\mathsf{u}}$. When $\rho_{\mathsf{u}}\approx0$, the resolution is considered perfect, meaning that the spatial channels for the users are orthogonal. In this scenario, near-field beamforming can focus energy on a specific location, facilitating the implementation of RDMA. Conversely, when $\rho_{\mathsf{u}}\approx1$, the resolution is considered poor. In this case, different users share the same near-field beam, facilitating the implementation of NOMA. When $\rho_{\mathsf{u}}$ is between $0$ and $1$, both schemes experience performance losses. Under our setup, we can demonstrate that the channel correlation degrades to zero as the number of antenna elements tends to infinity, i.e., achieving perfect resolution. The study in \cite{ding2023resolution} further identifies conditions under which the resolution of near-field beamforming is imperfect. This imperfect resolution implies that one user's near-field beam can still be useful to other users, motivating the feasibility of applying NOMA. The work in \cite{ding2023resolution} is limited to ULAs, with further extensions presented in \cite{rao2024general}. In this latter work, a general framework for calculating channel resolution is proposed, which also applies to UPAs. Additionally, the authors in \cite{rao2024general} explored the relationship between beamforming resolution and typical system parameters, such as the number of antennas and users' locations.
\subsubsection{Near-Field Beamforming Optimization}
In the previous section, we discussed the benefits of near-field beamfocusing, notably its ability to limit interference over distances when focusing on the LoS path. However, we encountered some limitations. First, effective beamfocusing is confined to a small part of the near-field. Second, beamfocusing is not flawless and still causes a non-negligible interference region in the range domain. Third, the NLoS paths caused by scattering generally exist in real-world systems. Although the strength of NLoS paths is much lower than that of the LoS paths in high-frequency bands, they may still have a non-negligible impact on system performance \cite{han2014multi}. Hence, beamforming optimization is still crucial for achieving optimal multi-user performance in the near-field region, rather than just relying on the simple matched filtering technique.
Despite these challenges, our findings suggest that enlarging the antenna array can significantly reduce interference and create semi-orthogonal channels in both angular and range domains. This is particularly beneficial in environments with minimal scattering. However, the complexity of beamforming optimization also increases with the size of the antenna array.

To mitigate this complexity, we now introduce a closed-form approach using the weighted minimum mean-squared error (WMMSE) method for optimizing near-field multi-user beamforming, which has a practical complexity for large-scale systems. To exemplify this approach, we consider a downlink near-field multi-user communication system with an $M$-antenna BS and $K$ single-antenna users. The ULA is assumed at the BS. The signal model for user $k$ is given by 
\begin{equation}
  y_k = \mathbf{h}_k^{\mathsf{T}} \mathbf{w}_k s_k + \sum_{i=1, i \neq k}^K \mathbf{h}_k^{\mathsf{T}} \mathbf{w}_i s_i + n_k,
\end{equation}    
where $\mathbf{h}_k \in \mathbbmss{C}^{M \times 1}$, $\mathbf{w}_k \in \mathbbmss{C}^{M \times 1}$, and $s_k \in \mathbbmss{C}$ denote the channel vector, beamformer, and unit-power random information symbols for user $k$, respectively. $n_k \sim \mathcal{CN}(0, \sigma_k^2)$ denote the additive Gaussian white noise with a power of $\sigma_k^2$. Here, $\mathbf{h}_k$ can be either a LoS channel or a multipath channel.  
The classical weighted sum-rate (WSR) maximization problem for beamforming design is given by 
\begin{subequations} \label{WSR_opt}
  \begin{align}
    \max_{\mathbf{w}_k} \quad &\sum_{k=1}^K w_k \log_2 \left( 1 + \frac{|\mathbf{h}_k^{\mathsf{T}} \mathbf{w}_k|^2}{\sum_{i\neq k} |\mathbf{h}_k^{\mathsf{T}} \mathbf{w}_i|^2 + \sigma_k^2 } \right) \\
    \mathrm{s.t.} \quad & \sum_{k=1}^K \|\mathbf{w}_k\|_2^2 \le P,
  \end{align}
\end{subequations}
where $w_k$ denotes weight for user $k$ and $P$ denote the transmit power budget. The weights can be obtained according to different strategies. For example, to maximize the sum rate, the weights can be set to $w_k = 1, \forall k$. To ensure fairness among users, the weights can be obtained using fairness-oriented scheduling methods, such as proportional fair scheduling (PFS). There have been many methods to solve this WSR maximization, represented by the WMMSE method \cite{christensen2008weighted, shi2011iteratively} and the fractional programming method \cite{shen2018fractional}. Although these methods can effectively transform the non-convex WSR objective function into an equivalent convex form by introducing auxiliary variables, the power-constrained optimization usually requires searching for dual variables, resulting in high computational complexity. The key idea of the closed-form WMMSE is removing the power constraint. This idea was initially introduced in \cite{christensen2008weighted} but was theoretically proven in \cite{zhao2023rethinking} recently. The closed-form WMMSE adopts the fact that the optimal beamformers that maximize the WSR must satisfy the power constraint with equality, i.e., $\sum_{k=1}^{K} \|\mathbf{w}_k\|^2 = P$ \cite{zhao2023rethinking}. Hence, we can consider the following unconstrained optimization problem:
\begin{align} \label{new_WSR}
  \max_{\bar{\mathbf{w}}_k} \sum_{k=1}^K w_k \log_2 \left( 1 + \frac{|\mathbf{h}_k^{\mathsf{T}} \bar{\mathbf{w}}_k|^2}{\sum_{i\neq k} |\mathbf{h}_k^{\mathsf{T}} \bar{\mathbf{w}}_i|^2 + \frac{\sigma_k^2}{P} \|\bar{\mathbf{W}}\|_F^2 } \right),
\end{align}
where $\bar{\mathbf{W}} = [\bar{\mathbf{w}}_1,...,\bar{\mathbf{w}}_K]$ and thus $\|\bar{\mathbf{W}}\|_F^2 = \sum_{k=1}^{K} \|\bar{\mathbf{w}}_k\|^2$. It is easy to show that for any optimal solution $\bar{\mathbf{w}}_k, \forall k,$ to problem \eqref{new_WSR}, the solution 
\begin{equation}
  \mathbf{w}_k = \sqrt{\frac{P}{\|\bar{\mathbf{W}}\|_F^2}} \bar{\mathbf{w}}_k, \forall k,
\end{equation}
must be the optimal solution to problem \eqref{WSR_opt} satisfying the power constraint with equality. Now, we shift our focus to solving problem \eqref{new_WSR}. The classical WMMSE method \cite{christensen2008weighted} can still be applied to transform this problem into the following equivalent form:
\begin{align} \label{new_WSR_WMMSE}
  \min_{\bar{\mathbf{W}}, u_k, v_k} \sum_{k=1}^K w_k \left(u_k e_k(\bar{\mathbf{W}}, v_k) - \log_2(u_k) \right),
\end{align}
where
\begin{align}
  &e_k(\bar{\mathbf{W}}, v_k) \nonumber \\
  &= \left|1 - v_k^* \mathbf{h}_k^{\mathsf{T}} \bar{\mathbf{w}}_k\right|^2  + |v_k|^2 \left(\sum_{i \neq k} |\mathbf{h}_k^{\mathsf{T}} \bar{\mathbf{w}}_i|^2 + \frac{\sigma_k^2}{P} \|\bar{\mathbf{W}}\|_F^2\right), \nonumber \\
  &= |v_k|^2 \mathrm{tr}\left( \mathbf{G}_k \bar{\mathbf{W}} \bar{\mathbf{W}}^{\mathsf{H}} \right) - \mathrm{Re}\left\{ v_k^* \mathbf{h}_k^{\mathsf{T}} \bar{\mathbf{w}}_k \right\} + 1,
\end{align}
where $e_k(\bar{\mathbf{W}}, v_k)$ denotes the MSE realized by the receiver $v_k$ at user $k$, $u_k$ is the weight for MSE, and $\mathbf{G}_k = \mathbf{h}_k^* \mathbf{h}_k^{\mathsf{T}} + \frac{\sigma_k^2}{P} \mathbf{I}$.

Problem \eqref{new_WSR_WMMSE} is not block-wise convex with respect the block variables $\bar{\mathbf{W}}$, $\{u_k\}$, and $\{v_k\}$, respectively. Thus, it can be effectively solved using the block coordinate descent (BCD) method. In particular, the optimal solution of $\{u_k\}$, $\{v_k\}$, and $\bar{\mathbf{W}}$ can be easily obtained by checking the first-order optimality condition as follows:
\begin{align}
  u_k = &\frac{|\mathbf{h}_k^{\mathsf{T}} \bar{\mathbf{w}}_k|^2}{\mathrm{tr}\left( \mathbf{G}_k \bar{\mathbf{W}} \bar{\mathbf{W}}^{\mathsf{H}} \right) - |\mathbf{h}_k^{\mathsf{T}} \bar{\mathbf{w}}_k|^2 } + 1, \\
  v_k = &\frac{\mathbf{h}_k^{\mathsf{T}} \bar{\mathbf{w}}_k}{\mathrm{tr}\left( \mathbf{G}_k \bar{\mathbf{W}} \bar{\mathbf{W}}^{\mathsf{H}} \right)}, \\
  \label{update_W}
  \bar{\mathbf{W}} = & \left( \sum_{k=1}^{K} w_k u_k |v_k|^2 \mathbf{G}_k  \right)^{-1} \mathbf{H} \mathsf{diag}(\mathbf{p}),
\end{align}
where $\mathbf{H} = [\mathbf{h}_1^*,\dots,\mathbf{h}_K^*]$ and $\mathbf{p} = [w_1 u_1 v_1^*,...,w_K u_K v_K^*]^{\mathsf{T}}$. Now, we have the closed-form updates of each block variable, and the solution to problem \eqref{new_WSR_WMMSE} can be obtained by updating these variables iteratively. 

\begin{figure}[!t]
  \centering
 \setlength{\abovecaptionskip}{0pt}
 \includegraphics[width=0.48\textwidth]{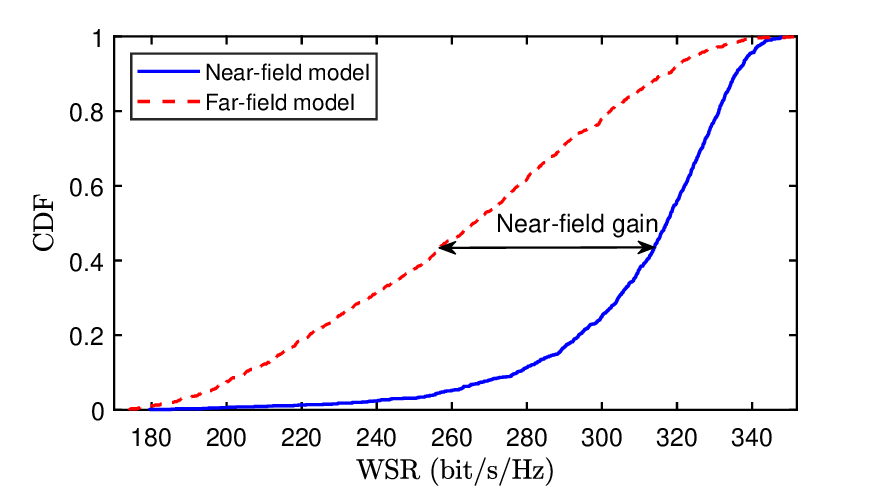}
 \caption{Cumulative distribution function of WSR.}
 \label{NF_CDF}
 \vspace{-10pt}
\end{figure}

However, when the number of antennas is large, the computational complexity of the matrix inversion in \eqref{update_W} may be unacceptable. In this case, the gradient-descent method can be exploited to update $\bar{\mathbf{W}}$, which avoids the matrix inversion operation. More specifically, for fixed $u_k$ and $v_k$, the optimal $\bar{\mathbf{W}}$ minimizes the following function:
\begin{equation}
    g(\bar{\mathbf{W}}) = \mathrm{tr}\left(\bar{\mathbf{W}}^{\mathsf{H}} \mathbf{A} \bar{\mathbf{W}} \right) - 2 \mathrm{Re}\left\{ \mathrm{tr}(\bar{\mathbf{W}}^{\mathsf{H}} \mathbf{B}) \right\},
\end{equation}
where $\mathbf{A} = \sum_{k=1}^{K} w_k u_k |v_k|^2 \mathbf{G}_k$ and $\mathbf{B} = \mathbf{H} \mathsf{diag}(\mathbf{p})$. The conjugate gradient of function $g(\bar{\mathbf{W}})$ is given by 
\begin{equation}
  \nabla_{\bar{\mathbf{W}}^*} g(\bar{\mathbf{W}}) = \mathbf{A} \bar{\mathbf{W}} - \mathbf{B}.
\end{equation}  
The gradient descent process to obtain the optimal $\bar{\mathbf{W}}$ that minimizes $g(\bar{\mathbf{W}})$ is given as follows:
\begin{align}
  \bar{\mathbf{W}}_{t+1} =  \bar{\mathbf{W}}_t - \alpha_t \nabla_{\bar{\mathbf{W}}^*} g(\bar{\mathbf{W}}_t),
\end{align}  
where $\alpha_t \ge 0$ is the learning rate at step $t$. Furthermore, the optimal learning rate can be obtained by 
\begin{align}
  \alpha_t = \arg \min_{\alpha} g\left(\bar{\mathbf{W}}_t - \alpha \nabla_{\bar{\mathbf{W}}^*} g(\bar{\mathbf{W}}_t)  \right) = \frac{ \mathrm{tr}( \mathbf{\Psi}_t \mathbf{\Psi}_t^{\mathsf{H}} )}{\mathrm{tr}( \mathbf{A} \mathbf{\Psi}_t \mathbf{\Psi}_t^{\mathsf{H}} )},
\end{align}
where $\mathbf{\Psi}_t = \nabla_{\bar{\mathbf{W}}^*} g(\bar{\mathbf{W}}_t) = \mathbf{A} \bar{\mathbf{W}}_t - \mathbf{B}$.

Fig. \ref{NF_CDF} depicted the cumulative distribution function of WSR achieved in a downlink near-field multi-user system operating at $28$ GHz. Here, we consider a BS equipped with a ULA of $512$ antennas, with antenna spacing of half a wavelength, and $20$ users randomly distributed within a range of $10 \sim 20$ m from the BS in the direction $85^\circ \sim 95^\circ$. The weight of each user is set to $w_k = 1, \forall k$. The results are obtained using the closed-form WMMSE method using the exact near-field model and approximated far-field model, respectively. It can be seen that the near-field model achieves significant WSR gains compared to the far-field model. This is due to the fact that the range domain interference region is finite in the near-field and infinite in the far-field. 

\begin{figure}[!t]
  \centering
    \subfigure[ULA.]{
        \includegraphics[width=0.3\textwidth]{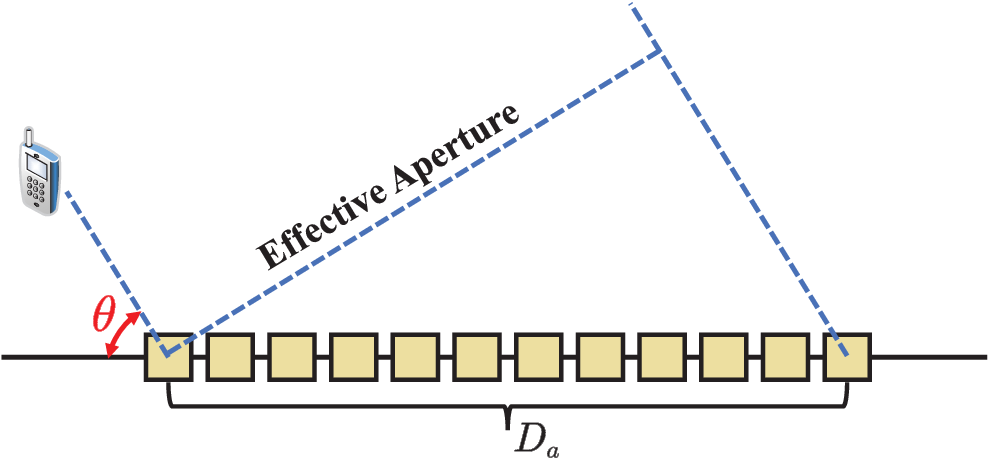}
        \label{Figure: ULA_UCA: ULA}	
    }
    \subfigure[UCA.]{
        \includegraphics[width=0.3\textwidth]{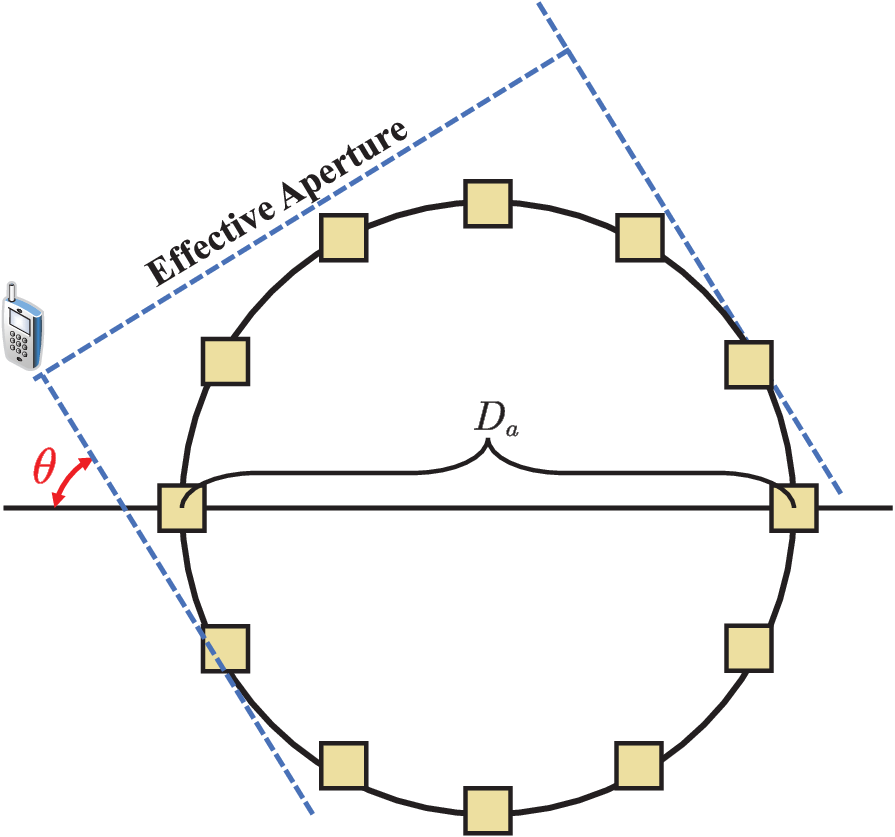}
        \label{Figure: ULA_UCA: UCA}	
    }
  \caption{Comparison between the ULA and the UCA.}
  \label{Figure: ULA_UCA}
\end{figure}

The discussions presented primarily focus on ULAs, as illustrated in {\figurename} {\ref{Figure: ULA_UCA: ULA}}. From our analysis, it is evident that the near-field effect can effectively mitigate IUI, even when users are aligned in the same direction. Consequently, it can be inferred that system throughput increases when the BS array generates a larger near-field region. However, the geometry of the ULA may constrain the aperture of the antenna array, thus limiting the near-field region. More specifically, the ULA exhibits an ellipsoidal near-field region that is largest for $\theta=\frac{\pi}{2}$ and gradually diminishes to zero as $\theta$ approaches $0$ or $\pi$ \cite{liu2023nearfield}. This phenomenon arises because the linear antenna arrangement results in varying effective apertures in different directions. Therefore, when users are located at angles where $\theta$ approaches $0$ or $\pi$, the near-field region is quite limited, thereby reducing system throughput. One approach to address this issue is to utilize uniform circular arrays (UCAs), as shown in {\figurename} {\ref{Figure: ULA_UCA: UCA}}. A UCA, with its rotationally symmetric geometry, maintains the same effective aperture in all directions, leading to an enlarged near-field region \cite{wu2023enabling}. Consequently, a user can benefit from near-field beamforming even when $\theta$ approaches $0$ or $\pi$. 

\begin{figure}[!t]
  \centering
 \setlength{\abovecaptionskip}{0pt}
 \includegraphics[width=0.48\textwidth]{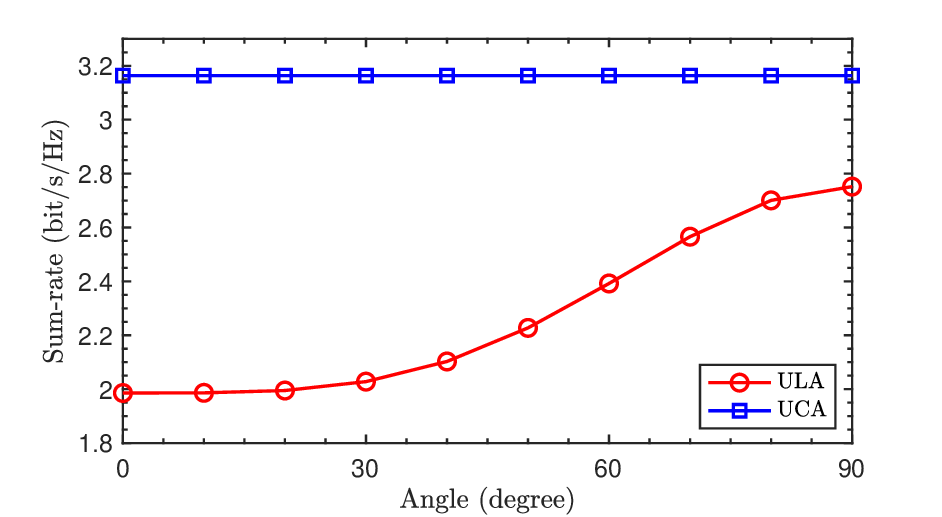}
 \caption{Comparison of the downlink sum-rates achieved by ULAs and UCAs under maximal-ratio transmission. $K=2$, $r_1=20$ m, $r_2=30$ m, $\lambda = 0.0107$ m, $M=256$, $D_a=(M-1)\frac{\lambda}{2}$, $\sigma_1^2=\sigma_2^2=1$, $A=\frac{\lambda^2}{4\pi}$, and the transmit SNR for each user is $20$ dB. The two users are located in the same direction with angle $\theta$.}
 \label{UCA2}
 \vspace{-10pt}
\end{figure}

{\figurename} {\ref{UCA2}} plots the sum-rate achieved by ULAs and UCAs as a function of the angle $\theta$. As observed, the UCA consistently achieves a higher sum-rate than the ULA across all angle ranges. Additionally, the sum-rate for the UCA remains virtually unchanged for different angles. This consistency is attributed to the UCA's rotationally symmetric geometry, which maintains a constant effective aperture in all directions. In contrast, the effective aperture of the ULA gradually increases as $\theta$ increass from $0$ to $\frac{\pi}{2}$, thereby enlarging the near-field region and consequently improving the sum-rate. However, beyond $\theta=\frac{\pi}{2}$, the effective aperture and the near-field region diminish, resulting in a lower sum-rate. In summary, the UCA is more robust to variations in user direction and achieves a higher sum-rate than the ULA. For further comparisons between ULAs and UCAs, refer to \cite{wu2023enabling}.

\subsection{Wavenumber-domain Beamforming}
\begin{figure}[!t]
  \centering
 \setlength{\abovecaptionskip}{0pt}
 \includegraphics[width=0.48\textwidth]{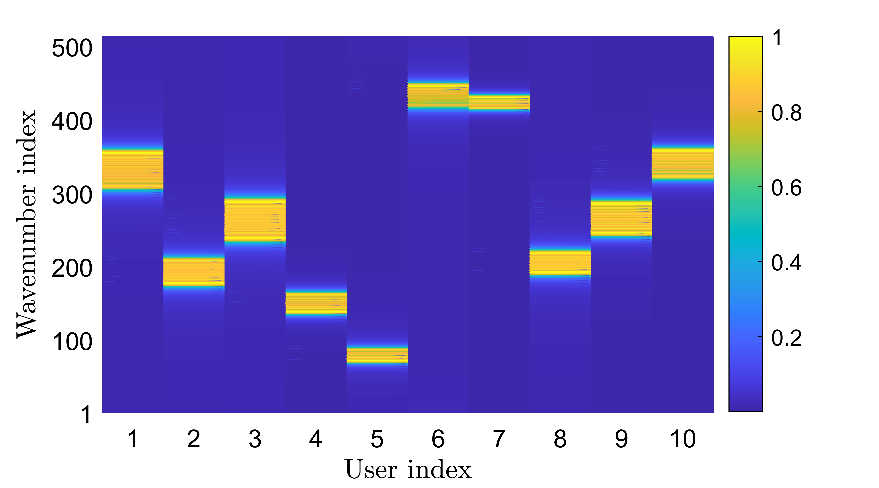}
 \caption{Illustration of normalized channel magnitudes of different users in the wavenumber-domain.}
 \label{chanel_sparsity}
 \vspace{-10pt}
\end{figure}

In the previous section, we unveiled the benefits of the near-field effect for NGMA in the spatial domain. Now, we investigate the near-field multiple access in the wavenumber domain. We still consider a multi-user-MISO system where $K$ single-antenna users are served by an $M$-antenna ULA.  Based on the theory of Fourier plane-wave transform introduced earlier in this paper, the near-field spatial-domain channel vector $\mathbf{h}_k$ of user $k$ can be transformed into the wavenumber-domain as follows:
\begin{equation} \label{zw_wavenumber_transform}
  \mathbf{h}_k^{\mathsf{T}}  = \mathbf{h}_{a,k}^{\mathsf{T}} \mathbf{\Phi}_{\mathsf{S}}^{\mathsf{H}},
\end{equation}
where $\mathbf{h}_{a,k} \in \mathbbmss{C}^{n_{\mathsf{S}} \times 1}$ denote the wavenumber-domain channel vector of user $k$ and $\mathbf{\Phi}_{\mathsf{S}} \in \mathbbmss{C}^{M \times n_{\mathsf{S}}}$ denotes the semi-unitary transformation matrix satisfying $\mathbf{\Phi}_{\mathsf{S}}^{\mathsf{H}} \mathbf{\Phi}_{\mathsf{S}} = \mathbf{\Phi}_{\mathsf{S}}^{\mathsf{T}} \mathbf{\Phi}_{\mathsf{S}}^* = \mathbf{I}$. By exploiting this semi-unitary property, we have 
\begin{equation}
  \mathbf{h}_{a,k} = \mathbf{\Phi}_{\mathsf{S}}^{\mathsf{T}} \mathbf{h}_k.
\end{equation}
Compared to the spatial-domain channel vector, the wavenumber-domain channel vector exhibits the following characteristics:
\begin{itemize}
  \item \emph{The dimension $n_{\mathsf{S}}$ of the wavenumber-domain depends on the aperture of the antenna array instead of the number of antennas in the antenna array.} More specifically, according to \eqref{25(a)}, we have $n_{\mathsf{S}} \approx \frac{2 L_{\mathsf{S}}}{\lambda}$ for ULAs, where $L_{\mathsf{S}} = M d$ denotes the aperture of ULAs. It is easy to see that $n_{\mathsf{S}} \le M$ if $d \le \frac{\lambda}{2}$. Although the dimension of the channel vector cannot be significantly reduced in the wavenumber domain for conventional ULAs with half-wavelength antenna spacing, the wavenumber-domain transformation does hold great promise for emerging holographic MIMO systems that exploit the metasurface antenna arrays. Compared to the conventional antenna arrays, metasurface antenna arrays typically have an antenna spacing of $\frac{\lambda}{10} \sim \frac{\lambda}{5}$ \cite{smith2017analysis}, resulting in $n_{\mathsf{S}} \ll M$. The gap between $n_{\mathsf{S}}$ and $M$ becomes even much larger for UPAs because of the two-dimensional deployment of antennas.    
  \item \emph{The wavenumber-domain channel vector is sparse in scenarios without rich scattering.} The Fourier plane-wave transform essentially represents the spherical-wave near-field channel as a superposition of plane waves in different directions. In scenarios without rich scattering, the user receives signals from a limited number of directions, resulting in the sparse structure of the wavenumber-domain channel vector, as depicted in Fig. \ref{chanel_sparsity}. 
  \item \emph{The transformation from the spatial domain to the wavenumber domain does not change the orthogonality of near-field channel vectors.} Based on the semi-unitary property of the transformation matrix $\mathbf{\Phi}_{\mathsf{S}}$, it is easy to prove that $|\mathbf{h}_k^{\mathsf{H}} \mathbf{h}_i| = | \mathbf{h}_{a,k}^{\mathsf{H}} \mathbf{\Phi}_{\mathsf{S}}^{\mathsf{T}} \mathbf{\Phi}_{\mathsf{S}}^* \mathbf{h}_{a,i} | = |\mathbf{h}_{a,k}^{\mathsf{H}} \mathbf{h}_{a,i}|$. Therefore, the design for multiple-access can be efficiently implemented within the wavenumber domain, which is referred to as WDMA.
\end{itemize} 

Based on the above favorable properties, the WDMA beamforming design exhibits lower complexity and negligible performance loss compared to SDMA beamforming design. Let us also consider the WDMA beamforming design for maximizing WSR as an example. Building upon \eqref{zw_wavenumber_transform}, the achievable rate of user $k$ in \eqref{WSR_opt} can be represented as 
\begin{align} \label{zw_wavenumber_rate}
  \mathsf{R}_k = &\log_2 \left( 1 + \frac{|\mathbf{h}_{a,k}^{\mathsf{T}} \mathbf{\Phi}_{\mathsf{S}}^{\mathsf{H}} \mathbf{w}_k|^2}{\sum_{i\neq k} |\mathbf{h}_{a,k}^{\mathsf{T}} \mathbf{\Phi}_{\mathsf{S}}^{\mathsf{H}} \mathbf{w}_i|^2 + \sigma_k^2 } \right), \nonumber \\
  \overset{(a)}{=} &\log_2 \left( 1 + \frac{|\mathbf{h}_{a,k}^{\mathsf{T}} \mathbf{\Phi}_{\mathsf{S}}^{\mathsf{H}} \mathbf{\Phi}_{\mathsf{S}} \mathbf{w}_{a,k}|^2}{\sum_{i\neq k} |\mathbf{h}_{a,k}^{\mathsf{T}} \mathbf{\Phi}_{\mathsf{S}}^{\mathsf{H}} \mathbf{\Phi}_{\mathsf{S}} \mathbf{w}_{a,i}|^2 + \sigma_k^2 } \right), \nonumber \\
  \overset{(b)}{=} & \log_2 \left( 1 + \frac{|\mathbf{h}_{a,k}^{\mathsf{T}} \mathbf{w}_{a,k}|^2}{\sum_{i\neq k} |\mathbf{h}_{a,k}^{\mathsf{T}} \mathbf{w}_{a,i}|^2 + \sigma_k^2 } \right),
\end{align}
where step $(a)$ is obtained by defining $\mathbf{w}_k = \mathbf{\Phi}_{\mathsf{S}} \hat{\mathbf{w}}_{a,k}$ with $\hat{\mathbf{w}}_{a,k} \in \mathbbmss{C}^{n_{\mathrm{S}} \times 1}$ being the wavenumber-domain beamformer for user $k$ and step $(b)$ stems from the semi-unitary property of $\mathbf{\Phi}_{\mathsf{S}}$. Furthermore, the expression of transmit power can be converted to
\begin{align}
  \sum_{k=1}^K \|\mathbf{w}_k\|_2^2  = \sum_{k=1}^K \hat{\mathbf{w}}_{a,k}^{\mathsf{H}} \mathbf{\Phi}_{\mathsf{S}}^{\mathsf{H}}  \mathbf{\Phi}_{\mathsf{S}} \hat{\mathbf{w}}_{a,k}  = \sum_{k=1}^K \|\hat{\mathbf{w}}_{a,k}\|_2^2.
\end{align}
Thus, the WSR maximization problem in \eqref{WSR_opt} can be transformed into the wavenumber domain as follows:
\begin{subequations} \label{WSR_opt_wavenumber}
  \begin{align}
    \max_{\mathbf{w}_{a,k}} \quad &\sum_{k=1}^K w_k \log_2 \left( 1 + \frac{|\mathbf{h}_{a,k}^{\mathsf{T}} \mathbf{w}_{a,k}|^2}{\sum_{i\neq k} |\mathbf{h}_{a,k}^{\mathsf{T}} \mathbf{w}_{a,i}|^2 + \sigma_k^2 } \right) \\
    \mathrm{s.t.} \quad & \sum_{k=1}^K \|\mathbf{w}_{a,k}\|_2^2 \le P,
  \end{align}
\end{subequations}
In \eqref{WSR_opt_wavenumber}, the dimension of beamformer is reduced from $M$ to $n_{\mathsf{S}}$. Additionally, the channel vector is no longer a $M$-dimensional full vector, but a $n_{\mathsf{S}}$-dimensional sparse vector. 

Given the sparsity of the wavenumber-domain channel vector, the optimal wavenumber-domain must also be sparse, which can be proven as follows. For an arbitrary set of feasible non-sparse beamforming vectors $\hat{\mathbf{w}}_{a,k}, \forall k$, we can always construct a new set of beamforming vectors $\mathbf{w}_{a,k}, \forall k,$ whose $n$th entry satisfies   
\begin{equation}
[\mathbf{w}_{a,k}]_n = \begin{cases}
  [\hat{\mathbf{w}}_{a,k}]_n, &\text{if } [\mathbf{h}_{a,k}]_n \neq 0, \\
  0, &\text{otherwise}.
\end{cases}
\end{equation}  
These new beamforming vectors have the following properties:
\begin{align}
|\mathbf{h}_{a,k}^{\mathsf{T}} \mathbf{w}_{a,k}|^2 &= |\mathbf{h}_{a,k}^{\mathsf{T}} \hat{\mathbf{w}}_{a,k}|^2, \forall k, \\
|\mathbf{h}_{a,k}^{\mathsf{T}} \mathbf{w}_{a,i}|^2 &\le |\mathbf{h}_{a,k}^{\mathsf{T}} \hat{\mathbf{w}}_{a,i}|^2, \forall i \neq k, \\
\sum_{k=1}^K \|\mathbf{w}_{a,k}\|_2^2 &\le \sum_{k=1}^K \|\hat{\mathbf{w}}_{a,k}\|_2^2 \le P.
\end{align}
The above results indicate that $\mathbf{w}_{a,k}$ is a feasible sparse beamforming vector that achieves higher WSR than $\hat{\mathbf{w}}_{a,k}$. The proof is thus completed. As a consequence, the optimal wavenumber-domain beamformers must be sparse for the sparse wavenumber-domain channel vectors. The indexes of the non-zero entries in beamformers are the same as those in the corresponding channel vectors. When the perfect channel state information (CSI) is available, these indexes can be easily obtained based on the magnitude values of the wavenumber-domain channel vectors. In this case, only the non-zero entries in the beamforming vectors need to be optimized, further reducing the number of optimization variables.

Nevertheless, when the CSI is imperfect, the estimated channel vector may be non-sparse anymore due to the estimation error. In particular, the estimated spatial-domain channel vector $\hat{\mathbf{h}}_k $ of user $k$ can be modeled as 
\begin{equation}
  \hat{\mathbf{h}}_k = \mathbf{h}_k + \Delta \mathbf{h}_k,
\end{equation}
where $\tilde{\mathbf{h}}_k$ denotes the channel estimation error that has an average power of $\mathbb{E}[ \|\Delta\mathbf{h}_k\|_2^2  ] = e_k \|\Delta\mathbf{h}_k\|_2^2$. Then, the corresponding wavenumber-number domain channel vector is given by 
\begin{equation}
  \hat{\mathbf{h}}_{a,k} = \mathbf{\Phi}_{\mathsf{S}}^{\mathsf{T}} \mathbf{h}_k + \mathbf{\Phi}_{\mathsf{S}}^{\mathsf{T}}  \Delta \mathbf{h}_k = \mathbf{h}_{a,k} + \Delta \mathbf{h}_{a,k},
\end{equation}
where $\Delta\mathbf{h}_{a,k} = \mathbf{\Phi}_{\mathsf{S}}^{\mathsf{T}}  \Delta \mathbf{h}_k$ is the estimation error in the wavenumber domain that satisfies $\mathbb{E}[ \|\Delta\mathbf{h}_{a,k}\|_2^2] = \mathbb{E}[ \|\Delta\mathbf{h}_k\|_2^2  ]$. Although the estimated channel vector $\hat{\mathbf{h}}_{a,k}$ may not be sparse, we can still exploit the a priori knowledge of the sparse structure of the optimal beamformer in the wavenumber domain to achieve robust beamforming. To this end, we can consider the following $\ell_1$-norm regularized WSR maximization problem in the wavenumber-domain \cite{feng2015sparse}:
\begin{subequations} \label{WSR_opt_wavenumber_robust}
  \begin{align}
    \max_{\mathbf{w}_{a,k}} \quad &\sum_{k=1}^K w_k R_k(\hat{\mathbf{h}}_{a,k}, \mathbf{w}_{a,k}) - \rho_0 \sum_{k=1}^K \|\mathbf{w}_{a,k}\|_1, \\
    \mathrm{s.t.} \quad & \sum_{k=1}^K \|\mathbf{w}_{a,k}\|_2^2 \le P.
  \end{align}
\end{subequations}
Here, the $\ell_1$-norm $\|\mathbf{w}_{a,k}\|_1$ is a convex approximation of the $\ell_0$-norm $\|\mathbf{w}_{a,k}\|_0$ that reflects the sparsity of the wavenumber-doamin channel vector. The factor $\rho_0$ is to regulate the degree of sparsity in the variables. In this $\ell_1$-norm regularized problem, the beamformer $\mathbf{w}_{a,k}$ is optimized to maximize the WSR using the estimated channel vectors with guaranteed sparsity, which is a property of the optimal wavenumber-domain beamformer with perfect CSI. 

\begin{figure}[!t]
  \centering
 \setlength{\abovecaptionskip}{0pt}
 \includegraphics[width=0.48\textwidth]{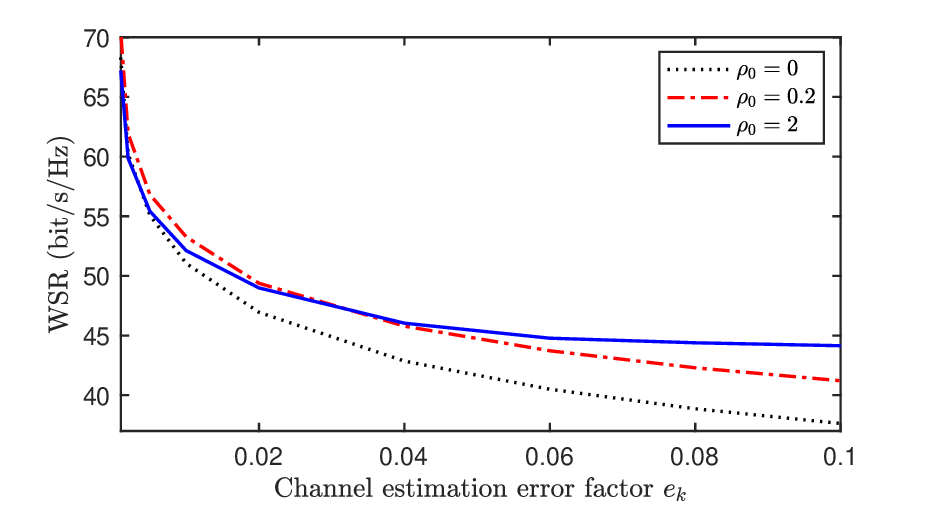}
 \caption{WSR performance under different levels of channel estimation error.}
 \label{NF_robust}
 \vspace{-10pt}
\end{figure}

In Fig. \eqref{NF_robust}, we show the performance of this simple $\ell_1$-norm regularisation method in robust beamforming. Here, we consider a BS equipped with a $256$-antenna ULA serving $4$ near-field users. It can be observed that higher WSR can be achieved using imperfect CSI by simply ensuring the sparsity of the wavenumber-domain beamformers.

\subsection{Beamforming Design for CAP Arrays}
The preceding discussion on near-field beamforming is based on the premise that the BS utilizes an SPD array. Since employing more antennas can increase the spatial DoFs, we now shift our focus to an extreme scenario wherein the BS employs a CAP array. To maintain consistency with the discourse in Section \ref{Section: Near-Field Beamfocusing for RDMA}, we assume that the CAP array is two-dimensional, situated along the $x$-axis, and centered at the origin. The aperture size, i.e., the length of the CAP array, is determined by $D=(M-1)d\approx Md$. Moreover, let $\mathcal{A}_{\mathsf{S}}$ and $\mathcal{A}_{\mathsf{R}}^{k}$ represent the apertures of the BS and each user $k$ ($\forall k$), respectively. For the sake of simplicity, we assume that $\lvert\mathcal{A}_{\mathsf{R}}^{k}\rvert=\lvert\mathcal{A}_{\mathsf{R}}\rvert\ll \lvert\mathcal{A}_{\mathsf{S}}\rvert$.
\subsubsection{Spatial-domain Beamforming}
Unlike the SPD array, which generates finite-dimensional signal vectors, the CAP array facilitates a continuous distribution of source currents within the transmit aperture. Let ${\mathsf{j}}(\mathbf{s})\in{\mathbbmss{C}}$ denote the continuous distribution of source currents at point $\mathbf{s}$, where $\mathbf{s}=[x,0]^{\mathsf{T}}\in{\mathbbmss{R}}^{2\times1}$ represents the source point within the transmit aperture ${\mathcal{A}}_{\mathsf{S}}=\{[x,0]^{\mathsf{T}}|x\in[-\frac{D}{2},\frac{D}{2}]\}$. Following the transmission model outlined in \eqref{CAP_SU_Basic_Model}, the electric radiation field observed at user $k$ is expressed as follows:
\begin{align}
\mathsf{y}_k(\mathbf{r})
=\int_{\mathcal{A}_{\mathsf{S}}}-{\rm{j}}\eta_0\frac{2\pi}{\lambda}\mathsf{g}(\mathbf{r},\mathbf{s}){\mathsf{j}}(\mathbf{s}){\rm{d}}\mathbf{s}
+\mathsf{n}_k(\mathbf{r})
\end{align}
for ${\mathbf{r}}\in\mathcal{A}_{\mathsf{R}}^{k}$, where $\mathsf{n}_k(\mathbf{r})$ denotes the random noise field that is modeled as a zero-mean complex Gaussian process with ${\mathbbmss{E}}\{{\mathsf{n}}_k(\mathbf{r}){\mathsf{n}}_k^{\mathsf{H}}(\mathbf{r}')\}=\overline{\sigma}_k^2\delta(\mathbf{r}-\mathbf{r}')$, and $\mathsf{g}(\mathbf{r}_k,\mathbf{s})=\frac{{\rm{e}}^{-{\rm{j}}\frac{2\pi}{\lambda}\lVert \mathbf{r}_k - \mathbf{s}\rVert}}{4\pi \lVert \mathbf{r}_k - \mathbf{s}\rVert}$ is the scalar Green's function. Let ${\mathsf{j}}_k(\mathbf{s})\in{\mathbbmss{C}}$ represent the source currents conveying the normalized data symbol $s_k$ to user $k$. Therefore, we have ${\mathsf{j}}(\mathbf{s})=\sum_{k=1}^{K}s_k {\mathsf{j}}_k(\mathbf{s})$, which is subject to a radiating power budget $\sum_{k=1}^{K}\int_{{\mathcal{A}}_{\mathsf{S}}}\lvert{\mathsf{j}}_k(\mathbf{s})\rvert^2{\rm{d}}{\mathbf{s}}\leq P_{\mathsf{rad}}$. Following the derivation steps in obtaining \eqref{Multiple_User_SINR_Step2}, the SINR at user $k$ is given by
\begin{equation}\label{Received_SINR_CAP}
\gamma_k^{\mathsf{CAP}}=\frac{\left\lvert\int_{{\mathcal{A}}_{\mathsf{S}}}{\mathsf{g}}(\mathbf{r}_k,{\mathbf{s}}){\mathsf{j}}_{k}({\mathbf{s}}){\rm{d}}{\mathbf{s}}\right\rvert^2}
{\frac{\overline{\sigma}^2}{(\eta_0\frac{2\pi}{\lambda})^2\lvert\mathcal{A}_{\mathsf{R}}\rvert}+\sum_{k'\ne k}\left\lvert\int_{{\mathcal{A}}_{\mathsf{S}}}{\mathsf{g}}(\mathbf{r}_k,{\mathbf{s}}){\mathsf{j}}_{k'}({\mathbf{s}}){\rm{d}}{\mathbf{s}}\right\rvert^2}.
\end{equation}

It can be observed from \eqref{Received_SINR_CAP} that CAP array-based near-field SDMA relies on the design of the continuous currents $\{{\mathsf{j}}_k(\mathbf{s})\}_{\forall k}$, which is in contrast to the discrete beamformer design in SPD array-based SDMA, as shown in \eqref{WSR_opt}. It is worth noting that optimizing a continuous current distribution poses a significantly more challenging task than optimizing a discrete vector. Therefore, our next step is to explore additional properties of near-field SDMA to simplify the beamforming design. Our particular focus lies in determining whether the property of ``near-field beamfocusing'' presented in \eqref{interference_region} holds or approximately holds for CAP arrays. The inner product of the normalized Green’s function is defined as follows:
\begin{align}\label{Correlation_Factor_LOS_NFC_CAP}
\overline{\rho}_{k,k'}=\frac{\left\lvert\int_{{\mathcal{A}}_{\mathsf{S}}}{\mathsf{g}}(\mathbf{r}_k,\mathbf{s})
{\mathsf{g}}^{*}(\mathbf{r}_{k'},\mathbf{s}){\rm{d}}\mathbf{s}\right\rvert}
{\sqrt{\int_{{\mathcal{A}}_{\mathsf{S}}}\lvert{\mathsf{g}}(\mathbf{r}_k,\mathbf{s})\rvert^2{\rm{d}}\mathbf{s}
\int_{{\mathcal{A}}_{\mathsf{S}}}\lvert{\mathsf{g}}(\mathbf{r}_{k'},\mathbf{s},)\rvert^2{\rm{d}}\mathbf{s}}}.
\end{align}
However, this expression is computationally challenging. To simplify computation, we make the assumption that each near-field user is positioned beyond the uniform-power distance. Under this assumption, the Green's function can be approximated as follows (c.f. \eqref{Green_Function_Simplify}):
\begin{align}\label{Favorable_Propagation_LOS_NFC_CAP}
\mathsf{g}(\mathbf{r}_k,\mathbf{s})\approx \frac{1}{4\pi r_k} {\rm{e}}^{-{\rm{j}}\frac{2\pi}{\lambda}
(r_k-x\cos{\theta_k}+\frac{x^2}{2r_k}\sin^2{\theta_k})}.
\end{align}
Inserting \eqref{Favorable_Propagation_LOS_NFC_CAP} into \eqref{Correlation_Factor_LOS_NFC_CAP} gives
\begin{equation}\label{Correlation_Factor_LOS_NFC_CAP}
\overline{\rho}_{k,k'}\approx \left| \frac{1}{D}\int_{\frac{-D}{2}}^{\frac{D}{2}} {\rm{e}}^{{\rm{j}}(x a_1+b_1x^2) } {\rm{d}}x \right|
=\left| \int_{\frac{-1}{2}}^{\frac{1}{2}} {\rm{e}}^{{\rm{j}}(x Da_1+D^2b_1x^2) } {\rm{d}}x \right|,
\end{equation}
where $a_1 = \frac{2\pi}{\lambda}  \left( \cos (\theta_k + \Delta \theta) - \cos \theta_k \right)$, $b_1 = \frac{2 \pi}{\lambda} \left( \frac{\sin^2 \theta_k}{2 r_k} - \frac{\sin^2 (\theta_k + \Delta \theta)}{2 (r_k + \Delta r)} \right) $, and $\theta_k+\Delta_k$ represents the direction of user $k'$. 

By comparing \eqref{Correlation_Factor_LOS_NFC_CAP} with \eqref{Correlation_Faactor_Approximate_Exp} and considering the approximation $D\approx Md$, we observe that $\overline{\rho}_{k,k'}$ presents the same expression as its SPD counterpart shown in \eqref{Correlation_Faactor_Approximate_Exp}. \emph{Based on this comparison, we conclude that the property of near-field beamfocusing also holds for CAP arrays.} \emph{This observation suggests that the SDMA in near-field SDMA for CAP arrays is essentially RDMA}, akin to the SPD case. This beneficial characteristic is also a consequence of spherical-wave propagation, which effectively mitigates IUI and enhances the system's spectral efficiency. By leveraging this property, we can utilize matched-filtering-based beamforming to design the source currents, which results in
\begin{align}
{\mathsf{j}}_k(\mathbf{s})=\sqrt{\overline{p}_{\mathsf{rad}}^{(k)}}\frac{{\mathsf{g}}^{*}(\mathbf{r}_k,\mathbf{s})}
{\sqrt{\int_{{\mathcal{A}}_{\mathsf{S}}}\lvert{\mathsf{g}}(\mathbf{r}_k,\mathbf{s})\rvert^2{\rm{d}}\mathbf{s}}},
\end{align}
with $\overline{p}_{\mathsf{rad}}^{(k)}$ representing the power allocation factor. As discussed in Section II, near-field channels are generally LoS dominated and sparsely scattered, which makes matched-filtering-based beamforming generally effective for NFC relying on CAP arrays.
\subsubsection{Wavenumber-domain Beamforming}
To enhance the system sum-rate further, we must address the optimization problem given by
\begin{subequations} \label{WSR_opt_CAP_space}
  \begin{align}
    \max_{{\mathsf{j}}_k(\mathbf{s})} \quad &\sum_{k=1}^K w_k \log_2 \left( 1 + \gamma_k^{\mathsf{CAP}} \right) \\
    \mathrm{s.t.} \quad & \sum_{k=1}^{K}\int_{{\mathcal{A}}_{\mathsf{S}}}\lvert {\mathsf{j}}_k(\mathbf{s}) \rvert^2{\rm{d}}\mathbf{s}\leq P_{\mathsf{rad}},
  \end{align}
\end{subequations}
where $w_k$ denotes weight for user $k$, and $P_{\mathsf{rad}}$ accounts for the total radiated power. However, due to the continuous nature of the current distributions, problem \eqref{WSR_opt_CAP_space} involves $K$ infinite-dimensional variables, which presents a significant computational challenge. In the sequel, we consider a special case where the spatial response between the user and the BS can be precisely described using the 4D Fourier plane-wave representation in \eqref{4DFPE_Stadard}. Under this case, spatial information is captured by its wavenumber counterpart. We further assume that the BS is equipped with a linear CAP array with length $D$ and is placed on the $x$-axis. The spatial response between each user $k$ and the BS can be described as follows:
 \begin{equation}
\begin{split}
&{{\mathsf{g}}(\mathbf{r}_k,\mathbf{s})}\approx\frac{1}{\sqrt{D}}{\sum_{\ell_x=-\lceil D/\lambda\rceil}^{\lceil D/\lambda\rceil}} \check{H}_k(\ell_x){\rm{e}}^{-{\rm{j}}(\frac{2\pi \ell_x}{D}{\mathbf{s}}_{x})},
\end{split}
\end{equation}
where $\check{H}_k(\ell_x)$ denotes the wavenumber-domain response of user $k$, and $\mathbf{s}$ is a point located in the aperture of the BS with its $x$-coordinate given by ${\mathbf{s}}_{x}$. By leveraging the orthogonality property of the source responses $\{{\rm{e}}^{-{\rm{j}}(\frac{2\pi \ell_x}{D}{\mathbf{s}}_{x})}\}$ over the transmit aperture, i.e., $\int_{-\frac{D}{2}}^{\frac{D}{2}}{\rm{e}}^{-{\rm{j}}(\frac{2\pi \ell_x}{D}x)}{\rm{e}}^{{\rm{j}}(\frac{2\pi \ell_x'}{D}x)}{\rm{d}}x=0$ ($\ell_x\ne \ell_x'$), we can design the source current as
\begin{align}
{\mathsf{j}}(\mathbf{s})={\sum_{\ell_x=-\lceil D/\lambda\rceil}^{\lceil D/\lambda\rceil}}\frac{1}{\sqrt{D}} {\rm{e}}^{{\rm{j}}(\frac{2\pi \ell_x}{D}{{\mathbf{s}}_{x}})}\sum_{k=1}^{K}w_{\ell_x,k}s_k,
\end{align}
with
\begin{align}
{\mathsf{j}}_k(\mathbf{s})={\sum_{\ell_x=-\lceil D/\lambda\rceil}^{\lceil D/\lambda\rceil}}\frac{1}{\sqrt{D}} {\rm{e}}^{{\rm{j}}(\frac{2\pi \ell_x}{D}{{\mathbf{s}}_{x}})}w_{\ell_x,k}.
\end{align}
Under this scenario, we obtain
\begin{align}
\int_{{\mathcal{A}}_{\mathsf{S}}}{{\mathsf{g}}(\mathbf{r}_k,\mathbf{s})}{\mathsf{j}}_{k'}(\mathbf{s}){\rm{d}}\mathbf{s}
={\sum_{\ell_x=-\lceil D/\lambda\rceil}^{\lceil D/\lambda\rceil}} \check{H}_k(\ell_x)w_{\ell_x,k'}
=\mathbf{h}_{a,k}^{\mathsf{T}} \mathbf{w}_{a,k'},
\end{align}
where $\mathbf{h}_{a,k}=[\check{H}_k(-\lceil D/\lambda\rceil),\ldots,\check{H}_k(\lceil D/\lambda\rceil)]^{\mathsf{T}}$ denotes the wavenumber-domain channel of user $k$, and $\mathbf{w}_{a,k}=[w_{-\lceil D/\lambda\rceil,k},\ldots,w_{\lceil D/\lambda\rceil,k}]^{\mathsf{T}}$ is the wavenumber-domain beamformer used to convey his information. Based on this transformation, problem \eqref{WSR_opt_CAP_space} can be converted into a classical multiuser MISO beamforming problem as follows:
\begin{subequations}\label{WDMA_CAP_Final_eQUAL}
  \begin{align}
    \max_{\mathbf{w}_{a,k}} \quad &\sum_{k=1}^K w_k \log_2 \left( 1 + \frac{|\mathbf{h}_{a,k}^{\mathsf{T}} \mathbf{w}_{a,k}|^2}{\sum_{i\neq k} |\mathbf{h}_{a,k}^{\mathsf{T}} \mathbf{w}_{a,i}|^2 + \frac{\overline{\sigma}^2}{(\eta_0\frac{2\pi}{\lambda})^2\lvert\mathcal{A}_{\mathsf{R}}\rvert} } \right) \\
    \mathrm{s.t.} \quad & {\sum_{\ell_x=-\lceil D/\lambda\rceil}^{\lceil D/\lambda\rceil}}\sum_{k=1}^{K} \lvert w_{\ell_x,k}\rvert^2 \le P_{\mathsf{rad}}.
  \end{align}
\end{subequations}
Compared with problem \eqref{WSR_opt_CAP_space}, problem \eqref{WDMA_CAP_Final_eQUAL} involves finite dimensions, making it significantly easier to solve. This simplification essentially transforms near-field SDMA or RDMA into near-field WDMA. 

For scenarios where the spatial response deviates from the standard outlined in \eqref{4DFPE_Stadard}, an alternative approach involves expanding the current and Green's function in the wavenumber domain to obtain their wavenumber-domain representations. Subsequently, further design can be conducted in the dimension-reduced wavenumber domain. Specifically, for a more general CAP array with an arbitrary aperture, one could use the Fourier basis function to transform the continuous spatial response and the currents into their wavenumber-domain counterparts. 

Given the transmit aperture ${\mathcal{A}}_{\mathcal{S}}$, the Fourier basis function is constructed as follows:
\begin{align}
\psi_{\bm{\ell}}({\mathbf{s}})=\frac{1}{\sqrt{\lvert{\mathcal{A}}_{\mathcal{S}}\rvert}}{\rm{e}}^{{\rm{j}}2\pi(\frac{\ell_x}{L_x}({\mathbf{s}}_x-\frac{L_x}{2})+
\frac{\ell_y}{L_y}({\mathbf{s}}_y-\frac{L_y}{2})+\frac{\ell_z}{L_z}({\mathbf{s}}_z-\frac{L_z}{2}))},
\end{align}  
where ${\bm{\ell}}=[\ell_x,\ell_y,\ell_z]^{\mathsf{T}}\in{\mathbbmss{Z}}^{3\times1}$, ${\mathbf{s}}=[{\mathbf{s}}_x,{\mathbf{s}}_y,{\mathbf{s}}_z]^{\mathsf{T}}\in{\mathcal{A}}_{\mathcal{S}}$, and $L_x$, $L_y$, and $L_z$ denote the maximum projection lengths of ${\mathcal{A}}_{\mathcal{S}}$ on the $x$-, $y$-, and $z$-axis of three-dimensional (3D) coordinate system, respectively. Note that we have
\begin{align}
\int_{{\mathcal{A}}_{\mathcal{S}}}\psi_{\bm{\ell}}({\mathbf{s}})\psi_{\bm{\ell}}^{\mathsf{H}}({\mathbf{s}}){\rm{d}}{\mathbf{s}}=
\frac{1}{{\lvert{\mathcal{A}}_{\mathcal{S}}\rvert}}\int_{{\mathcal{A}}_{\mathcal{S}}}{\rm{d}}{\mathbf{s}}=1,\forall{\bm{\ell}}\in{\mathbbmss{Z}}^{3\times1}
\end{align} 
and 
\begin{align}
\int_{{\mathcal{A}}_{\mathcal{S}}}\psi_{\bm{\ell}}({\mathbf{s}})\psi_{{\bm{\ell}}'}^{\mathsf{H}}({\mathbf{s}}){\rm{d}}{\mathbf{s}}=0,\forall {\bm{\ell}}\ne {\bm{\ell}}'.
\end{align} 
Using the orthogonal basis constructed by $\{\psi_{\bm{\ell}}({\mathbf{s}})\}_{{\bm{\ell}}\in{\mathbbmss{Z}}^{3\times1}}$, we approximate the spatial response as follows:
\begin{align}
{{\mathsf{g}}(\mathbf{r}_k,\mathbf{s})}\approx \sum_{\ell_x=-\hat{L}_x}^{\hat{L}_x}\sum_{\ell_y=-\hat{L}_y}^{\hat{L}_y}\sum_{\ell_z=-\hat{L}_z}^{\hat{L}_z}
h_{k,\bm{\ell}}\psi_{\bm{\ell}}({\mathbf{s}}),
\end{align}
where $h_{k,\bm{\ell}}=\int_{{\mathcal{A}}_{\mathcal{S}}}{{\mathsf{g}}(\mathbf{r}_k,\mathbf{s})}\psi_{\bm{\ell}}^{\mathsf{H}}({\mathbf{s}}){\rm{d}}{\mathbf{s}}$, and $\{\hat{L}_x,\hat{L}_y,\hat{L}_z\}$ are complexity-vs-accuracy tradeoff parameters. Accordingly, the source current for each user $k$ is designed as follows:
\begin{align}
{\mathsf{j}}_k(\mathbf{s})=\sum_{\ell_x=-\hat{L}_x}^{\hat{L}_x}\sum_{\ell_y=-\hat{L}_y}^{\hat{L}_y}\sum_{\ell_z=-\hat{L}_z}^{\hat{L}_z}
w_{k,\bm{\ell}}\psi_{\bm{\ell}}^{\mathsf{H}}({\mathbf{s}}),\forall k.
\end{align} 
Taken together, problem \eqref{WSR_opt_CAP_space} can be converted as follows: 
\begin{subequations}\label{WDMA_CAP_Final_eQUAL_General}
  \begin{align}
    \max_{\mathbf{w}_{f,k}} \quad &\sum_{k=1}^K w_k \log_2 \left( 1 + \frac{|\mathbf{h}_{f,k}^{\mathsf{T}} \mathbf{w}_{f,k}|^2}{\sum_{i\neq k} |\mathbf{h}_{f,k}^{\mathsf{T}} \mathbf{w}_{f,i}|^2 + \frac{\overline{\sigma}^2}{(\eta_0\frac{2\pi}{\lambda})^2\lvert\mathcal{A}_{\mathsf{R}}\rvert} } \right) \\
    \mathrm{s.t.} \quad & \sum_{\ell_x=-\hat{L}_x}^{\hat{L}_x}\sum_{\ell_y=-\hat{L}_y}^{\hat{L}_y}\sum_{\ell_z=-\hat{L}_z}^{\hat{L}_z}\sum_{k=1}^K \lvert w_{k,{\bm\ell}}\rvert^2 \le P_{\mathsf{rad}},
  \end{align}
\end{subequations}
where $\mathbf{h}_{f,k}=[h_{k,{\bm\ell}}]_{\ell_x=-\hat{L}_x,\ell_y=-\hat{L}_y,\ell_z=-\hat{L}_z}^{\hat{L}_x,\hat{L}_y,\hat{L}_z}$ is a column vector that stores all the wavenumber-domain channel coefficients, and $\mathbf{w}_{f,k}=[w_{k,{\bm\ell}}]_{\ell_x=-\hat{L}_x,\ell_y=-\hat{L}_y,\ell_z=-\hat{L}_z}^{\hat{L}_x,\hat{L}_y,\hat{L}_z}$ is a column vector that stores all the wavenumber-domain beamforming coefficients. It is observed that problem \eqref{WDMA_CAP_Final_eQUAL_General} can be treated as an extension of problem \eqref{WDMA_CAP_Final_eQUAL_General} in the 3D scenario.

While the above method offers a relatively tractable solution, it remains quite complex. Initial efforts to approximate the solution of problem \eqref{WSR_opt_CAP_space} using this approach have been documented in \cite{sanguinetti2022wavenumber,zhang2023pattern,huang2024holographic}. However, a general framework for addressing \eqref{WSR_opt_CAP_space} under arbitrary near-field transmission propagation environments is yet to be established, making it a potential avenue for future research.

\section{Conclusions}\label{Sections: Conclusions}
This paper provided a tutorial-based primer on NFC in terms of multiuser communications and MA. We commenced by introducing several commonly used spherical-wave propagation-based channel models and their simplifications. Emphasis was placed on the low dimensionality and sparsity of a novel Fourier plane-wave propagation-based method. Moreover, we analyzed the information-theoretic capacity limits of near-field MACs and BCs and presented the sum-rate capacity and capacity regions. We demonstrated that NFC-MA can approach the capacity upper bound of spatial multiple-antenna systems, even when users are located in the same direction. Furthermore, we investigated and illustrated near-field multiuser beamforming design in both the conventional spatial domain and the novel wavenumber domain. We offered low-complexity designs by leveraging near-field transmission properties in these domains. Throughout the tutorial, we have highlighted the superiority of near-field RDMA over far-field ADMA in terms of flexible interference management, which is attributed to the additional range dimensions introduced by spherical-wave propagation. With these characteristics, NFC is positioned to pave the way towards realizing NGMA.
\bibliographystyle{IEEEtran}
\bibliography{ref}
\end{document}